\documentclass[mnsc,nonblindrev]{informs3_nojournal}
\usepackage{fix-cm}
\OneAndAHalfSpacedXI
\usepackage[utf8]{inputenc}
\usepackage{algorithmic}
\usepackage{algorithm}
\usepackage{subcaption}
\usepackage{csquotes}
\usepackage[section]{placeins}
\usepackage{dsfont}
\usepackage{enumitem}
\usepackage{fontenc}    % use 8-bit T1 fonts
\usepackage{url}            % simple URL typesetting
\usepackage{booktabs}       % professional-quality tables
\usepackage{amsfonts}       % blackboard math symbols
\usepackage{nicefrac}       % compact symbols for 1/2, etc.
\usepackage{natbib}
 \bibpunct[, ]{(}{)}{,}{a}{}{,}%
 \def\bibfont{\small}%
\allowdisplaybreaks

\TheoremsNumberedThrough     % Preferred (Theorem 1, Lemma 1, Theorem 2)
\ECRepeatTheorems

\EquationsNumberedThrough    % Default: (1), (2), ...
\MANUSCRIPTNO{}

\begin{document}
%%%%%%%%%%%%%%%%

% Outcomment only when entries are known. Otherwise leave as is and
%   default values will be used.
%\setcounter{page}{1}
%\VOLUME{00}%
%\NO{0}%
%\MONTH{Xxxxx}% (month or a similar seasonal id)
%\YEAR{0000}% e.g., 2005
%\FIRSTPAGE{000}%
%\LASTPAGE{000}%
%\SHORTYEAR{00}% shortened year (two-digit)
%\ISSUE{0000} %
%\LONGFIRSTPAGE{0001} %
%\DOI{10.1287/xxxx.0000.0000}%

% Author's names for the running heads
% Sample depending on the number of authors;
% \RUNAUTHOR{Jones}
% \RUNAUTHOR{Jones and Wilson}
% \RUNAUTHOR{Jones, Miller, and Wilson}
% \RUNAUTHOR{Jones et al.} % for four or more authors
% Enter authors following the given pattern:
\RUNAUTHOR{Elmachtoub, Kanoria, and Tan}

% Title or shortened title suitable for running heads. Sample:
% \RUNTITLE{Bundling Information Goods of Decreasing Value}
% Enter the (shortened) title:
\RUNTITLE{Joint Service Depth and Priority Design in Queueing Systems}

% Full title. Sample:
% \TITLE{Bundling Information Goods of Decreasing Value}
% Enter the full title:
\TITLE{Joint Service Depth and Priority Design in Queueing Systems}

% Block of authors and their affiliations starts here:
% NOTE: Authors with same affiliation, if the order of authors allows,
%   should be entered in ONE field, separated by a comma.
%   \EMAIL field can be repeated if more than one author
\ARTICLEAUTHORS{%
\AUTHOR{Adam N. Elmachtoub}
\AFF{Department of Industrial Engineering and Operations Research and Data Science Institute, Columbia University, New York, NY 10027, \EMAIL{adam@ieor.columbia.edu}}
\AUTHOR{Yash Kanoria}
\AFF{Graduate School of Business, Columbia University, New York, NY 10027, \EMAIL{ykanoria@columbia.edu}}
\AUTHOR{Jonathan Y. Tan}
\AFF{Department of Industrial Engineering and Operations Research, Columbia University, New York, NY 10027, \EMAIL{jyt2123@columbia.edu}}
% Enter all authors
} % end of the block

\ABSTRACT{In many service systems, including data centers serving AI jobs, the server can choose the amount of processing effort, which we call the service depth. Deeper reasoning can improve output quality, but it also consumes computing resources for longer and increases congestion. The realized processing time depends on the chosen service depth, but remains random because AI models are probabilistic and tasks vary in complexity. Motivated by these applications, we study the joint design of service depth and priority in a single-server queue with customers who differ in delay sensitivities, under a welfare-maximization objective. The decision maker determines, for each customer type, both a service priority and a service depth, where service times are exponentially distributed with means equal to the chosen service depths.
    
We consider a setting with two types of customers and a setting with a continuum of types of customers. In the setting with two types, we characterize the optimal joint service depth and priority assignment. We find that, in the centralized setting, counterintuitively, the decision maker's optimal policy prioritizes the patient type if a sufficiently large fraction of customers are patient. In the decentralized setting, the optimal incentive compatible price menu never prioritizes the patient type. Consequently, there can be a welfare gap between the centralized decision maker’s optimal control policy and the optimal decentralized pricing mechanism. In the setting with a continuum of types, we show that the centralized optimum is achieved by a service depth assignment that allocates weakly smaller service depth and gives priority to more delay-sensitive customers. As a result, the optimal allocation can be implemented by an incentive compatible pricing menu under strategic customers with private delay sensitivities. We then calibrate an agentic AI service system using published coding-agent trials from DeepSWE and numerically examine priority inversion, the value of joint design, and the welfare gain from personalization.
}

% Fill in data. If unknown, outcomment the field
%\HISTORY{This paper was submitted on XXXX, 2017.}

\maketitle
% \vspace{-25pt}

% Text of your paper here

\section{Introduction}

Many modern service systems decide not only \emph{which} customer to serve next, but also \emph{how much} service to provide for each job. Completion is not governed by a fixed service requirement; instead, the provider has discretion over how much time, attention, or computation to devote to processing the job. In such systems, service effort is both value-producing and capacity-consuming: a deeper diagnosis, a more careful customization, or a longer computational search generates greater value for the customer, but it also occupies the service capacity and delays other customers \citep{deshmukh1977capacity, anand2011quality}. We refer to this controllable processing effort as the \emph{service depth}. Service depth directly enters customers' utility, so it can also affect customers' strategic behavior and contract choices. 

AI-agent service systems provide a contemporary motivation for this problem. With the rapid growth of Large Language Models (LLMs) and Generative AI (GenAI), AI agents can now act as domain experts in specialized fields and be applied in many Internet service systems, ranging from customer support to software development \citep{huang2024understanding, sapkota2025ai}. These systems offer a new form of operational flexibility. Unlike human service providers, whose skills, availability, and task assignments are usually difficult and costly to adjust, it is possible for AI-agent service systems to adjust the service depth (reasoning effort) and switch between different types of tasks at lower cost, enabling service systems to respond differently to changing task requirements.

However, AI agents introduce several important operational challenges. LLM requests are heterogeneous and difficult to predict in advance, making it difficult for centralized planning and control \citep{mitzenmacher2025queueing}. In addition, the quality of LLM output is highly dependent on the processing time, introducing a trade-off between the service quality and the congestion of the system: faster responses may reduce waiting times and operating costs, whereas deeper reasoning requires longer processing time but yields higher accuracy and better service outcomes \citep{wei2022chain, snell2025scaling}.

Classical priority rules rank jobs using fixed primitives, such as delay costs and service requirements, which may be suboptimal when service depth is endogenous. This motivates a joint design problem that has received comparatively little attention:
\begin{center}
    \emph{How can we develop efficient policies to control the service depth and priority in service systems?}
\end{center}

We investigate this question through a stylized queueing and mechanism design model. Specifically, we model the system as a preemptive queue in which both service depth and service priority are controllable variables. Customers arrive to a single-server queue according to a Poisson process and differ in their delay sensitivities, represented by the delay cost they experience. For each customer type, the decision maker determines the service depth to provide and the priority with which its jobs are served. Service depth assignments are type-dependent and do not depend on the current state of the queue. Conditional on a chosen service depth $x$, the realized service time is exponentially distributed with mean $x$, and the service value generated is $V(x)$. A larger service depth generates a greater service value, but also increases workload and delays other customers. Thus, service depth captures the server's ability to adjust processing effort, while preemption captures scheduling flexibility. This framework isolates the key operational primitives that arise in AI-agent services and many other service systems: heterogeneous demand, flexible service allocation, and endogenous service depth.

Throughout the paper, the objective is to maximize long-term average social welfare, defined as service value net of delay costs. We study two informational regimes. In the centralized benchmark, the decision maker observes each customer's type and directly assigns service depth and priority. In the decentralized setting, the delay sensitivity of a customer is private information. The provider then offers a menu of service depth, priority, price options, and customers choose the options that maximize their expected surplus from the menu. Therefore, we focus on finding the optimal menu that satisfies the incentive compatibility and individual rationality constraints.

\subsection{Main Contributions}
Our key findings can be summarized as follows:
\begin{enumerate}
    \item In a two-type model, we characterize the welfare-maximizing policy for the centralized decision maker. For linear service values, the optimal policy has three regimes: the more delay-sensitive type may be excluded, the more delay-sensitive type may be prioritized, or the less delay-sensitive type may be prioritized. The last regime is counterintuitive: when a sufficiently large fraction of customers are patient, the centralized decision maker optimally assigns them a smaller service depth and serves them first. Thus, the optimal priority order is not determined by delay sensitivities alone; it also depends on the prevalence of each type. We further show that such a phenomenon exists for nonlinear concave value functions as well.
    \item In the two-type model with linear service values, we characterize the optimal menu when delay sensitivity is private information. Incentive compatibility requires the more delay-sensitive type to receive a weakly shorter sojourn time, and any service policy that satisfies this property is implementable in an incentive compatible way. Therefore, a centralized allocation that prioritizes the less delay-sensitive type cannot be implemented by a self-selecting price menu. Hence, there is a welfare gap between the centralized and decentralized optimal policies. We further derive an upper bound of the welfare gap based on worst-case analysis: the optimal decentralized menu achieves at least $75\%$ of the centralized welfare, and this bound is tight.
    
    \item In a continuous-type model, we show that the welfare-maximizing policy for the centralized decision maker has a monotone structure: more delay-sensitive customers receive weakly smaller service depth and higher priority. This monotonicity is exactly the ordering required for incentive compatibility: more delay-sensitive customers also receive weakly shorter sojourn times. Hence, whenever the centralized optimum is attained, it can be implemented by an incentive-compatible pricing menu. This contrasts with the two-type case and shows that finer control over service depth assignment can align the operational priority order with decentralized implementation.

    \item We complement the theory with a case study calibrated from GPT-5.6 Sol trials on DeepSWE at five reasoning-effort levels. The two-type experiments exhibit a welfare loss from incentive compatibility of up to $53.4\%$. With many customer types, the experiments show how the value of priority and depth changes with congestion: priority alone can match joint design at lower load, while personalized depth sustains service near capacity. Personalization also produces an implementable allocation with higher welfare than the coarse centralized optimum.
\end{enumerate}

Although we focus on maximizing social welfare in this paper, our analysis in the two-type model and for regular distributions in the continuous-type model also applies to revenue maximization. The general idea is, under incentive compatibility, the revenue objective can be rewritten as a welfare-like objective in which the true delay sensitivities are replaced by virtual delay sensitivities. We provide an analysis for revenue maximization in Section~\ref{appendix:revenue_maximization}.

The remainder of this paper is organized as follows. Section~\ref{subsection:literature_review} reviews the literature. Section~\ref{section:model} introduces our model and states our two different settings: the setting where delay sensitivities are known to the decision maker, and the setting where delay sensitivities are customers' private information. Section~\ref{section:two-type setting} studies the two-type model where delay sensitivities follow a two-point distribution. Section~\ref{section:continuous-type setting} studies the continuous-type model where delay sensitivities follow a continuous distribution. Section~\ref{section:numerical_results} presents some numerical examples on different value functions. Section~\ref{section:case_study} presents a case study of an agentic AI service system calibrated from published software-engineering trials. Section~\ref{section:extensions} presents some extensions of the model. Section~\ref{section:conclusion} concludes the paper and provides directions for future work.

\section{Literature Review}
\label{subsection:literature_review}

Our study bridges three topics in queueing systems: service rate control, priority control and pricing, and queueing for AI systems.

\paragraph{Service rate control.}
This paper contributes to the queueing literature on service rate control. Classical service rate control studies how a decision maker should adjust server speed, service capacity, or effort in response to congestion. \citet{weber1987optimal} analyze optimal service rate control in queueing networks, showing that the optimal service rate is monotone with respect to the congestion level. This is further extended by a number of papers establishing similar monotonicity results \citep{stidham1989monotonic, veatch1992monotone, george2001dynamic}. 

A related line of work studies service control and pricing when customers are self-interested. \citet{ha1998incentive} provides an incentive compatible pricing mechanism via steady-state analysis in a setting where customers choose their service times, and this framework is further extended to a multiclass setting in which customers are heterogeneous \citep{ha2001optimal}. Regarding the real-time control setting, \citet{ata2006dynamic} study dynamic control of arrival and service rates and characterize the optimal pricing policy with respect to the length of the queue.

It is worth mentioning that these prior works differ fundamentally from our setting. In the existing literature, faster service is obtained by adding capacity at an operating cost, and the service rate is primarily a tool for managing system congestion. In contrast, our model has no explicit cost of changing service depth. Instead, choosing faster service entails an opportunity cost: it lowers the service depth and therefore reduces the value delivered to the customer. Service depth directly affects the utility received by customers, affecting their behavior and creating a trade-off between service quality and congestion. More closely related to our setting, \citet{anand2011quality} consider the model in which adjusting service speed affects customers' utility, and \citet{lin2023wait} study wait-time-based pricing in queues with customer-chosen service times, where customer utility depends on service time.
 
\paragraph{Priority control and design.}
Priority control studies how the decision maker should schedule the service order of different tasks in congested systems. \citet{cobham1954priority} is one of the earliest papers on priority assignment in waiting-line systems. \citet{kleinrock1965conservation} derives a conservation law for a broad class of queueing disciplines, showing how expected waiting times can be redistributed across customer classes under work-conserving policies. \citet{coffman1980characterization, shanthikumar1992multiclass} further characterize the achievable-region of waiting times in queueing system. Regarding the optimal priority allocation, \citet{smith1956various} shows that $c\mu$-rule is optimal in a deterministic, static environment, and is extended to different queueing settings by \citet{klimov1975time, van1995dynamic}. For more general settings in M/G/1 queues, \citet{scully2021gittins} provide a general proof showing that the Gittins Policy is optimal.

When customers have private information, priority auctions or pricing policies have been developed to construct efficient priority allocations. \citet{kleinrock1967optimum} is one of the first papers to study priority auction. Each customer bids a price, and will be placed into a position in front of all customers with lower bid and behind all customers with higher bid. \citet{lui1985equilibrium} extended the model by deriving customers' bid functions in equilibrium. \citet{afeche2004pricing} further study social welfare- and revenue-maximizing auctions under generalized delay cost, and \citet{kittsteiner2005priority} study priority auctions when the job processing time is private information. More recently, \citet{yang2017trading} study how to design mechanisms for customers to trade their positions in the queue. Regarding priority pricing in queues, \citet{mendelson1990optimal} provide an optimal mechanism assuming customers with private information on expected service times and waiting sensitivities. Following this, \citet{van2000price} develops incentive compatible scheduling rules under more general delay-cost curves. \citet{afeche2013incentive, afeche2016optimal} consider revenue maximizing policies, showing that strategically delaying patient customers can increase the seller's revenue under certain conditions. \citet{lin2024dynamic} show strategic delay can also increase the revenue in a dynamic setting. \citet{freund2024dedicated} analyze the Dedicated Docket in U.S. immigration courts, which creates a separate fast-track queue for certain immigration proceedings. \citet{afeche2019pricing} show that demand-frequency heterogeneity can make it optimal to prioritize one customer class over another even when the two classes have identical waiting costs. Their results provide a similar insight to ours: delay sensitivity is not the only driver of priority. 

The key difference between our model and priority design literature is that service requirements are typically primitives, whereas in our model they are chosen by the decision-maker. As a result, the priority order is endogenous: changing service depth changes both the congestion externality and the priority order.

\paragraph{Queueing for AI Systems.}
There is growing interest in LLM scheduling and queueing theory for AI systems. \citet{zheng2023response, fu2024efficient, bari2025optimal} develop algorithms to efficiently schedule different jobs for LLM inference. \citet{mitzenmacher2025queueing} survey recent work on queueing systems with predictions and discuss open problems in LLM scheduling, emphasizing predicted service times, variable inference durations, memory constraints, and preemption. \citet{li2025throughput, ao2025optimizing} model LLM inference into a queueing problem, providing algorithms to achieve optimal or near-optimal throughput. \citet{lee2024design, legros2026human, yang2026design} consider human-AI interaction in queueing systems, providing insights on how AI impacts congestion.

Recent work at the intersection of AI service operations and pricing studies how providers route requests, allocate service capacity, and design contracts under quality uncertainty and congestion. \citet{guo2026congestion} formulate the allocation of requests across LLM models as a routing problem and characterize the optimal routing policy. \citet{lobel2026pricing} study pricing AI as a service, and compare two different pricing mechanisms: conversation-based pricing and outcome-based pricing. \citet{feldman2026pricing} further consider the pricing problem in human-AI services, showing the relationship between congestion, human capacity, AI capability and AI observability. More closely related to our setting, \citet{abdolmaleki2026pricing} study a queueing model for agentic AI service systems in which customers differ in their valuations of service. They show that, under the optimal mechanism, higher-value jobs receive weakly longer sojourn times and more inference-time compute. Our model differs from theirs in several important aspects. First, we model customer heterogeneity through delay sensitivities rather than service valuations. Second, they assume that service quality is determined directly by the realized service time, whereas we introduce service depth as a decision variable and model the realized service time as a random outcome conditional on the chosen service depth. Finally, they use a fluid approximation to study the structure of the optimal policy, whereas we derive exact structural characterizations through steady-state analysis.

\section{Problem Formulation}
\label{section:model}

\subsection{Queueing Framework}
\label{subsection:queueing_framework}

We consider a single-server queue where customers arrive according to a Poisson process with arrival rate $\lambda>0$. Each customer brings a task that will be processed by the server. Each customer is assigned a service depth $x\in[0,\infty)$, representing the amount of effort devoted to processing the task. A larger service depth corresponds to more intensive processing and potentially higher service quality. We assume that when $x>0$, the actual service time is exponentially distributed with a service rate $\mu=\frac{1}{x}$, so that the expected service time equals $x$. The case $x=0$ represents no service: the customer leaves the system immediately and receives zero service.

A customer assigned service depth $x$ generates service value $V(x)$ upon completion. Notice that the value function $V(x)$ depends on the service depth, not the actual service time of a customer. We assume that $V:[0,\infty)\to\mathbb{R}_{+}$ is (weakly) increasing, (weakly) concave, continuous and twice-differentiable on $(0,\infty)$, with $V(0)=0$. These assumptions capture that service value is nonnegative and that larger service depth generates weakly higher service value, with weakly diminishing returns.

Customers are price- and delay-sensitive and heterogeneous in their delay sensitivities. A customer’s type is represented by $\theta$, which denotes the cost per unit time while the customer is present in the system. We assume that $\theta$ is drawn independently from a known distribution $F$. Let $\Theta$ denote the set of possible delay sensitivities among customers. In Section~\ref{section:two-type setting}, we assume that $\Theta=\{\theta_1,\theta_2\}$, and $\theta$ follows a two-point distribution, while in Section~\ref{section:continuous-type setting}, we assume that $\Theta=[\underline{\theta},\bar{\theta}]$, and $\theta$ follows a continuous distribution.

In this paper, we focus on the social welfare maximization problem. Social welfare is the long-term average service value generated by the system, net of customers' delay costs. For any feasible policy, let $x(\theta)$ and $w(\theta)$ denote the service depth and expected sojourn time received by type $\theta$ under that policy. If type $\theta$ does not enter the system, we set $x(\theta)=w(\theta)=0$. The social-welfare objective is then
\begin{equation*}
    \mathcal{W}:=\lambda\mathbb{E}_{\theta}\big[V(x(\theta))-\theta w(\theta)\big].
\end{equation*}
Welfare maximization chooses a stable and operationally achievable allocation to maximize $\mathcal{W}$. We assume that the arrival rate $\lambda$, the value function $V(x)$ and the distribution of customers' types $F$ are public information.

\subsection{Centralized Benchmark}
We first study a centralized benchmark in which the decision maker directly observes customer types and implements a stationary policy along two dimensions:
\begin{enumerate}
    \item \textbf{Service depth assignment.} The centralized decision maker selects a type-dependent service depth to each type of customers. We restrict attention to assignments that depend only on customer type and not on the current queue state.
    \item \textbf{Priority rule.} The centralized decision maker selects a scheduling policy governing which customer is served at each point in time. We allow a broad class of preemptive scheduling policies, including pure priority rules, randomized scheduling policies, and non-work-conserving policies that may intentionally idle the server or strategically delay certain customer types.
\end{enumerate}
We emphasize that our formulation of the priority rule is more general than pure priority orderings. For example, when two customer types are present, the decision maker may implement FCFS, randomize over which customer type receives priority, or intentionally delay serving one type even when the server is available. Pure policies that always prioritize one type over the other are special cases of our formulation. Note that under our assumption of service time being exponential in $x$, the remaining service times for all jobs of a given customer type are i.i.d., and without loss of generality we assume that the server serves them in the first-come, first-served order. 

Although the primitive decision includes the priority rule, the priority rule affects the objective only through the expected sojourn times it induces. Therefore, the decision maker's allocation can be represented by two type-dependent functions:
\begin{equation*}
    x:\Theta \to [0,\infty), \quad w:\Theta \to [0,\infty).
\end{equation*}
The function $x(\cdot)$ maps a customer with delay sensitivity $\theta$ to a chosen service depth $x(\theta)$, and $w(\cdot)$ maps a customer with delay sensitivity $\theta$ to their steady-state expected sojourn time $w(\theta)$. This representation can be interpreted as follows. The decision maker directly chooses the service depth assignment $x(\cdot)$ and a priority rule. Together, these two choices generate a sojourn time function $w(\cdot)$. For social welfare, the detailed identity of the priority rule matters only through this induced function, so we can equivalently optimize over pairs $(x,w)$, provided that $w$ is consistent with some priority rule under the chosen service depths.

Under this representation, we must ensure that the pair $(x,w)$ is feasible. First, the service depth assignment must satisfy the stability condition
\begin{equation*}
    \lambda \mathbb{E}_{\theta}[x(\theta)] < 1,
\end{equation*}
so that the long-term workload brought to the server is less than its capacity. We also require the second moment
\begin{equation*}
    \mathbb{E}_{\theta}[x^2(\theta)] < \infty
\end{equation*}
to ensure sojourn times for every type of customers are finite. Such conditions are stated as Assumption~\ref{assumption:continuous_types} for the continuous-type model. Moreover, the sojourn-time function $w(\cdot)$ must be operationally feasible: there must exist a priority rule such that, under the service-depth assignment $x(\cdot)$, the induced expected sojourn time of type-$\theta$ customers is exactly $w(\theta)$. We characterize this by the achievable region approach developed by \citet{coffman1980characterization}.

\subsection{Decentralized Mechanism}
We next consider a decentralized setting in which a customer's type is private information. In this setting, the decision maker can set service prices to influence customers' behavior. For a type-$\theta$ customer, let $w$ denote the expected sojourn time of the customer and $p$ denote the service price paid by the customer. Given a service depth $x$, the customer receives expected surplus
\begin{equation*}
    V(x)-\theta w-p.
\end{equation*}
We allow the payment $p$ to be negative, representing a transfer from the server to the customer. Prices do not enter the welfare objective because they are transfers between customers and the service provider.

When the decision maker does not observe $\theta$ directly, we adopt a direct-mechanism approach. The decision maker offers a menu of contracts $\mathcal{M}=\{(p(\theta),x(\theta),w(\theta)):\theta\in\Theta\}$ from which customers select. Upon arrival, the customer observes the menu and reports their delay-sensitivity $\hat\theta$, and then the server assigns the corresponding option $(p(\hat\theta),x(\hat\theta),w(\hat\theta))$ to them. We assume that customers do not observe the current state of the queue and make their decisions solely based on their expected surplus.

Similar to the centralized policy, the menu must be operationally achievable. Moreover, the menu must be truthfully implementable. We impose these requirements as follows. First, under truthful reporting, the induced queue must be stable. Second, under truthful reporting, the announced sojourn time $w(\theta)$ in the menu must coincide with the steady-state expected sojourn time of customers who report $\theta$. Third, truthful reporting must be an equilibrium of the induced steady-state system. More precisely, customers make their decisions based on the announced sojourn times, and these decisions must be the best responses to the induced steady state, in which the announced sojourn times are exactly the steady-state expected sojourn times generated by the given service-depth assignment and priority rule. Such consistency requirements are standard in the literature \citep{afeche2013incentive, afeche2016optimal}. To ensure participation and truthful reporting, the menu must satisfy the individual rationality (IR) and incentive compatibility (IC) constraints:
\begin{equation*}
    \begin{aligned}
        &V(x(\theta))-\theta w(\theta) - p(\theta)\geq 0,\quad \forall \theta\in\Theta,\\
        &V(x(\theta))-\theta w(\theta) - p(\theta)\geq V(x(\hat\theta))-\theta w(\hat\theta) - p(\hat\theta), \quad \forall \theta,\hat\theta\in\Theta.
    \end{aligned}
\end{equation*}
If a customer is indifferent among multiple reports, we assume that they report their type truthfully.

\section{Two-Type Model}
\label{section:two-type setting}

In this section, we assume there are two types of customers in the system, indexed by $i\in\{1,2\}$. Type $1$ is more delay sensitive than type $2$, with delay cost parameters $\theta_1>\theta_2>0$; equivalently, type $1$ can be interpreted as the impatient type and type $2$ as the patient type. For notational simplicity, we write $x_i=x(\theta_i)$, $w_i=w(\theta_i)$, $p_i=p(\theta_i)$ for $i=1,2$. Customer types are i.i.d., with probability $q$ of being the impatient type and probability $1-q$ of being the patient type. Accordingly, the type-specific arrival rates are
\begin{equation*}
    \lambda_1=q\lambda, \quad \lambda_2=(1-q)\lambda.
\end{equation*}
All proofs in this section are deferred to Appendix~\ref{appendix:section_3}.

First, we characterize all feasible policies in the two-type model. For a given service depth assignment $(x_1,x_2)$, let $\rho_i := \lambda_ix_i$ denote the workload generated by type $i$ customers. The service depth assignment is stable only if $\rho_1+\rho_2<1$. To mathematically characterize all the priority rules, we directly apply Theorem~$2$ from \cite{coffman1980characterization} that gives all the operationally feasible sojourn times $(w_1, w_2)$:
\begin{equation*}
    \begin{aligned}
        & w_i\geq\frac{x_i}{1-\rho_i},\quad i=1,2,\\
        & \rho_1 w_1+\rho_2 w_2 \geq \frac{\rho_1x_1+\rho_2x_2}{1-\rho_1-\rho_2}.
    \end{aligned}
\end{equation*}

\subsection{Centralized Optimum}
The centralized welfare-maximization problem can be written as
\begin{equation}
    \label{eq:welfare_maximization}
    \begin{aligned}
        \max_{x_1,x_2,w_1,w_2}\quad & \lambda_1\left(V(x_1)-\theta_1 w_1\right)+\lambda_2\left(V(x_2)-\theta_2 w_2\right) \\
        \text{s.t.}\quad
        &w_i\geq\frac{x_i}{1-\rho_i},\quad i=1,2,\\
        & \rho_1 w_1+\rho_2 w_2 \geq \frac{\rho_1x_1+\rho_2x_2}{1-\rho_1-\rho_2}, \\
        & \rho_1+\rho_2<1,
    \end{aligned}
\end{equation}
where the first three constraints ensure the sojourn time is operationally feasible and the last constraint ensures the queue is stable.

Lemma~\ref{lemma:cmu_rule} states the classic result that $c\mu$-rule is the optimal priority rule for an M/M/1 queue with fixed service depth and delay sensitivities, which was first developed by \citet{cox1961queues}.
\begin{lemma}[$c\mu$-rule]
    \label{lemma:cmu_rule}
    Fix an admissible service depth assignment $(x_1,x_2)$. Among all feasible  priority rules, an optimal rule is work-conserving and gives priority to the type with the smaller ratio $x_i/\theta_i$.
\end{lemma}
The classic result applies because the only thing remaining once the service depths are fixed is to minimize delay costs. It is classical in form but new in its implication here. In standard priority models, service requirements are primitives, so the index ranks fixed types. In our setting, Lemma~\ref{lemma:cmu_rule} implies that, once $(x_1,x_2)$ is fixed, the priority rule is pinned down endogenously to one of the two pure priority orderings, and the expected sojourn times are functions of $(x_1, x_2)$. Hence the centralized problem can be analyzed by comparing the two candidate priority regimes induced by the service depth assignment. For simplicity, when the two types are tied under the $c\mu$-rule, we assume type $1$ customers receive priority. 

If service depths were fixed exogenously, the $c\mu$-rule would prioritize the type with smaller $x_i/\theta_i$. Since type~$1$ customers have larger delay sensitivity, one would expect type~$1$ customers to receive priority whenever service depths are comparable. However, this intuition no longer holds when service depth is an endogenous decision variable. Proposition~\ref{proposition:welfare_maximization_centralized_case} shows the characterization of the welfare-maximizing policy when the value function $V(x)$ is linear.

\begin{proposition}
    \label{proposition:welfare_maximization_centralized_case}
    Suppose $V(x)=ax$ with $a>\theta_1>\theta_2$. The centralized optimum has the following structure:
    \begin{enumerate}
        \item If $\theta_1^2\ge a\theta_2$, then it is optimal to set $x_1^*=0$ and $x_2^*=\frac{1-\sqrt{\theta_2/a}}{\lambda_2}$.
        \item If $\theta_1^2<a\theta_2$, then both types receive strictly positive and finite service depth. Moreover, there exists a unique threshold $c\in(0,+\infty)$ such that when $\frac{\lambda_2}{\lambda_1}\le c$, the optimal service depth assignment $(x_1^*, x_2^*)$ satisfies
        \begin{equation}
            \label{equation:prop1}
            \begin{aligned}
                &a\theta_2(1-\lambda_1x_1^*)\left(\left(1+\frac{\lambda_2}{\lambda_1}\right)(1-\lambda_1x_1^*)^2-\frac{\lambda_2}{\lambda_1}\right)^2=\theta_1^2\left(\lambda_1\lambda_2{x_1^*}^2+(1-\lambda_1x_1^*)^2\right),\\
                &x_2^*=\frac{1}{\lambda_2}\left[1-\frac{\theta_2}{\theta_1}+\left(2\frac{\theta_2}{\theta_1}(\lambda_1+\lambda_2)-\lambda_1\right)x_1^*-\frac{\theta_2}{\theta_1}\lambda_1(\lambda_1+\lambda_2){x_1^*}^2\right],
            \end{aligned}
        \end{equation}
        and type $1$ customers are prioritized; when $\frac{\lambda_2}{\lambda_1}>c$, the optimal service depth assignment $(x_1^*, x_2^*)$ satisfies
        \begin{equation*}
            \begin{aligned}
                &a\theta_1(1-\lambda_2x_2^*)\left(\left(1+\frac{\lambda_1}{\lambda_2}\right)(1-\lambda_2x_2^*)^2-\frac{\lambda_1}{\lambda_2}\right)^2=\theta_2^2\left(\lambda_1\lambda_2{x_2^*}^2+(1-\lambda_2x_2^*)^2\right),\\
                &x_1^*=\frac{1}{\lambda_1}\left[1-\frac{\theta_1}{\theta_2}+\left(2\frac{\theta_1}{\theta_2}(\lambda_1+\lambda_2)-\lambda_2\right)x_2^*-\frac{\theta_1}{\theta_2}\lambda_2(\lambda_1+\lambda_2){x_2^*}^2\right],
            \end{aligned}
        \end{equation*}
        and type $2$ customers are prioritized. Moreover, the threshold $c$ is decreasing with respect to $a$, and $\lim_{a\rightarrow \theta_1^2/\theta_2}c=\infty$.
    \end{enumerate}
\end{proposition}

\begin{figure}[htbp]
\centering
\caption{Visualization of Proposition~\ref{proposition:welfare_maximization_centralized_case}.}

\begin{subfigure}[t]{0.48\textwidth}
    \centering
    \caption{Different Regions of the Optimal Policy.}
    \includegraphics[width=\textwidth]{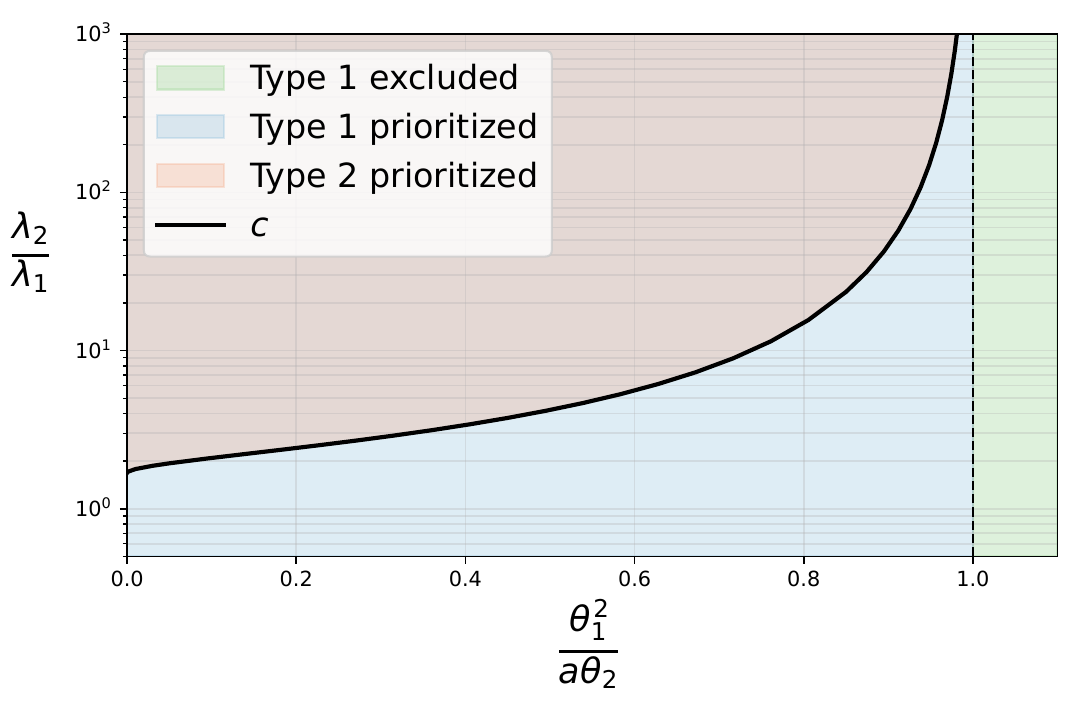}
    \label{figure:prop1_1}
    \vspace{0.3em}
    \begin{minipage}{\textwidth}
    \footnotesize
    {Note: The parameters are $\lambda=1.0$, $\theta_1=0.7$ and $\theta_2=0.6$. We adjust $\frac{\theta_1^2}{a\theta_2}$ by controlling $a$, and adjust $\frac{\lambda_2}{\lambda_1}$ by using different $\lambda_1$.}
    \end{minipage}
\end{subfigure}
\hfill
\begin{subfigure}[t]{0.48\textwidth}
    \centering
    \caption{Welfare under Different Priority Rules.}
    \includegraphics[width=\textwidth]{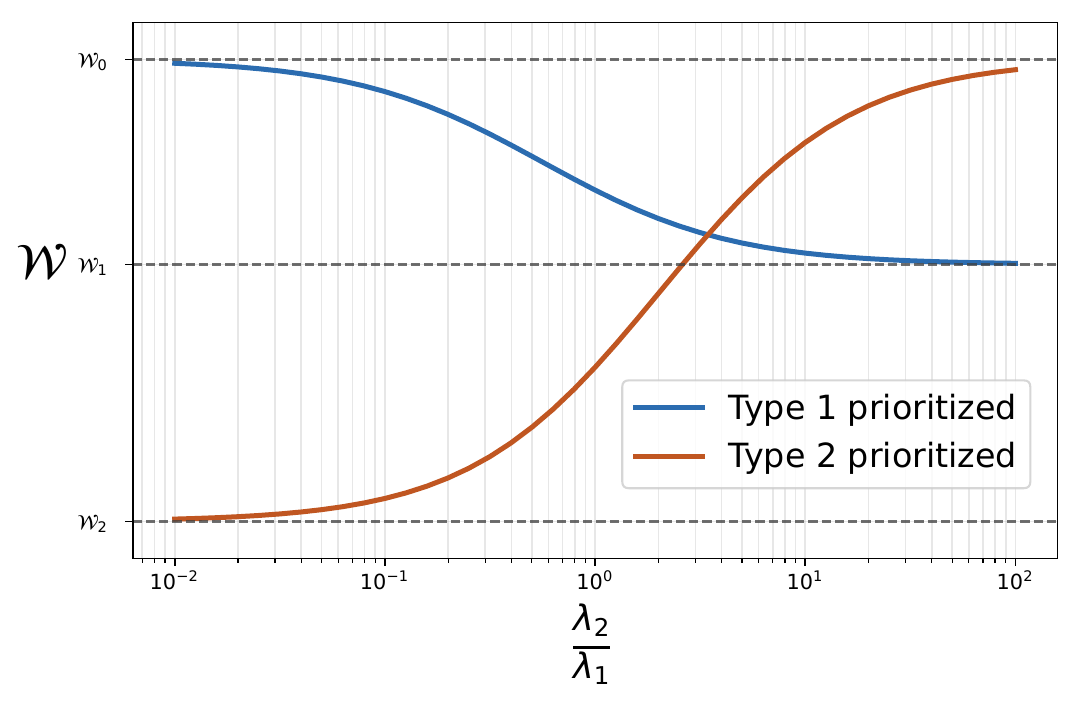}
    \label{figure:prop1_2}
    \vspace{0.3em}
    \begin{minipage}{\textwidth}
    \footnotesize
    Note: The parameters are $\lambda=1.0$, $a=2.0$, $\theta_1=0.7$ and $\theta_2=0.6$. We adjust $\frac{\lambda_2}{\lambda_1}$ by using different $\lambda_1$. Here, $\mathcal{W}_0=a+\theta_1+\theta_2-3\sqrt[3]{a\theta_1\theta_2}=0.469$, $\mathcal{W}_1=a+\theta_2-2\sqrt{a\theta_2}=0.409$, $\mathcal{W}_2=a+\theta_1-2\sqrt{a\theta_1}=0.334$.
    \end{minipage}
\end{subfigure}

\label{fig:prop1}

\end{figure}

Proposition~\ref{proposition:welfare_maximization_centralized_case} characterizes three different regions of the optimal policy, which is also illustrated in Figure~\ref{figure:prop1_1}. When the delay sensitivity of impatient customers is too high (the green region in Figure~\ref{figure:prop1_1}), it is no longer valuable to keep them in the system, therefore type $1$ customers receive zero service depth; when the delay sensitivity of impatient customers is lower than a certain threshold, it is still valuable to keep them in the system. In this regime, however, the resulting priority structure splits into two distinct cases. When patient customers are scarce (the blue region in Figure~\ref{figure:prop1_1}), the intuitive ranking prevails: the impatient type receives priority. However, when patient customers are sufficiently abundant (the red region in Figure~\ref{figure:prop1_1}), the decision maker optimally gives the patient type smaller service depth and serves them first. Figure~\ref{figure:prop1_2} further illustrates the reason why such a threshold exists when $\theta_1^2<a\theta_2$. Each curve represents the maximum social welfare achievable under a given priority rule, where service depths are chosen optimally. As $\frac{\lambda_2}{\lambda_1}$ increases, the value of prioritizing type $2$ customers increases, while the value of prioritizing type $1$ customers decreases. Consequently, when $\frac{\lambda_2}{\lambda_1}$ is above a certain threshold, social welfare is larger if patient customers are prioritized.

Proposition~\ref{proposition:welfare_maximization_centralized_case} shows that optimal policy is not only determined by the delay sensitivities alone, but also highly related to the size of each customer type. The intuition is that the population size of each customer type plays a crucial role in determining congestion externalities. Increasing the service depth of a high-volume type generates a large increase in aggregate delay throughout the system, i.e., there will be a lot of customers with larger service depth assignments waiting in the queue. As a result, the decision maker has a greater tendency to assign lower service depth to high-volume types. If the patient type is prevalent enough, this service-depth reduction can make its $c\mu$-index larger than that of the impatient type, leading the planner to prioritize the patient type.

Proposition~\ref{proposition:welfare_maximization_centralized_case} assumes linearity of the value function, but provides insight into when patient customers are prioritized in the optimal centralized policy: impatient customers are valuable to receive service but have relatively small population. Proposition~\ref{proposition:welfare_maximization_centralized_case_general_u} shows that, for nonlinear value functions, similar structure holds.

\begin{proposition}
    \label{proposition:welfare_maximization_centralized_case_general_u}
    Let $\hat{\rho}$ denote the welfare-maximizing workload when all the customers have delay sensitivity $\theta_2$, i.e.,
    \begin{equation*}
        \hat{\rho}=\argmax_{0\leq\rho<1}\left\{\lambda V\left(\frac{\rho}{\lambda}\right)-\theta_2\frac{\rho}{1-\rho}\right\}.
    \end{equation*}
    If the asymptotic value per unit depth $\lim_{x\rightarrow\infty} \frac{V(x)}{x}$ satisfies
    \begin{equation}
        \label{equation:prop2}
        \lim_{x\rightarrow\infty} \frac{V(x)}{x}>\frac{\theta_1}{1-\hat\rho},
    \end{equation}
    then there exists $c\in(0,+\infty)$ such that if $\frac{\lambda_2}{\lambda_1}>c$, the centralized optimum gives priority to type $2$ customers.
\end{proposition}

Since $V(x)$ is non-negative, weakly increasing and weakly concave, $V(x)/x$ is non-negative and weakly decreasing. Therefore, $ \lim_{x\rightarrow\infty} V(x)/x$ exists. Condition~\eqref{equation:prop2} imposes a lower bound on this limit and can be interpreted as requiring that serving type-$1$ customers remains socially valuable, even when very large service depths are assigned. In the linear value case, Condition~\eqref{equation:prop2} becomes $a>\frac{\theta_1}{\sqrt{\theta_2/a}}$, which is equivalent to the condition $\theta_1^2<a\theta_2$ in which type-$1$ customers are not excluded.

\subsection{Optimal Decentralized Mechanism}
When delay sensitivities are private information and customers are self-interested, the server offers a menu of prices and service allocations to maximize the long-term average welfare. In the two-type setting, the menu $\mathcal{M}$ needs to contain at most two different options $(p_1, x_1, w_1)$ and $(p_2, x_2, w_2)$, where option $i$ is intended for customers with delay sensitivity $\theta_i$. Here $x_i$ is the service depth, $w_i$ is the steady-state expected sojourn time, and $p_i$ is the price paid by a customer who chooses option $i$. We allow either option to be the outside option $(0,0,0)$. The definition of incentive compatibility directly yields Lemma~\ref{lemma:welfare_maximization_incentive_compatibility}, which characterizes the set of all implementable menus.
\begin{lemma}
    \label{lemma:welfare_maximization_incentive_compatibility}
    Fix an operationally feasible allocation with service depth and sojourn time pairs $(x_1, w_1)$ and $(x_2, w_2)$. There exists an incentive compatible price menu with corresponding payments $(p_1, p_2)$ if and only if $w_1\leq w_2$.
\end{lemma}

The decentralized welfare-maximization problem can be written as
\begin{equation}
    \label{eq:welfare_maximization_decentralized}
    \begin{aligned}
        \max_{x_1, x_2, w_1, w_2}&\quad \lambda_1(V(x_1)-\theta_1w_1)+\lambda_2(V(x_2  )-\theta_2w_2)\\
        \textbf{s.t.}&\quad w_i\geq\frac{x_i}{1-\rho_i},\quad i=1,2,\\
        &\quad \rho_1w_1+\rho_2w_2\geq\frac{\rho_1x_1+\rho_2x_2}{1-\rho_1-\rho_2},\\
        &\quad \rho_1+\rho_2<1,\\
        &\quad w_1\leq w_2.
    \end{aligned}
\end{equation}

Compared to \eqref{eq:welfare_maximization}, one incentive compatibility constraint is added in \eqref{eq:welfare_maximization_decentralized}. Any feasible solution of \eqref{eq:welfare_maximization_decentralized} corresponds to a menu $\mathcal{M}=\{(p_1,x_1,w_1), (p_2,x_2,w_2)\}$, where the prices $p_1, p_2$ satisfy
\begin{equation}
    \label{equation:payment}
    \begin{aligned}
        &p_1=V(x_1)-\theta_1w_1,\\
        &p_2=V(x_2)-\theta_2w_2-(\theta_1-\theta_2)w_1.
    \end{aligned}
\end{equation}
If we fix the payment rule to be \eqref{equation:payment}, we can characterize the optimal decentralized policy. Proposition~\ref{proposition:welfare_maximization_decentralized_case} presents the structure of the optimal menu under linear value function and compares it with the optimal centralized policy.

\begin{proposition}
    \label{proposition:welfare_maximization_decentralized_case}
    Suppose $V(x)=ax$, where $a>\theta_1>\theta_2$. Let $c$ denote the priority threshold defined in Proposition~\ref{proposition:welfare_maximization_centralized_case}.
    \begin{enumerate}
        \item If $\theta_1^2\geq a\theta_2$, then the optimal menu is
        \begin{equation*}
            \begin{aligned}
                &(p_1^*,x_1^*, w_1^*)=(0,0,0),\\
                &(p_2^*, x_2^*, w_2^*)=\left(ax_2^*-\theta_2\frac{x_2^*}{1-\lambda_2x_2^*}, \frac{1-\sqrt{\theta_2/a}}{\lambda_2},\frac{x_2^*}{1-\lambda_2x_2^*}\right);
            \end{aligned}
        \end{equation*}
        \item If $\theta_1^2< a\theta_2$ then the optimal menu is
        \begin{equation*}
            \begin{aligned}
                &(p_1^*,x_1^*, w_1^*)=\left(ax_1^*-\theta_1 w_1^*, x_1^*,\frac{x_1^*}{1-\lambda_1x_1^*}\right),\\
                &(p_2^*, x_2^*, w_2^*)=\left(ax_2^*-\theta_2w_2^*-(\theta_1-\theta_2)w_1^*,x_2^*, \frac{x_2^*}{1-\lambda_1x_1^*-\lambda_2x_2^*}+\frac{\lambda_1{x_1^*}^2}{(1-\lambda_1x_1^*)(1-\lambda_1x_1^*-\lambda_2x_2^*)}\right),
            \end{aligned}
        \end{equation*}
        where $(x_1^*,x_2^*)$ satisfies condition~\eqref{equation:prop1}. Consequently, when $\frac{\lambda_2}{\lambda_1}>c$, the optimal price menu gives strictly lower long-term average social welfare than the optimal centralized policy.
    \end{enumerate}
    Moreover, the payment $p_1^*, p_2^*\geq 0$. 
\end{proposition}
Proposition~\ref{proposition:welfare_maximization_decentralized_case} shows an interesting structure of the optimal menu: type~$1$ customers are always prioritized. When the optimal centralized policy satisfies $w_1^*\leq w_2^*$, the same allocation rule is achievable by a price menu. When the optimal centralized policy prioritizes type~$2$ customers, it violates the incentive compatibility constraints. In this case, the decision maker has two natural incentive compatible alternatives. One is to optimize over allocations that prioritize type~$1$ customers. The other is to keep service depths for which type~$2$ customers would have higher priority under a centralized decision maker, but then distort the sojourn times to satisfy $w_1\leq w_2$, which forces the solution to the boundary $w_1=w_2$. Proposition~\ref{proposition:welfare_maximization_decentralized_case} shows that the first alternative always dominates the second. Thus, the optimal menu always gives impatient customers a lower service depth and higher priority.

Proposition~\ref{proposition:welfare_maximization_decentralized_case} reveals a new price of decentralization. It identifies the precise region in which operational efficiency and incentive compatibility conflict. A policy that prioritizes patient customers may be operationally efficient, but it is not incentive compatible as a menu: the decision maker cannot force impatient customers to choose a contract that results in a longer wait. Whenever such a policy is optimal from the centralized decision maker’s perspective, a welfare gap emerges between the centralized optimum and the optimal decentralized mechanism.

\begin{definition}[Welfare Gap]
    Let $\mathcal{W}_{C}$ denote the optimal social welfare achieved by the centralized decision maker, and let $\mathcal{W}_{D}$ denote the optimal social welfare of the server's menu. If $\mathcal{W}_{C}>0$, the welfare gap $\mathcal{G}$ is defined as the normalized difference between $\mathcal{W}_{C}$ and $\mathcal{W}_{D}$, i.e.,
    \begin{equation*}
        \mathcal{G}=\frac{\mathcal{W}_C-\mathcal{W}_D}{\mathcal{W}_C}.
    \end{equation*}
    Otherwise, define $\mathcal{G}=0$.
\end{definition}

In the case in which such a welfare gap exists (the red region in Figure~\ref{figure:prop1_1}), it remains important to understand how good the decentralized mechanism is. Proposition~\ref{proposition:social_welfare_gap} quantifies the welfare gap, showing that loss from the incentive compatibility constraint is bounded.
\begin{proposition}
    \label{proposition:social_welfare_gap}
    Suppose $V(x)=ax$, where $a>\theta_1>\theta_2$. Denote by $\mathcal{W}_{C}$ the optimal social welfare achieved by the centralized decision maker, and $\mathcal{W}_{D}$ the optimal social welfare of the server's menu. Then
    \begin{equation*}
        \mathcal{W}_{D}\geq \frac{3}{4}\mathcal{W}_{C}. 
    \end{equation*}
    Moreover, for any $\epsilon>0$, there exists an instance such that
    \begin{equation*}
        \mathcal{W}_{D}< \left(\frac{3}{4}+\epsilon\right)\mathcal{W}_{C}, 
    \end{equation*}
    thus the bound is tight.
\end{proposition}
Proposition~\ref{proposition:social_welfare_gap} provides a worst-case characterization of the welfare gap. In the decentralized problem, the optimal menu leads to the same welfare as the best policy conditional on prioritizing type~$1$ customers. Thus, when the centralized optimum instead prioritizes type~$2$ customers, the welfare gap is the loss from replacing the centralized allocation with the best allocation that prioritizes type~$1$. In the worst case in which type~$1$ customers are extremely rare, the decentralized mechanism behaves like the system that drops type~$1$ and serves only type~$2$: type~$1$ customers receive nearly zero service depth and most of the welfare comes from serving type~$2$. In contrast, in the centralized setting, the decision maker still extracts welfare from rare type~$1$ customers by giving them deep but slow service. The formal proof compares the limiting welfare of these two policies and establishes a tight bound that shows that, in the decentralized setting, the mechanism can lose up to $\frac{1}{4}$ of the social welfare compared to the centralized benchmark, i.e., the worst-case welfare gap $\mathcal{G}=\frac{1}{4}$.

\section{Continuous-Type Model}
\label{section:continuous-type setting}

In this section, we assume the customer type is drawn from a continuous distribution. A customer’s type is their delay-cost parameter $\theta$, now drawn independently from a commonly known continuous distribution $F$ supported on $[\underline{\theta},\bar{\theta}]$. We assume the probability density function satisfies $f(\theta)>0$ for all $\theta\in[\underline{\theta},\bar{\theta}]$. All proofs in this section are deferred to Appendix~\ref{appendix:section_4}.

A service depth assignment assigns a type-dependent deterministic service depth $x(\theta)$ for each $\theta\in[\underline{\theta},\bar{\theta}]$. Mathematically, the service depth is a measurable function
\begin{equation*}
    x:[\underline{\theta},\bar{\theta}]\to[0,\infty),
\end{equation*}
As before, $x(\theta)=0$ represents instantaneous completion with zero service value.

\begin{assumption}
\label{assumption:continuous_types}
A service depth assignment $x(\cdot)$ is admissible if
    \begin{equation*}
        \lambda\int_{\underline{\theta}}^{\bar{\theta}}x(\theta)dF(\theta)<1\quad\text{and}\quad\int_{\underline{\theta}}^{\bar{\theta}}x^2(\theta)dF(\theta)<\infty.
    \end{equation*}
\end{assumption}

The first condition is the stability condition. The second condition guarantees finite expected sojourn times under the  priority rules considered below. Notice that service times are still exponentially distributed conditional on the assigned mean \(x(\theta)\); what is deterministic here is the service depth assignment as a function of type.

For a given service depth assignment $x(\cdot)$ and a priority rule, let $w(\theta)$ denote the expected sojourn time of a type-$\theta$ customer. In the continuous-type setting, the finite-class achievable region result in \cite{coffman1980characterization} can be generalized to Lemma~\ref{lemma:feasible_region_continuous}, which has also been proved in \cite{abdolmaleki2026pricing}.

\begin{lemma}
    \label{lemma:feasible_region_continuous}
    Fix a stable service depth assignment $x(\cdot)$. A nonnegative measurable sojourn-time function $w(\cdot)$ with $\lambda\int x(\theta)w(\theta)dF(\theta)<\infty$ is operationally achievable by some priority rule if and only if for any measurable $\mathcal{S}\subseteq [\underline{\theta},\bar{\theta}]$
    \begin{equation*}
       \lambda\int_\mathcal{S} x(\theta)w(\theta)f(\theta)d\theta\geq \frac{\lambda\int_\mathcal{S} x^2(\theta)f(\theta)d\theta}{1-\lambda\int_\mathcal{S} x(\theta)f(\theta)d\theta}.
    \end{equation*}
\end{lemma}

\subsection{Centralized Optimum}

In the centralized benchmark, the decision maker observes each customer's delay sensitivity $\theta$, chooses a deterministic service depth assignment $x(\theta)$, and then schedules customers using a  priority rule. The decision maker's welfare-maximization problem can be written as
\begin{equation}
    \label{equation:continuous_types_welfare_maximization}
    \begin{aligned}
        \max_{x(\cdot), w(\cdot)}\quad&\lambda\int_{\underline{\theta}}^{\bar{\theta}}\left(V(x(\theta))-\theta w(\theta)\right)dF(\theta)\\
        \text{s.t.}\quad&\lambda\int_\mathcal{S} x(\theta)w(\theta)f(\theta)d\theta\geq \frac{\lambda\int_\mathcal{S} x^2(\theta)f(\theta)d\theta}{1-\lambda\int_\mathcal{S} x(\theta)f(\theta)d\theta}, \quad \mathcal{S}\subseteq [\underline{\theta},\bar{\theta}].
    \end{aligned}
\end{equation}
By the same $c\mu$-rule argument as in Lemma \ref{lemma:cmu_rule}, once the service depths are assigned, an optimal priority rule ranks customers in increasing order of $x/\theta$. We therefore focus on the service depth assignment problem. 

Proposition~\ref{proposition:continuous_types_welfare_maximization} shows that for a continuum of types of customers, the optimal policy will prioritize customers with larger delay sensitivities and the service depth assignment is non-increasing with respect to $\theta$.

\begin{proposition}
    \label{proposition:continuous_types_welfare_maximization}
    Fix any feasible centralized allocation $(x,w)$. There exists a feasible centralized allocation $(\tilde{x}, \tilde{w})$ with non-increasing $\tilde x:[\underline{\theta},\bar{\theta}]\to[0,\infty)$ that achieves weakly higher long-term average social welfare. Consequently, the optimization problem can be restricted without loss to deterministic non-increasing service depth assignments. Under any such assignment, the decision maker gives priority to customers with larger delay sensitivities.
\end{proposition}

Proposition~\ref{proposition:continuous_types_welfare_maximization} says that the decision maker should sort service depth against urgency, where more delay-sensitive customers receive the smaller service depth. Such a structure is fundamentally different from the two-type case. With only two positive-mass types, the decision maker may sometimes move the entire patient type ahead of the impatient type; however, in the continuous-type model, more delay-sensitive customers are always in front of less delay-sensitive customers. 

This contrast can be interpreted as the value of personalization. Suppose, for example, that a continuous distribution is concentrated around two regions of delay sensitivities. If the decision maker can still choose service depths separately for each realized type, Proposition~\ref{proposition:continuous_types_welfare_maximization} applies and the monotone rearrangement remains feasible. If instead the decision maker pools all customers in each region and assigns one common service depth to each pool, then it becomes a two-type model, and the pattern in Proposition~\ref{proposition:welfare_maximization_centralized_case} may appear. We provide a more detailed comparison between the two-type model and the continuous-type model and show that the rearrangement is available for the two-type model with randomized service depth in Section~\ref{subsection:randomized}.

Under a non-increasing assignment $x(\theta)$, the ratio $x(\theta)/\theta$ is also non-increasing in $\theta$. Hence the $c\mu$-rule is equivalent to giving higher priority to customers with larger delay sensitivities. The expected sojourn time of type $\theta$ is therefore
\begin{equation}
\label{equation:continuous_types_sojourn_time}
w(\theta)=\frac{\lambda\int_{\theta}^{\bar{\theta}}x^2(c)dF(c)}{\left(1-\lambda\int_{\theta}^{\bar{\theta}}x(c)dF(c)\right)^2}+\frac{x(\theta)}{1-\lambda\int_{\theta}^{\bar{\theta}}x(c)dF(c)}.
\end{equation}
The first term is the congestion created by customers with higher priority than type $\theta$. The second term is the effective service requirement of type $\theta$ after accounting for interruptions by higher-priority arrivals.

Substituting \eqref{equation:continuous_types_sojourn_time} into \eqref{equation:continuous_types_welfare_maximization}, the centralized problem becomes
\begin{equation}
\label{equation:continuous_types_final_problem}
\max_{x(\cdot)}\lambda\int_{\underline{\theta}}^{\bar{\theta}}\left[V(x(\theta))-\theta\left(\frac{\lambda\int_{\theta}^{\bar{\theta}}x^2(c)dF(c)}{\left(1-\lambda\int_{\theta}^{\bar{\theta}}x(c)dF(c)\right)^2}+\frac{x(\theta)}{1-\lambda\int_{\theta}^{\bar{\theta}}x(c)dF(c)}\right)\right]dF(\theta),
\end{equation}
where $x(\cdot)$ is an admissible and non-increasing function.

Problem~\eqref{equation:continuous_types_final_problem} optimizes over an infinite-dimensional space, making closed-form solutions difficult to obtain. A trivial case arises when $\lim_{x\rightarrow 0^+} V'(x)\leq \underline{\theta}$. For every type of customer, the expected sojourn time is at least the service depth. Moreover, by concavity of $V$, 
\begin{equation*}
    V(x)\leq \underline{\theta}x\leq \theta x\leq \theta w(\theta)
\end{equation*}
for every $\theta\in[\underline{\theta},\bar{\theta}]$. Hence, the maximum welfare is zero. Therefore, we focus on the nontrivial case in which $\lim_{x\rightarrow 0^+} V'(x)>\underline{\theta}$. Proposition~\ref{proposition:existence_of_optimality} provides conditions that characterize the existence of an optimal solution.

\begin{proposition}
    \label{proposition:existence_of_optimality}
    \begin{enumerate}
        \item If $V(x)=ax$, where $a>\underline{\theta}$, then the optimal solution of \eqref{equation:continuous_types_final_problem} is degenerate, and the long-term average social welfare supremum is $a-\underline{\theta}-\underline{\theta}\left(\ln{a}-\ln{\underline{\theta}}\right)$, but is not attained.
        \item If $V(x)$ is continuously differentiable and $\lim_{x\rightarrow\infty}V'(x)<\underline{\theta}$, then there exists a continuous and non-increasing optimal solution $x^*(\cdot)$ of \eqref{equation:continuous_types_final_problem}.
    \end{enumerate}
\end{proposition}

When the value function $V(x)$ is linear, serving different customers generates the same value per unit of service depth. As a result, it is always better to spend all service capacity to the most patient type of customer and have all other customers leave immediately. This leads to a degenerate service depth allocation: $x(\underline{\theta}) \rightarrow \infty$, while $x(\theta)=0$ for all $\theta>\underline{\theta}$. The degeneracy does not exist in the two-type model because both types have strictly positive arrival rates. If the decision maker assigns an infinite service depth to one type of customer, the resulting workload diverges and violates the stability constraint. On the other hand, when $\lim_{x\rightarrow\infty}V'(x)< \underline{\theta}$, the service depths of all types of customers are uniformly bounded. Therefore, the optimal service depth assignment exists.

\subsection{Optimal Decentralized Mechanism}

We next consider the decentralized setting, where $\theta$ is private information. The server offers a menu
\begin{equation*}
    M=\{(p(\theta),x(\theta),w(\theta)): \theta\in[\underline{\theta},\bar{\theta}]\}.
\end{equation*}

Lemma~\ref{lemma:continuous_types_incentive_compatibility} is the continuous-type counterpart of the two-type incentive compatibility result. It shows that incentive compatibility is governed entirely by the ordering of sojourn times.

\begin{lemma}
\label{lemma:continuous_types_incentive_compatibility}
Suppose type $\theta$ customers choose option $(p(\theta),x(\theta),w(\theta))$. There exists an incentive compatible pricing policy $p(\theta)$ if and only if $w(\theta)$ is non-increasing in $\theta$. Moreover, if $w(\theta)$ is non-increasing, one implementing payment rule is
\begin{equation}
    \label{equation:payment_continuous}
    p(\theta)=V(x(\theta))-\theta w(\theta)-\int_{\theta}^{\bar{\theta}} w(c)dc.
\end{equation}
\end{lemma}

Combining Proposition~\ref{proposition:continuous_types_welfare_maximization} and Lemma~\ref{lemma:continuous_types_incentive_compatibility} gives the main conclusion of the decentralized setting.

\begin{proposition}
\label{proposition:continuous_types_welfare_maximization_decentralized_case}
If the centralized welfare supremum in the continuous-type setting can be achieved by some admissible $x^*(\cdot)$, it can also be achieved by a decentralized pricing menu. In particular, if $x^*(\theta)$ is the optimal solution of \eqref{equation:continuous_types_final_problem}, then the menu
\begin{equation*}
    M^*=\{(p^*(\theta),x^*(\theta),w^*(\theta)): \theta\in[\underline{\theta},\bar{\theta}]\},
\end{equation*}
is optimal, where $w^*(\theta)$ is given by \eqref{equation:continuous_types_sojourn_time} and $p^*(\theta)=V(x^*(\theta))-\theta w^*(\theta)-\int_{\theta}^{\bar{\theta}}w^*(c)dc$. Moreover, $p^*(\theta)\geq 0$.
\end{proposition}

Proposition~\ref{proposition:continuous_types_welfare_maximization_decentralized_case} states that in the continuous-type setting, there is no welfare gap between the mechanism and the benchmark. The centralized allocation already has the correct monotone geometry. Queueing efficiency ranks customers by $x(\theta)/\theta$, while incentive compatibility requires sojourn time to be non-increasing in $\theta$. Under the optimal service depth assignment, both rankings agree: delay-sensitive customers receive shorter service, higher priority, and shorter sojourn time.

\section{Numerical Examples}
\label{section:numerical_results}

In this section, we provide numerical examples for both the two-type and continuous-type models, where closed-form optimal solutions are generally not available.

\subsection{Welfare Gap in the Two-Type Model}

In Section~\ref{section:two-type setting}, we develop theoretical analysis on the welfare gap between the centralized optimum and the optimal price menu, showing a worst-case bound for linear value functions. The numerical results below complement this analysis by examining the magnitude of the welfare gap under different value functions and model parameters. Importantly, the results show that the welfare gap is not an artifact of the linear value function: positive gaps also arise under strictly concave value functions.

Table~\ref{tab:welfare_gap_lambda} reports how the welfare gap varies with the population ratio $\lambda_2/\lambda_1$. Across all value functions, the gap is zero when $\lambda_2/\lambda_1\leq 1.0$. In this regime, type~$2$ customers are not sufficiently prevalent,  and the centralized optimum already satisfies the incentive compatibility constraints that the impatient type receives a weakly shorter sojourn time, so the centralized allocation can be implemented by a price menu. As $\lambda_2/\lambda_1$ increases, it becomes more attractive for the centralized decision maker to assign type~$2$ customers a smaller service depth and give them priority. When this priority reversal occurs, the centralized allocation violates incentive compatibility and a positive welfare gap can appear.  

The magnitude of the gap strongly depends on the value function. For the linear value function $V(x)=2x$, the welfare gap increases as $\lambda_2/\lambda_1$ gets larger, reaching $16.2\%$ when $\lambda_2/\lambda_1=32.0$. In contrast, for the concave value functions, the gap is much smaller and need not be monotone in $\lambda_2/\lambda_1$. Intuitively, for linear value functions, the marginal value of service is constant, so the decision maker can assign deep service to type~$1$ customers while using priority to reduce delay for type~$2$ customers. However, when the value function is strictly concave, the marginal benefit of assigning very deep service becomes small, reducing the benefit from prioritizing type~$2$ customers.   
\begin{table}[htbp]
    \centering
    \caption{Welfare gap across $\lambda_2/\lambda_1$ under different value functions}
    \label{tab:welfare_gap_lambda}
    \renewcommand{\arraystretch}{1.1} 

    \begin{tabular}{lccccccc}
    \toprule
    & \multicolumn{7}{c}{$\lambda_2/\lambda_1$} \\
    \cmidrule(lr){2-8}
    Value function & 0.5 & 1.0 & 2.0 & 4.0 & 8.0 & 16.0 & 32.0 \\
    \midrule
    $V(x)=2x$ & 0.0\% & 0.0\% & 3.7\% & 8.9\% & 12.5\% & 14.8\% & 16.2\%\\
    $V(x)=2\sqrt{x}$ & 0.0\% &  0.0\% & 0.0\% & 0.5\% & 0.6\% & 0.5\% & 0.3\%\\
    $V(x)=2\log (1+x)$ & 0.0\% &  0.0\% & 0.2\% & 1.5\% & 1.9\% & 1.6\% & 1.1\%\\
    $V(x)=2-2\exp (-x)$ & 0.0\% &  0.0\% & 0.0\% & 0.8\% & 1.0\% & 0.8\% & 0.5\%\\
    \bottomrule
    \\
    \end{tabular}
    \begin{minipage}{0.5\textwidth}
    \footnotesize
    Note: The parameters are $\lambda=1.0$, $\theta_1=0.51$, $\theta_2=0.49$.
    \end{minipage}
\end{table}

Table~\ref{tab:welfare_gap_theta} presents the welfare gap across different delay-sensitivity pairs $(\theta_1, \theta_2)$ when fixing the arrival rates of each customer type. When $\theta_1$ and $\theta_2$ are close, the benefit of serving the large patient type first dominates the relatively small difference in delay costs, so the centralized decision maker prefers a priority order that violates incentive compatibility. As the two types have a bigger difference in their delay sensitivities, prioritizing the impatient type becomes increasingly valuable for the
centralized decision maker, and the conflict between centralized efficiency and incentive compatibility disappears.
\begin{table}[htbp]
    \centering
    \caption{Welfare gap across $(\theta_1, \theta_2)$ under different value functions}
    \label{tab:welfare_gap_theta}
    \renewcommand{\arraystretch}{1.1} 

    \begin{tabular}{lcccc}
    \toprule
    & \multicolumn{4}{c}{$(\theta_1, \theta_2)$} \\
    \cmidrule(lr){2-5}
    Value function & $(0.51,0.49)$ & $(0.52,0.48)$ & $(0.55,0.45)$ & $(0.60,0.40)$ \\
    \midrule
    $V(x)=2x$ & 13.0\% & 11.1\% & 5.5\% & 0.0\%\\
    $V(x)=2\sqrt{x}$ & 0.6\% &  0.3\% & 0.0\% & 0.0\%\\
    $V(x)=2\log (1+x)$ & 1.9\% &  1.0\% & 0.0\% & 0.0\%\\
    $V(x)=2-2\exp (-x)$ & 1.0\% &  0.3\% & 0.0\% & 0.0\%\\
    \bottomrule
    \\
    \end{tabular}
    \begin{minipage}{0.5\textwidth}
    \footnotesize
    Note: The parameters are $\lambda=1.0$, $\lambda_1=0.1$, $\lambda_2=0.9$.
    \end{minipage}
\end{table}

\subsection{Examples for the Continuous-Type Model}
\label{subsec: examples for the continuous type model}

For a general concave value function $V(x)$, it is difficult to obtain the closed-form optimal service depth assignment. In this subsection, we numerically study the optimal service depth assignment of Problem~\eqref{equation:continuous_types_final_problem}. We consider a discrete version of \eqref{equation:continuous_types_final_problem}:
\begin{equation}
    \label{equation:continuous_type_discrete}
    \max_{x_1\geq \cdots\geq x_n\geq0}\quad\frac{\lambda}{n}\sum_{i=1}^n\left[V(x_i)-\theta_i\left(\frac{\frac{\lambda}{n}\sum_{j=i}^nx_j^2}{\left(1-\frac{\lambda}{n}\sum_{j=i}^nx_j\right)\left(1-\frac{\lambda}{n}\sum_{j=i+1}^nx_j\right)}+\frac{x_i}{1-\frac{\lambda}{n}\sum_{j=i+1}^nx_j}\right)\right]
\end{equation}
over the stability constraint $\frac{\lambda}{n}\sum_{j=1}^nx_j<1$, where $\theta_i=F^{-1}(\frac{i-1}{n-1})$ represents evenly spaced quantiles of a distribution with cumulative distribution function $F$. For instance, if $F$ is a uniform distribution on $[\underline{\theta},\bar{\theta}]$, then $\theta_i=\frac{i-1}{n-1}\bar\theta+\frac{n-i}{n-1}\underline{\theta}$. Problem~\eqref{equation:continuous_type_discrete} is equivalent to the problem in which there are $n$ different customer types with the same arrival rate and delay sensitivities $\theta_1,\cdots,\theta_n$, conditional on the service depth assignment being non-increasing. In the experiments, we set $n=200$.

\begin{figure}[htbp]
\centering
\caption{Optimal Solution for $V(x)=\sqrt{x}$.}

\begin{subfigure}[t]{0.24\textwidth}
    \centering
    \caption{Service Depth.}
    \includegraphics[width=\textwidth]{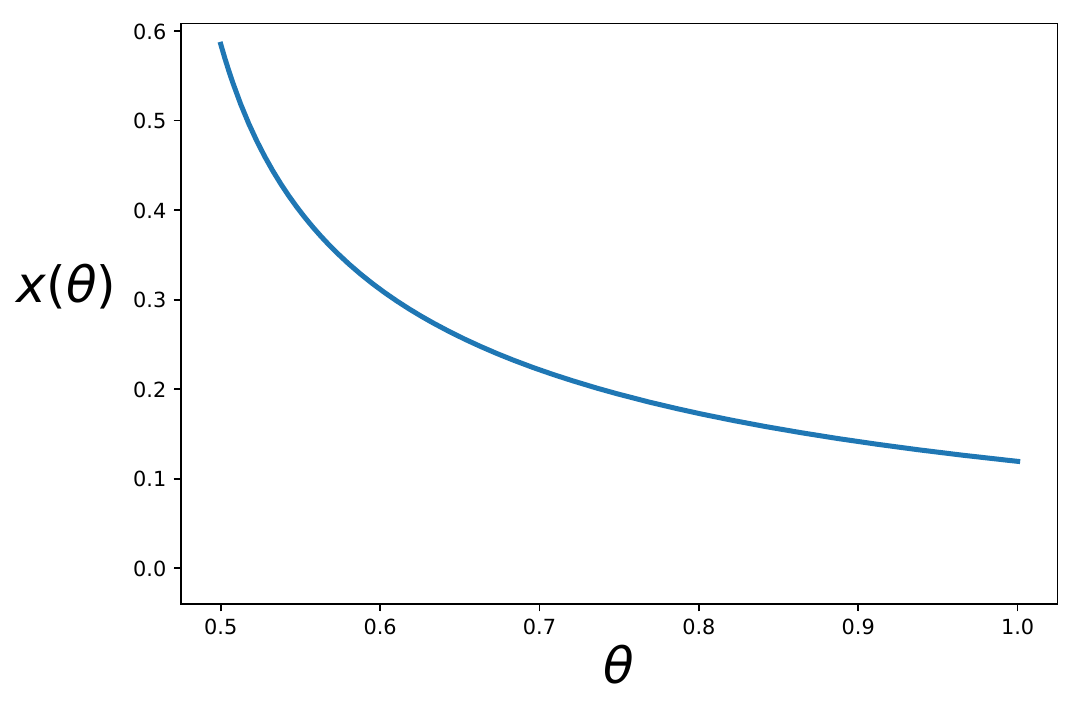}
    \label{figure:sqrt_opt}
\end{subfigure}
\begin{subfigure}[t]{0.24\textwidth}
    \centering
    \caption{Sojourn Time.}
    \includegraphics[width=\textwidth]{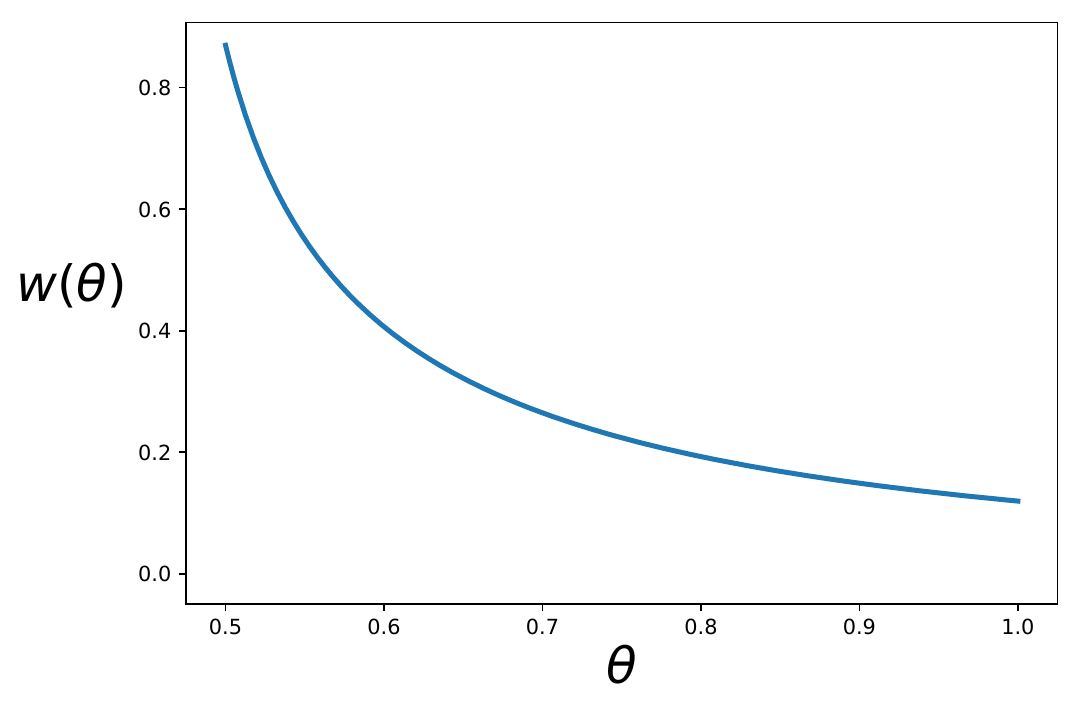}
    \label{figure:sqrt_w}
\end{subfigure}
\begin{subfigure}[t]{0.24\textwidth}
    \centering
    \caption{Welfare \& Surplus.}
    \includegraphics[width=\textwidth]{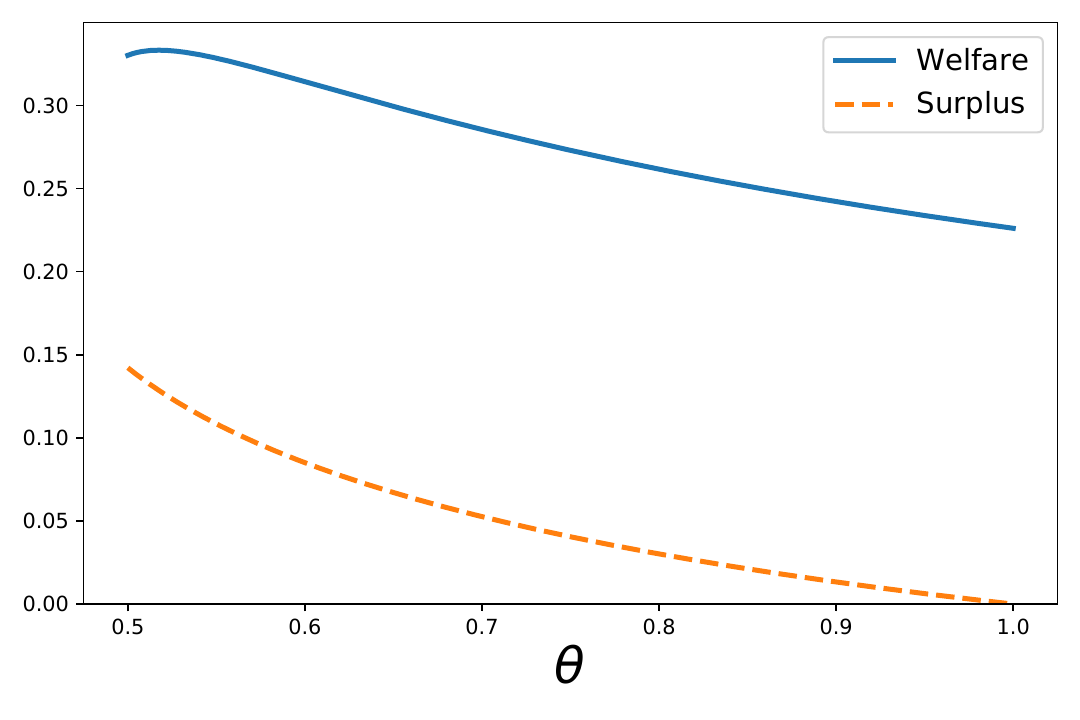}
    \label{figure:sqrt_welfare}
\end{subfigure}
\begin{subfigure}[t]{0.24\textwidth}
    \centering
    \caption{Payment.}
    \includegraphics[width=\textwidth]{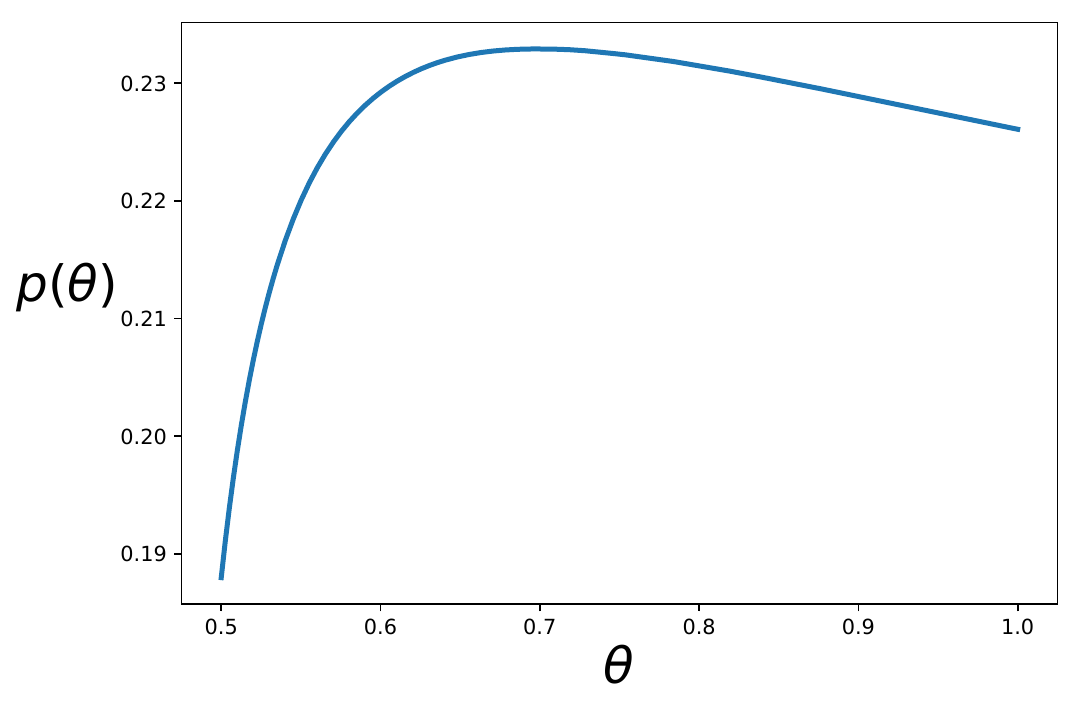}
    \label{figure:sqrt_p}
\end{subfigure}
\vspace{0.3em}
\begin{minipage}{0.5\textwidth}
\footnotesize
Note: The parameters are $\lambda=1.0$, $\theta\sim \text{Unif}[0.5,1.0]$.
\end{minipage}

\label{fig:example_sqrt}

\end{figure}

Figure~\ref{fig:example_sqrt} shows the optimal service depth assignment for a square root value function, where Figure~\ref{figure:sqrt_opt} presents the optimal solution, Figure~\ref{figure:sqrt_w} presents the corresponding sojourn time, Figure~\ref{figure:sqrt_welfare} plots the welfare contribution and customer surplus for each type and Figure~\ref{figure:sqrt_p} presents the corresponding payment according to \eqref{equation:payment_continuous}. In this case, the service depth $x(\theta)$ is strictly decreasing with respect to $\theta$. Moreover, all types of customers receive positive service depths. This is because $\lim_{x\rightarrow0}V'(x)=\infty$, so no matter how large $\theta$ is, the initial increment in service value for a new customer will dominate customers' delay cost. 

To understand how each type of customers contributes to the welfare, Figure~\ref{figure:sqrt_welfare} plots the welfare contribution $V(x(\theta))-\theta w(\theta)$ for each type of customers. Although less delay sensitive customers receive larger surplus under the implementing menu, the welfare contribution by each type of customer is not monotone in $\theta$. The intuition is that customers with larger delay sensitivities receive a bundle with two offsetting changes: they receive smaller service depth, which reduces service value, but they also receive shorter sojourn time, which decreases their delay cost. The value of the reduction in sojourn time dominates the loss in service value for less delay sensitive customers, so the welfare contribution initially increases with $\theta$. For more delay sensitive customers, the loss in service value dominates the value of the additional reduction in sojourn time, so the welfare contribution decreases with $\theta$.

Figure~\ref{figure:sqrt_p} shows that the payment is also not monotone with respect to $\theta$. The payment reflects the same two offsetting bundle effects: more delay sensitive customers receive smaller service depth, which pushes the payment downward, but they also receive shorter sojourn time, which pushes the payment upward. The non-monotonicity of welfare contribution and the non-monotonicity of payment are driven by the same underlying trade-off between service value and delay cost, but they are not identical. Since the payment equals the welfare contribution minus the surplus, it subtracts this information rent from the welfare contribution.

\begin{figure}[htbp]
\centering
\caption{Optimal Solution for $V(x)=\ln{(1+x)}$.}

\begin{subfigure}[t]{0.24\textwidth}
    \centering
    \caption{Service Depth.}
    \includegraphics[width=\textwidth]{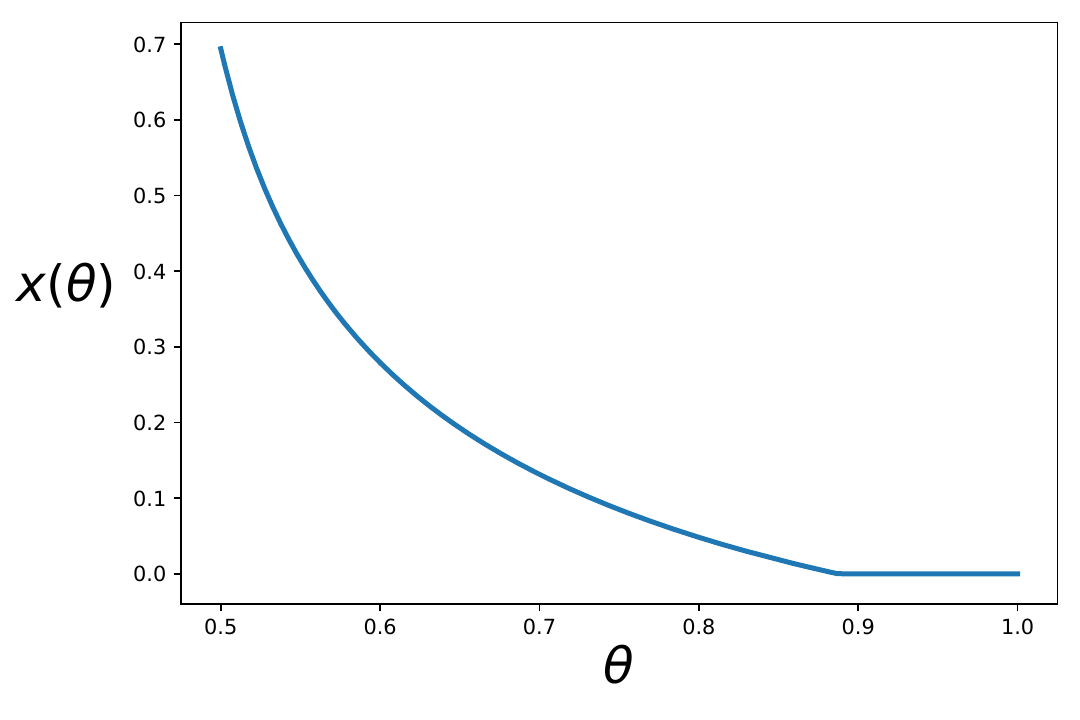}
    \label{figure:log_opt}
\end{subfigure}
\begin{subfigure}[t]{0.24\textwidth}
    \centering
    \caption{Sojourn Time.}
    \includegraphics[width=\textwidth]{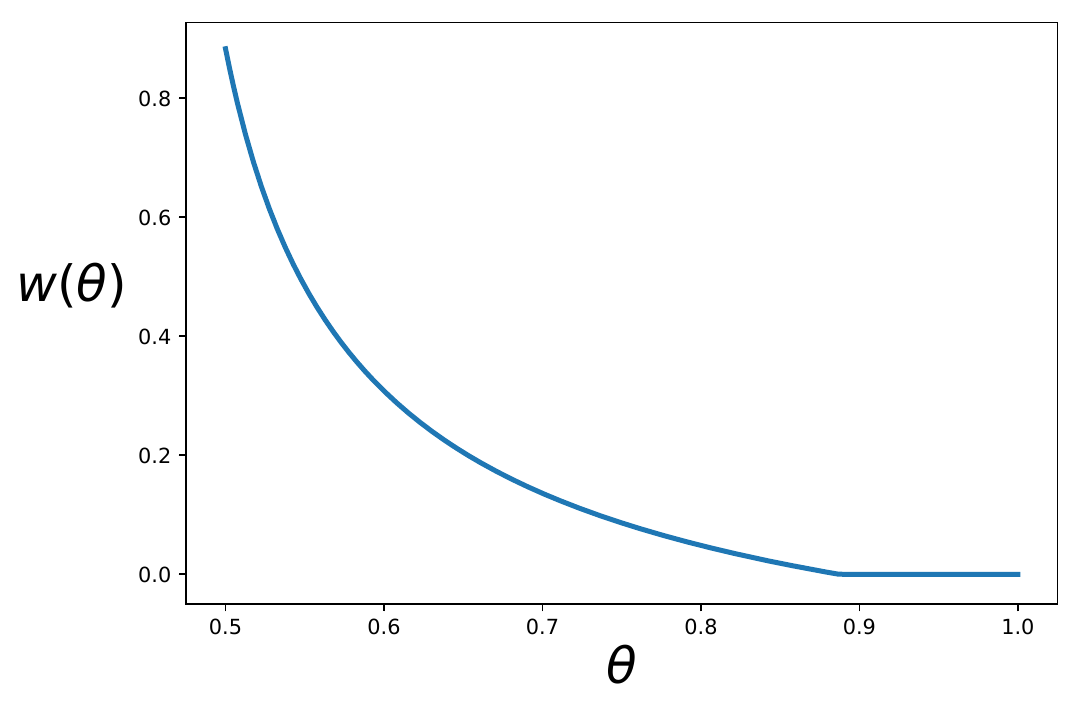}
    \label{figure:log_w}
\end{subfigure}
\begin{subfigure}[t]{0.24\textwidth}
    \centering
    \caption{Welfare \& Surplus.}
    \includegraphics[width=\textwidth]{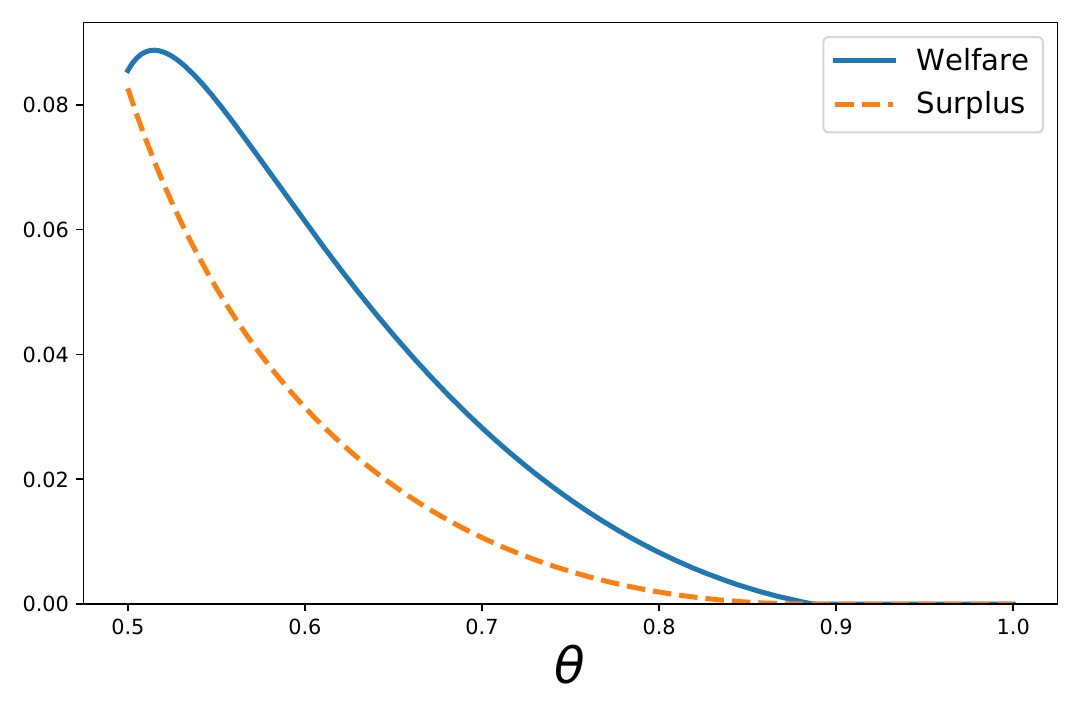}
    \label{figure:log_welfare}
\end{subfigure}
\begin{subfigure}[t]{0.24\textwidth}
    \centering
    \caption{Payment.}
    \includegraphics[width=\textwidth]{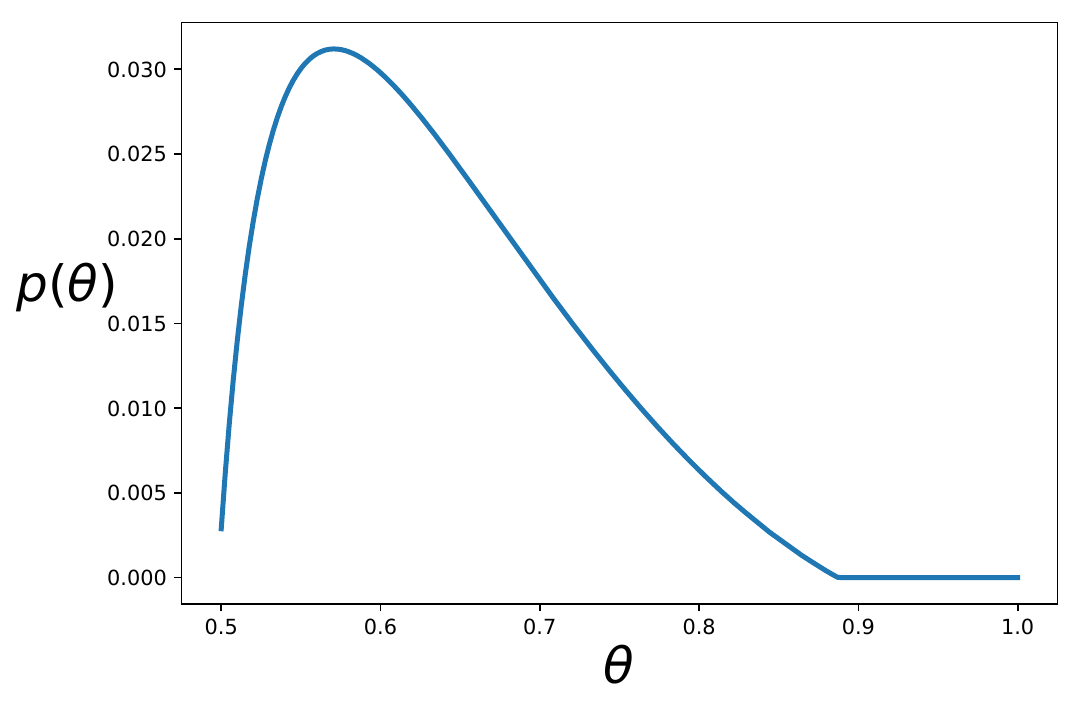}
    \label{figure:log_p}
\end{subfigure}
\vspace{0.3em}
\begin{minipage}{0.5\textwidth}
\footnotesize
Note: The parameters are $\lambda=1.0$, $\theta\sim \text{Unif}[0.5,1.0]$.
\end{minipage}

\label{fig:example_log}

\end{figure}

Figure~\ref{fig:example_log} shows the results for a logarithmic value function. In this case, we observe that when $\theta$ is above a certain threshold, the optimal policy allocates zero service depth for this type of customers. This is because $V'(0)=1$ is finite, so for customers whose delay sensitivities are high, it is not worthwhile to keep them in the system. Similar to the results for the square root function, the welfare contribution $V(x(\theta))-\theta w(\theta)$ and the payment $p(\theta)$ are not monotone with $\theta$.

\section{Case Study}
\label{section:case_study}
To complement the stylized numerical studies in Section~\ref{section:numerical_results}, we calibrate an agentic AI service system using published trials of a coding agent on software-engineering tasks. We use reasoning effort as service depth, task success as service value, and token-based expenditure as a proxy for service time. We then use these empirically measured quantities to build the queueing model and evaluate different service-depth and priority policies. This experiment provides a closer connection between our theoretical framework and its motivating application to agentic AI systems.

\subsection{Data}
We use the DeepSWE v1.1 benchmark \citep{huang2026deepswe}, which contains 113 software-engineering tasks across 91 repositories and five programming languages.\footnote{Benchmark and methodology: \url{https://deepswe.datacurve.ai/blog/deepswe}. The versioned trial export is \url{https://deepswe.datacurve.ai/artifacts/v1.1/trials.json}.} Each task asks the agent to modify a repository to implement specified behavior; executable verifiers assess the resulting code. For example, a task may require adding configuration-file support to a command-line library. This provides a setting in which a service request involves several steps of code inspection, modification, and testing.

We use the published GPT-5.6 Sol trials under the \texttt{mini-swe-agent} harness. The five service depths $\mathcal{D}=\{d_1,d_2,d_3,d_4,d_5\}$ correspond to \texttt{low}, \texttt{medium}, \texttt{high}, \texttt{xhigh}, and \texttt{max} reasoning effort; $d_0=0$ denotes no service. The same 113 tasks are evaluated four times at each depth, giving 452 trials per depth and 2,260 trials in total. Each trial records a binary success outcome, token usage, and monetary cost. Four non-verifiable trials were omitted from the publisher's score due to timeouts. We retain their recorded costs but assign them a service value of $0$. Thus every depth is calibrated on the same tasks and number of attempts.

\subsection{Empirical Calibration}
\begin{figure}[htbp]
\centering
\caption{Service Value and Mean Service Time by Reasoning Effort.}
\vspace{1em}
\begin{minipage}{\textwidth}
\begin{subfigure}[t]{0.48\textwidth}
    \centering
    \caption{Service Value.}
    \includegraphics[width=\textwidth]{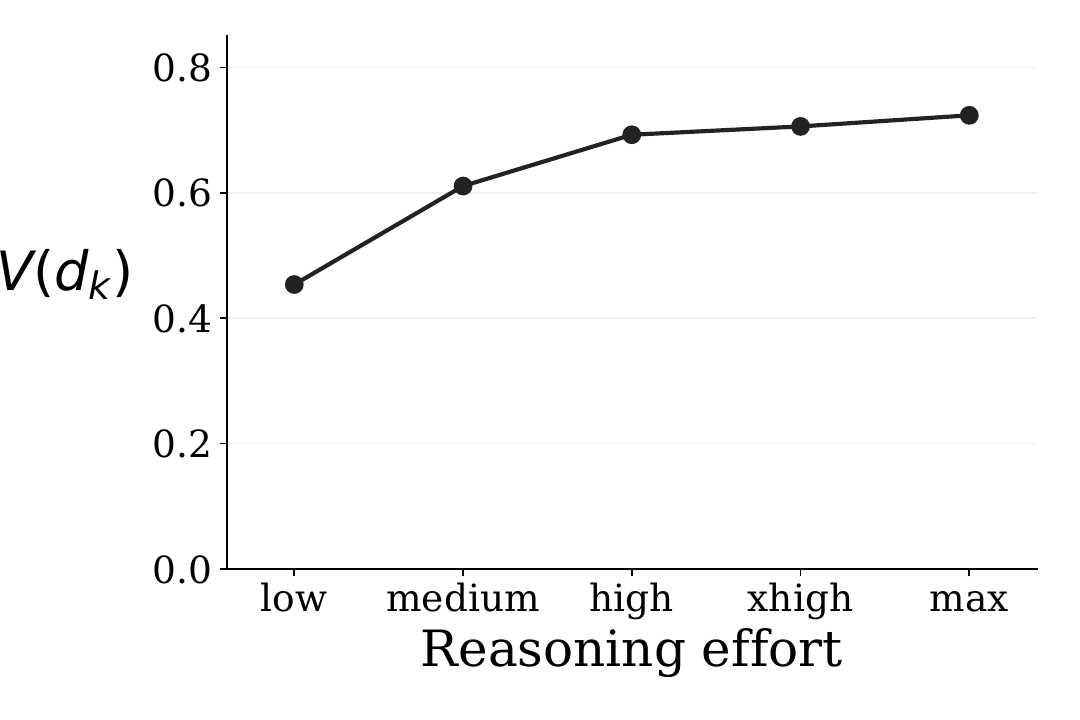}
    \label{fig:deepswe_calibration_value}
\end{subfigure}
\hfill
\begin{subfigure}[t]{0.48\textwidth}
    \centering
    \caption{Mean Service Time.}
    \includegraphics[width=\textwidth]{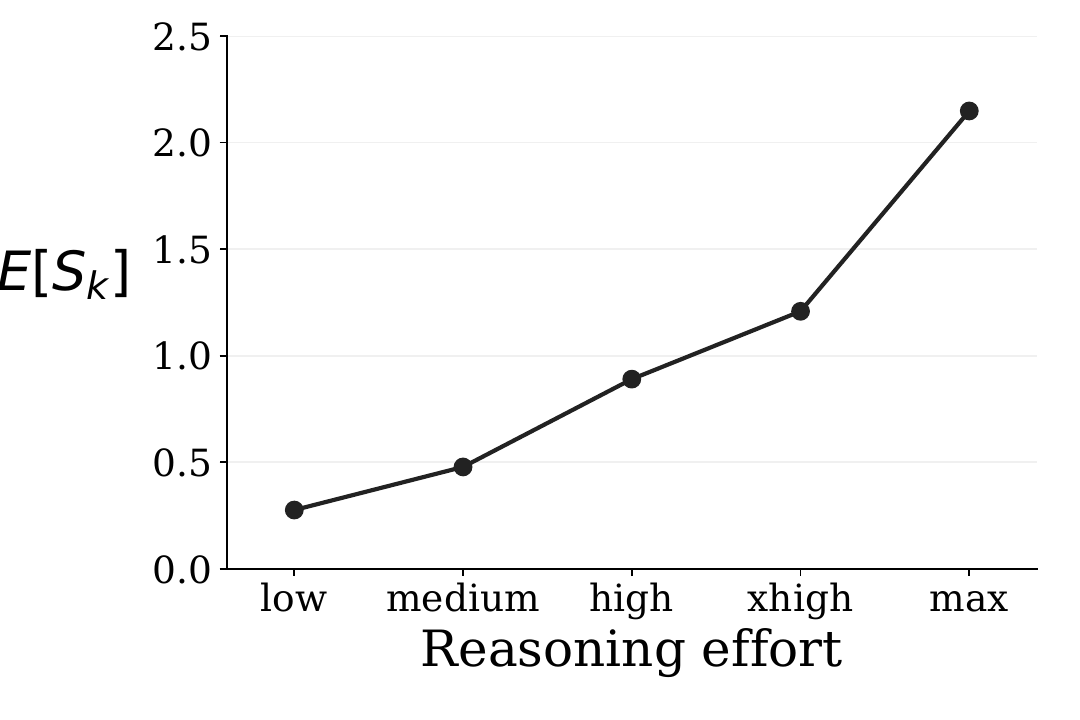}
    \label{fig:deepswe_calibration_time}
\end{subfigure}
\end{minipage}

\label{fig:deepswe_calibration}
\vspace{0.3em}
\begin{minipage}{0.9\textwidth}
\footnotesize
Note: Each point averages the 452 trials at that reasoning effort. Service time is $S_{jkr}=\kappa C_{jkr}$ with $\kappa\approx0.256657$.
\end{minipage}
\end{figure}

For task $j$, repetition $r$, and depth $d_k$, let $Y_{jkr}$ denote the indicator that the agent successfully solves the task. We estimate service value at depth $d_k$ by the empirical success rate
\begin{equation*}
    V(d_k)=\frac{1}{4N}\sum_{j=1}^{N}\sum_{r=1}^{4}Y_{jkr},\qquad N=113.
\end{equation*}
Let $C_{jkr}$ be the trial's recorded token-based monetary cost. We use this monetary cost to measure service time, denoted by $S_{jkr}$. This choice is motivated by the fact that token-based expenditure provides a consistent measure of the computational effort required to process a request. We therefore assume that service time is proportional to expenditure:
\begin{equation*}
    S_{jkr}=\kappa C_{jkr},\qquad \kappa\approx0.256657.
\end{equation*}
We choose $\kappa$ such that the mean service time across the 2,260 trials equals $1$. Table~\ref{tab:deepswe_calibration} and Figure~\ref{fig:deepswe_calibration} report the calibration. Increasing reasoning effort from \texttt{low} to \texttt{max}, success rises from $45.4\%$ to $72.3\%$, while mean service time increases from $0.276$ to $2.148$.

\begin{table}[htbp]
\centering
\caption{Empirical calibration by reasoning effort.}
\label{tab:deepswe_calibration}
\begin{tabular}{llcccc}
\toprule
Depth & Effort & $V(d_k)$ & Mean Cost (USD) & $E[S_k]$ & $E[S_k^2]$ \\
\midrule
$d_1$ & \texttt{low}    & 0.454 & 1.074 & 0.276 & 0.094 \\
$d_2$ & \texttt{medium} & 0.611 & 1.862 & 0.478 & 0.297 \\
$d_3$ & \texttt{high}   & 0.692 & 3.467 & 0.890 & 1.009 \\
$d_4$ & \texttt{xhigh}  & 0.706 & 4.708 & 1.208 & 1.842 \\
$d_5$ & \texttt{max}    & 0.723 & 8.370 & 2.148 & 6.632 \\
\bottomrule
\end{tabular}
\end{table}

\begin{figure}[htbp]
    \centering
    \caption{Empirical service-time distributions across reasoning
    depths.}
    \includegraphics[width=0.7\textwidth]
    {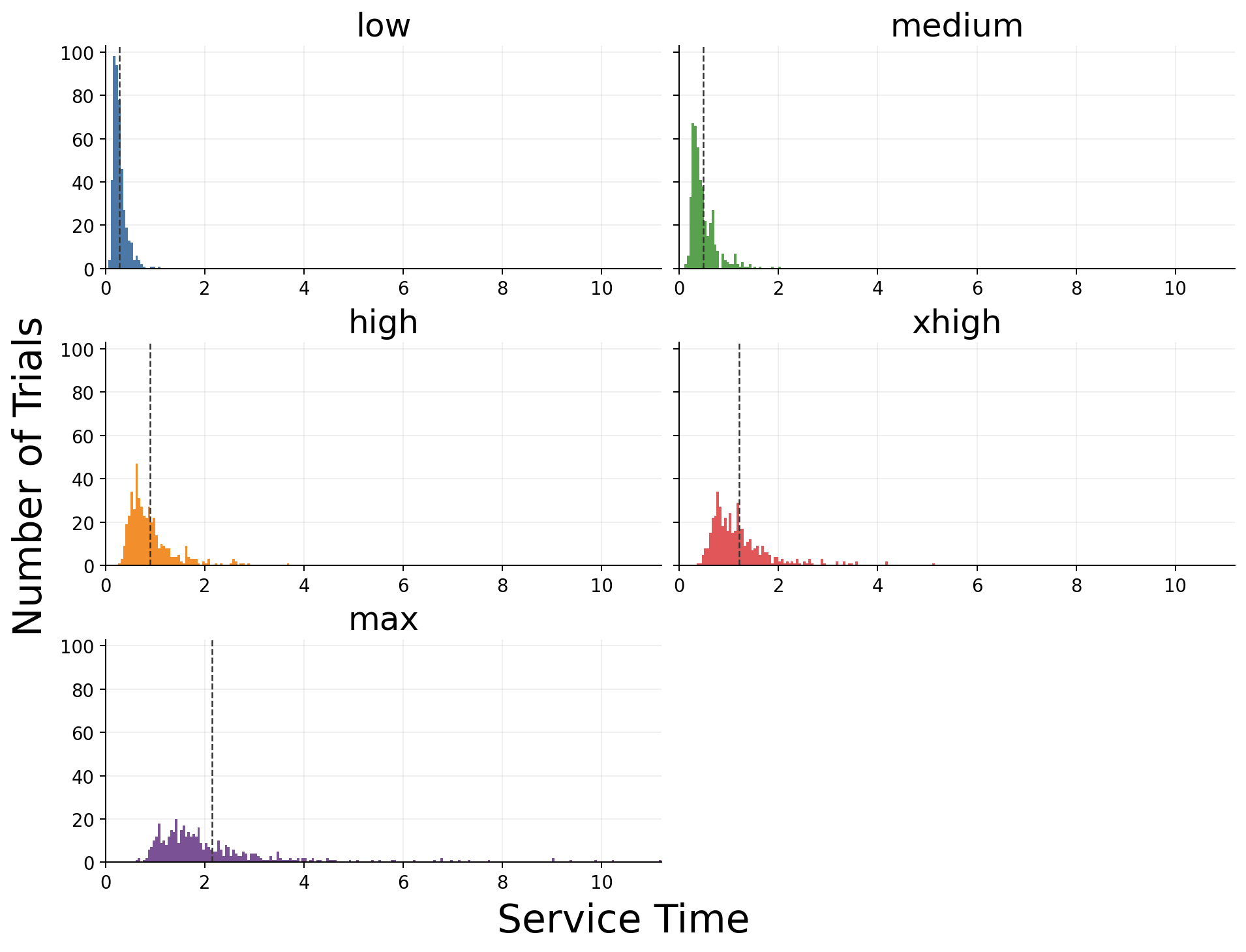}
    \vspace{0.5em}
    \begin{minipage}{0.65\textwidth}
    \footnotesize
    Note: Each panel contains 452 trials, with service
    time constructed as $S_{jkr}=\kappa C_{jkr}$. Dashed lines mark the means. Histograms use a common bin width of $0.05$.
    \end{minipage}
    \label{fig:empirical_service_times}
\end{figure}

Rather than assuming the service time follows an exponential distribution, we use the empirical distribution directly in the case study. Specifically, conditional on assigning service depth $d_k$, a customer's service time is drawn uniformly at random from the 452 observations, independently of customer delay sensitivity. Figure~\ref{fig:empirical_service_times} shows these distributions. 

We assume a job's realized service time is unknown until completion. As in the theoretical analysis, we restrict attention to deterministic policies that assign each customer type a service depth. Because the service time is unknown when processing begins, we restrict attention to scheduling rules that use only customer types: FCFS, and nonpreemptive priority order across types. These policies are easier to implement than policies that require tracking the progress of each job in service. Within each customer type, service follows FCFS.

\subsection{Two-Type Results}
\label{subsection:two type results}
We first consider two customer types, consistent with Section~\ref{section:two-type setting}. Each type receives a deterministic depth, and scheduling is FCFS or nonpreemptive priority to either type. We exhaustively evaluate $25\times3+10+1=86$ policies: 75 with both types served, 10 with only one type served, and the policy with neither type served. The optimal menu is the best of these policies satisfying incentive compatibility, which requires $w_1\leq w_2$ because type 1 is more delay-sensitive. We also construct prices and verify individual rationality and all incentive constraints.

\begin{figure}[htbp]
\centering
\caption{Comparative Statics with Respect to $\lambda$.}
\vspace{1em}
\begin{minipage}{\textwidth}
\begin{subfigure}[t]{0.33\textwidth}
    \centering
    \caption{Centralized Optimum.}
    \includegraphics[width=\textwidth]{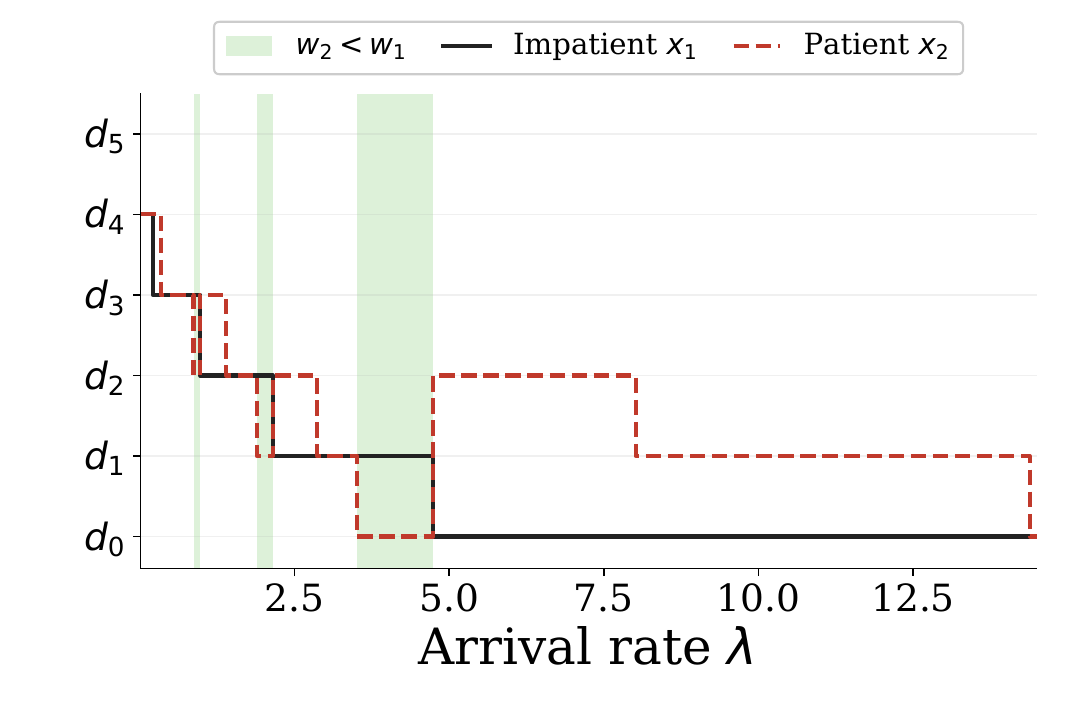}
    \label{fig:two_type_load_policy}
\end{subfigure}
\begin{subfigure}[t]{0.33\textwidth}
    \centering
    \caption{Optimal IC Menu.}
    \includegraphics[width=\textwidth]{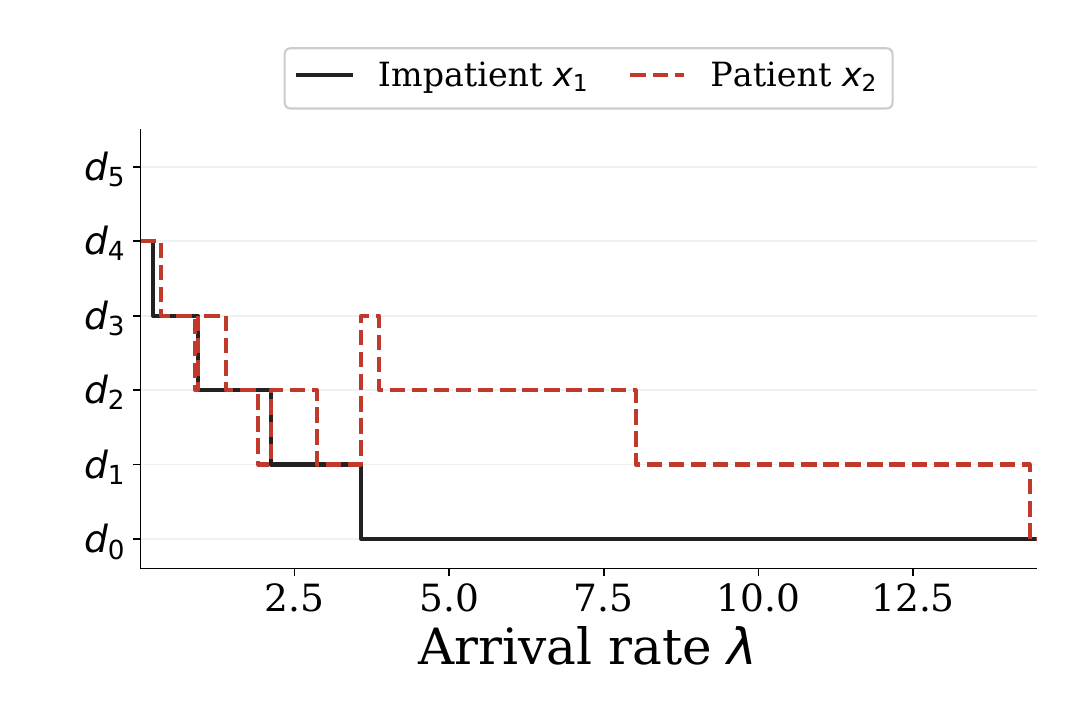}
    \label{fig:two_type_load_policy_ic}
\end{subfigure}
\begin{subfigure}[t]{0.33\textwidth}
    \centering
    \caption{Welfare.}
    \includegraphics[width=\textwidth]{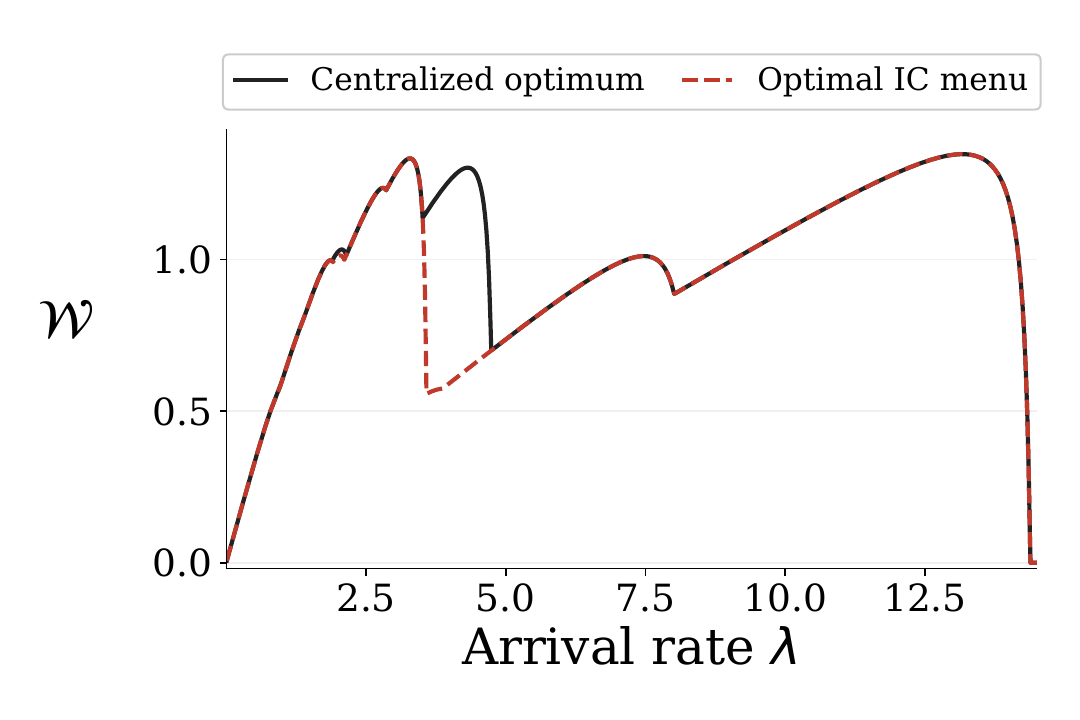}
    \label{fig:two_type_load_welfare}
\end{subfigure}
\end{minipage}

\label{fig:two_type_load}
\vspace{0.3em}
\begin{minipage}{0.9\textwidth}
\footnotesize
Note: $\theta_1=0.030$, $\theta_2=0.0225$, $\lambda_1=0.75\lambda$, and $\lambda_2=0.25\lambda$. The shaded regions in panel (a) indicate $w_2<w_1$, including assignments that exclude the patient type. When both types are served, these regions give priority to the patient type. Panel (b) gives the impatient type priority whenever both types are served. Priority is nonpreemptive throughout.
\end{minipage}

\end{figure}

Figure~\ref{fig:two_type_load} reports the optimal centralized policy, the optimal incentive-compatible menu, and the resulting social welfare as the arrival rate $\lambda$ varies. As demand increases, the decision maker first reduces reasoning effort, and then begins to exclude customers. In three windows, $0.87\leq\lambda\leq0.97$, $1.89\leq\lambda\leq2.15$ and $3.51\leq\lambda\leq4.74$, the centralized policy prioritizes patient customers. This is because the impatient type accounts for three quarters of the population and generates most of the service value, making it more valuable to serve this type with larger depths. When demand grows further, the impatient type itself becomes too congested, and the decision maker switches to serving the patient minority with larger depths. Therefore, welfare gaps exist in these windows and lead to a $53.4\%$ welfare loss in maximum.

\begin{figure}[htbp]
\centering
\caption{Comparative Statics with Respect to $\lambda_2/\lambda_1$.}
\vspace{1em}
\begin{minipage}{\textwidth}
\begin{subfigure}[t]{0.33\textwidth}
    \centering
    \caption{Centralized Optimum.}
    \includegraphics[width=\textwidth]{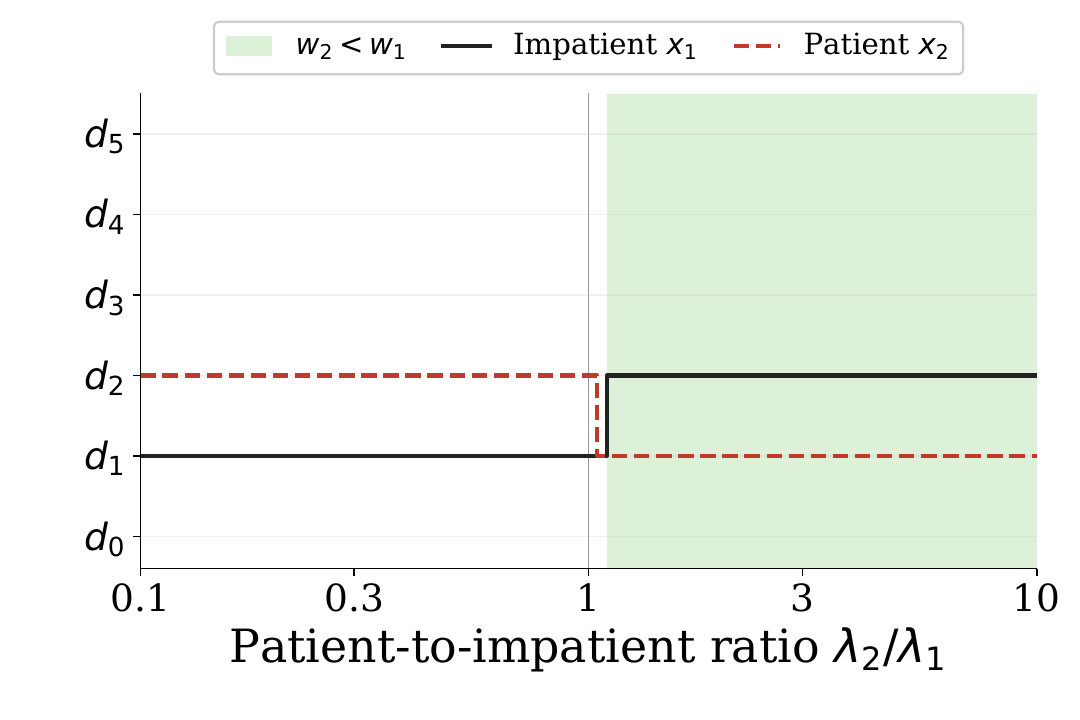}
    \label{fig:two_type_composition_policy}
\end{subfigure}
\begin{subfigure}[t]{0.33\textwidth}
    \centering
    \caption{Optimal IC Menu.}
    \includegraphics[width=\textwidth]{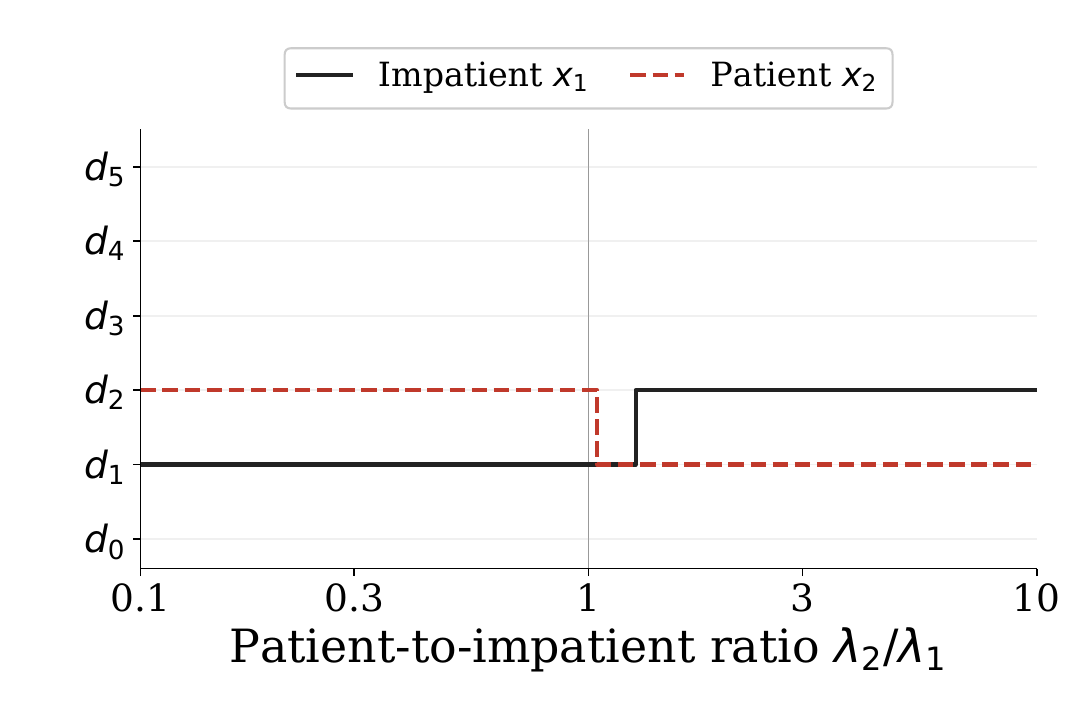}
    \label{fig:two_type_composition_policy_ic}
\end{subfigure}
\begin{subfigure}[t]{0.33\textwidth}
    \centering
    \caption{Welfare.}
    \includegraphics[width=\textwidth]{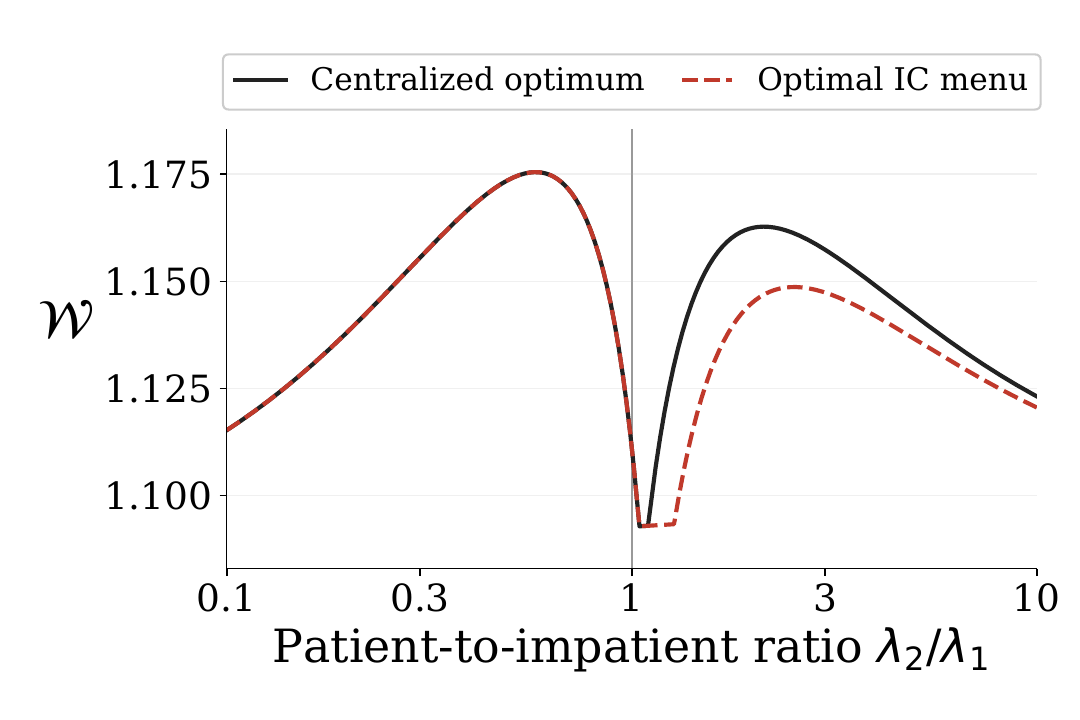}
\end{subfigure}
\end{minipage}

\label{fig:two_type_composition}
\vspace{0.3em}
\begin{minipage}{0.9\textwidth}
\footnotesize
Note: $\theta_1=0.030$, $\theta_2=0.0225$, and $\lambda_1+\lambda_2=2.5$. The ratio ranges from $0.1$ to $10$, symmetrically around $1$ on a logarithmic axis. The shaded region in panel (a) gives priority to the patient type; the impatient type has priority otherwise.
\end{minipage}

\end{figure}

Figure~\ref{fig:two_type_composition} varies the composition of demand at a fixed total arrival rate, with the same delay sensitivities as in Figure~\ref{fig:two_type_load}. Below a ratio of approximately $1.04$, impatient customers receive \texttt{low} effort and priority, while patients receive \texttt{medium} effort. Above a ratio of approximately $1.10$, the assignment reverses: the patient majority receives \texttt{low} effort and priority, and the impatient minority receives \texttt{medium} effort; in between, both types receive \texttt{low} effort. Serving everyone at \texttt{medium} would overload the server, and the decision maker can afford it for only one type. Giving the deeper service to the smaller group saves capacity; once patients form a sufficiently large majority, this saving outweighs the impatient type's higher delay cost. This reproduces the priority inversion in Proposition~\ref{proposition:welfare_maximization_centralized_case}.

The optimal IC menu continues to prioritize impatient customers. It can still give them deeper service, provided their sojourn remains shorter than that of patients. Such a policy leads to a welfare gap that reaches $3.28\%$. Together, the two experiments illustrate how population composition affects both whom to serve and whom to prioritize.

\subsection{Continuous-Type Results}
We next consider a continuum of customer types, following the theoretical analysis in Section~\ref{section:continuous-type setting}. Similar to Section~\ref{subsec: examples for the continuous type model}, we discretize the population to $n$ types with delay sensitivities $\theta_1<\cdots<\theta_n$ at quantiles of $F$, each arriving at rate $\lambda/n$. As in Section~\ref{subsection:two type results}, each type is assigned a service depth in $\{d_0,\ldots,d_5\}$ and a priority without preemption.

\begin{figure}[htbp]
\centering
\caption{Value of Joint Design with a Continuum of Types.}
\vspace{1em}
\begin{minipage}{\textwidth}
\begin{subfigure}[t]{0.33\textwidth}
    \centering
    \caption{Welfare.}
    \includegraphics[width=\textwidth]{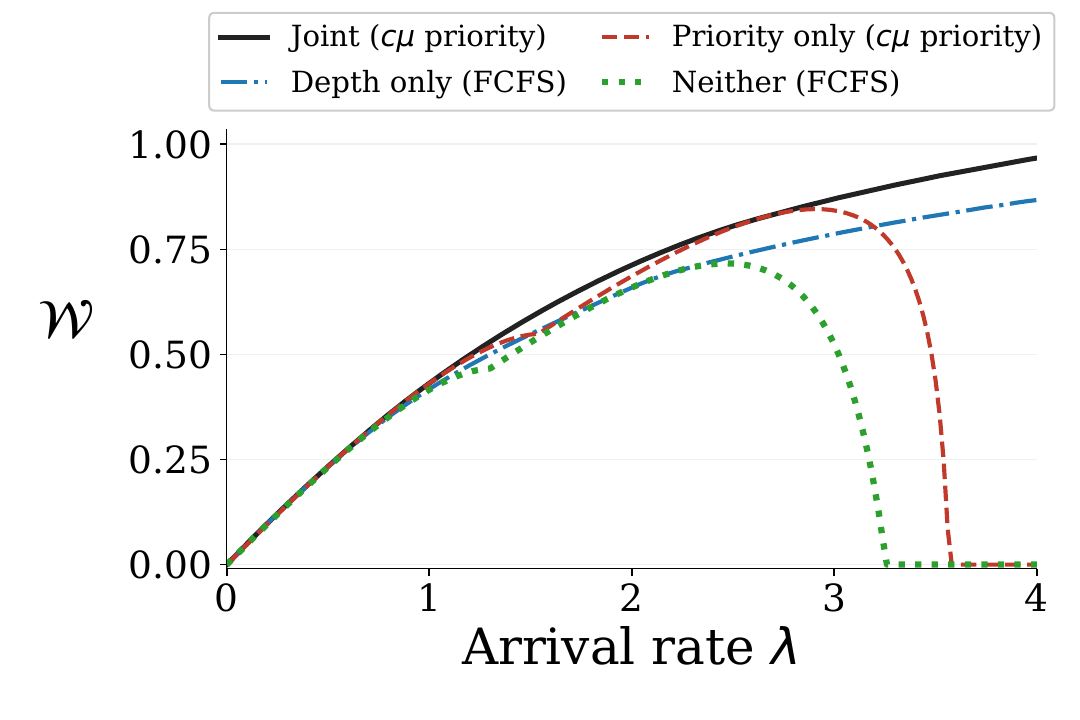}
    \label{fig:continuous_type_load_welfare}
\end{subfigure}
\begin{subfigure}[t]{0.33\textwidth}
    \centering
    \caption{Service Depth at $\lambda=1.5$.}
    \includegraphics[width=\textwidth]{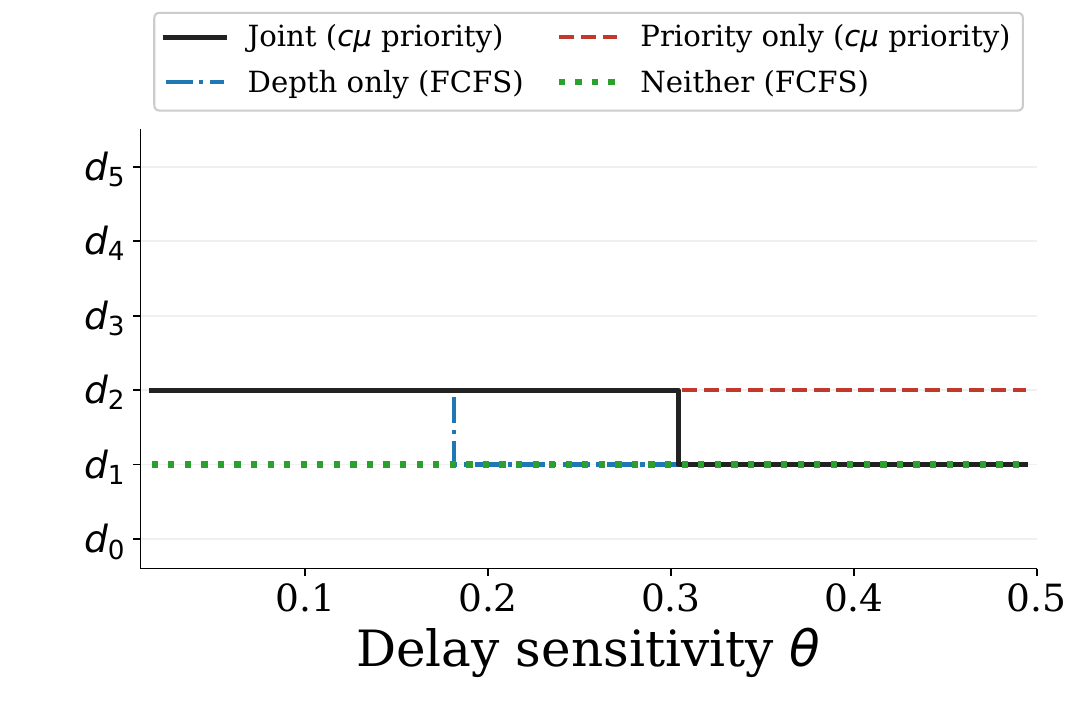}
    \label{fig:continuous_type_load_depth}
\end{subfigure}
\begin{subfigure}[t]{0.33\textwidth}
    \centering
    \caption{Sojourn Time at $\lambda=1.5$.}
    \includegraphics[width=\textwidth]{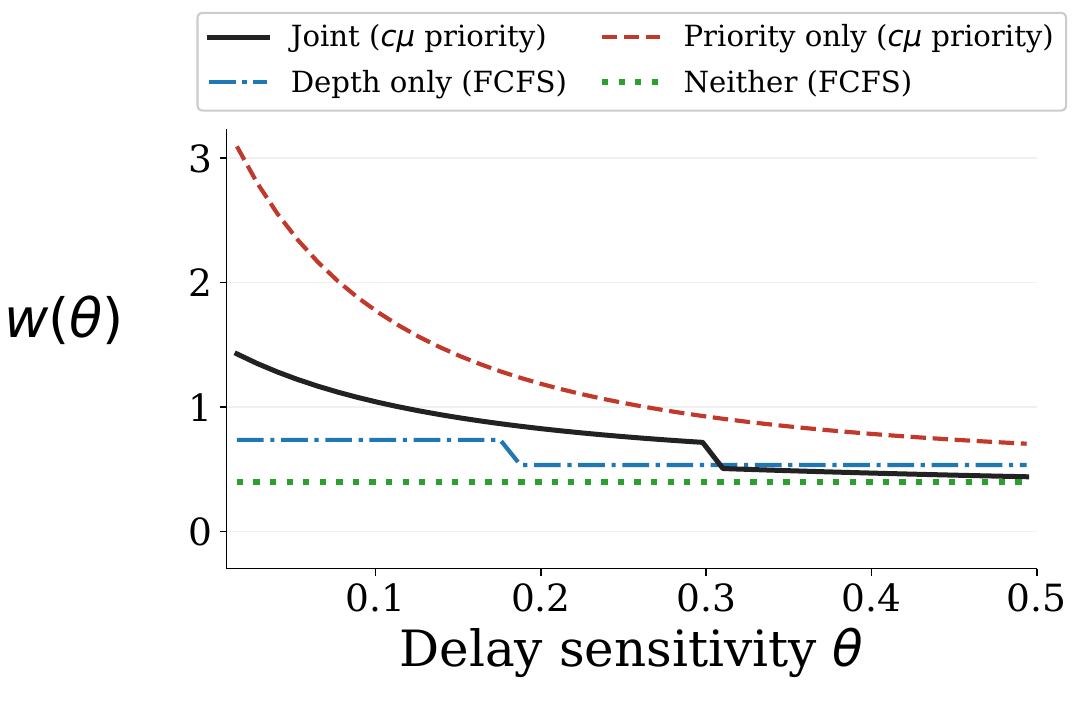}
    \label{fig:continuous_type_load_sojourn}
\end{subfigure}
\end{minipage}

\label{fig:continuous_type_load}
\vspace{0.3em}
\begin{minipage}{0.9\textwidth}
\footnotesize
Note: $n=40$ and $\theta\sim\mathrm{Unif}[0.01,0.5]$. Panel (a) displays $0\leq\lambda\leq4$; panels (b)--(c) use $\lambda=1.5$.
\end{minipage}
\end{figure}

We first compare the joint design with three benchmarks: depth-only design, which assigns type-dependent depths but serves customers FCFS; priority-only design, which assigns a common depth to all types and prioritizes by delay sensitivity; and a design that does neither. Figure~\ref{fig:continuous_type_load_welfare} reports the welfare of the four designs as the arrival rate varies. When the arrival rate is low, the four designs are nearly indistinguishable: the server is rarely congested, so neither decision has much to act on. For $0.66\leq\lambda\leq0.90$, priority becomes the more valuable decision. The joint design gives every type \texttt{medium} effort, so priority alone attains the welfare of the joint design. When the arrival rate is extremely high, service depth becomes the decision that matters. Designs with a common depth must serve every customer or none, whereas type-dependent depth keeps the system viable by withholding service from delay-sensitive customers.

Figures~\ref{fig:continuous_type_load_depth} and~\ref{fig:continuous_type_load_sojourn} show the mechanism at $\lambda=1.5$, where the four designs assign different service depths. The design that does neither gives every customer \texttt{low} effort, and priority-only gives every customer \texttt{medium} effort. Both personalized designs give \texttt{medium} effort to the relatively patient customers and \texttt{low} effort to the rest. Under FCFS, every served customer bears the same waiting time, so the congestion from deeper service falls on everyone, and depth-only design gives \texttt{medium} effort only to the $35\%$ most patient customers. Priority shields the most delay-sensitive customers from this congestion: the joint design gives \texttt{medium} effort to the $60\%$ most patient customers, while the sojourn time of the most delay-sensitive customers is $0.44$, compared with $0.53$ under depth-only design. The joint design attains welfare $0.589$, about $7\%$ above depth-only ($0.548$) and priority-only ($0.547$). At every arrival rate we evaluate, the joint design has nonincreasing sojourn times and is incentive compatible, in line with Proposition~\ref{proposition:continuous_types_welfare_maximization_decentralized_case}.

Finally, we quantify the value of personalization discussed in Section~\ref{section:continuous-type setting}.  We consider a population concentrated in two narrow clusters: $75\%$ of the population have delay sensitivities uniformly spaced on $[0.0295,0.0305]$ and the remaining $25\%$ on $[0.0220,0.0230]$. The cluster means match the two-type load experiment. We analyze this population in two ways. In the two-type treatment, each cluster is one customer type that receives a single service depth and priority position. In the continuum treatment, the population is discretized into $n=40$ types as above, so customers within a cluster may receive different service depths.

\begin{figure}[htbp]
\centering
\caption{Value of Personalization for a Two-Cluster Population.}
\vspace{1em}
\begin{minipage}{\textwidth}
\begin{subfigure}[t]{0.48\textwidth}
    \centering
    \caption{Welfare.}
    \includegraphics[width=\textwidth]{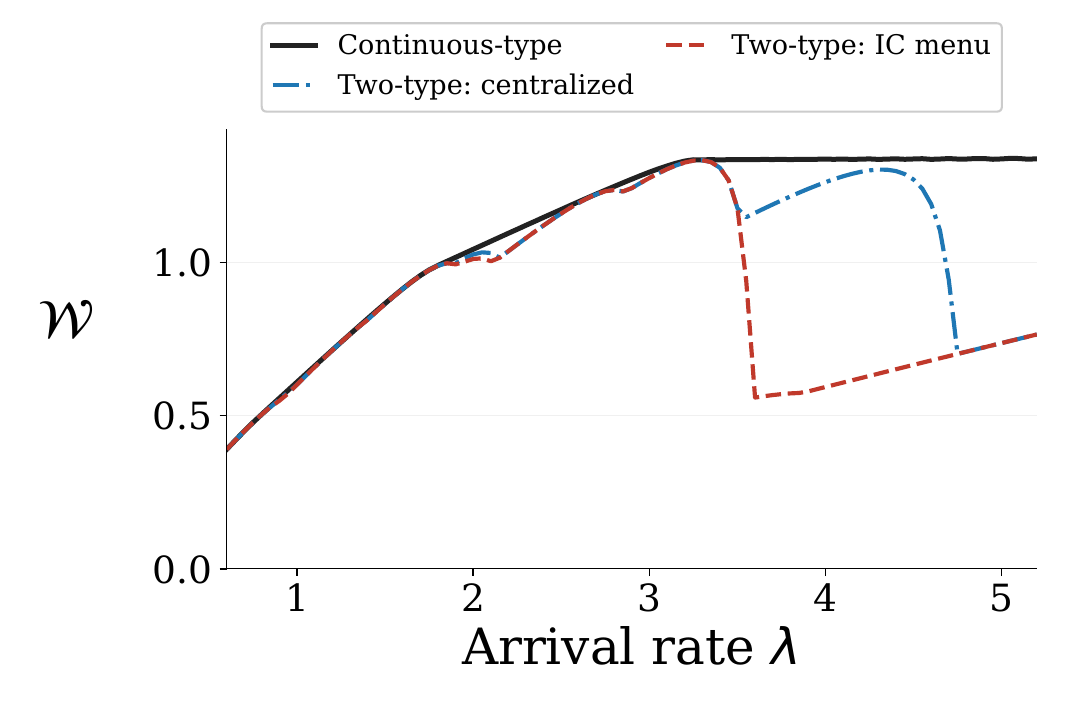}
    \label{fig:continuous_type_personalization_welfare}
\end{subfigure}
\hfill
\begin{subfigure}[t]{0.48\textwidth}
    \centering
    \caption{Service Depth at $\lambda=3.8$.}
    \includegraphics[width=\textwidth]{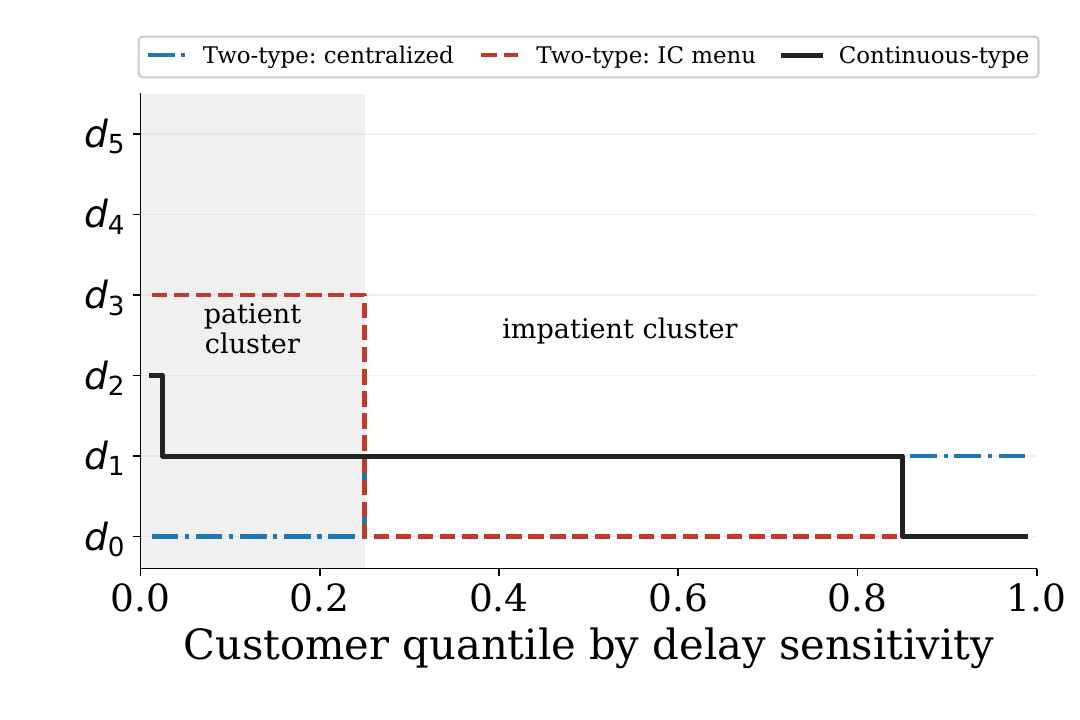}
    \label{fig:continuous_type_personalization_depth}
\end{subfigure}
\end{minipage}

\label{fig:continuous_type_personalization}
\vspace{0.3em}
\begin{minipage}{0.9\textwidth}
\footnotesize
Note: $n=40$, with $75\%$ of types at midpoint quantiles of $\mathrm{Unif}[0.0295,0.0305]$ and $25\%$ at midpoint quantiles of $\mathrm{Unif}[0.0220,0.0230]$. The two-type treatment uses the cluster means $\theta_1=0.030$ and $\theta_2=0.0225$, with $\lambda_1=0.75\lambda$ and $\lambda_2=0.25\lambda$. Panel (b) uses $\lambda=3.8$. Fine personalization allows both depths and priorities to vary within each cluster.
\end{minipage}
\end{figure}

Figure~\ref{fig:continuous_type_personalization_welfare} shows the resulting welfare. At $\lambda=3.8$, the coarse centralized policy excludes patients and attains welfare $1.215$, while the optimal coarse IC menu serves patients alone and attains $0.573$. Fine personalization instead serves every patient and $80\%$ of impatient customers, excluding only the most delay-sensitive tail (Figure~\ref{fig:continuous_type_personalization_depth}). This allocation is incentive compatible and attains welfare $1.336$, about $9.9\%$ above the coarse centralized optimum. Personalization thus removes the conflict between efficiency and incentive compatibility.

\section{Extensions}
\label{section:extensions}

\subsection{Revenue Maximization}
\label{appendix:revenue_maximization}

In this section, we focus on the revenue maximization problem. The revenue objective is
\begin{equation*}
    \mathcal{R}:= \lambda\mathbb{E}_{\theta}[p(\theta)],
\end{equation*}
and the feasibility constraints are the same as the welfare maximization problem.

In the two-type model, the revenue maximization problem is written as
\begin{equation}
    \label{eq:revenue_maximization_decentralized}
    \begin{aligned}
        \max_{x_1, x_2, w_1, w_2,p_1,p_2}&\quad \lambda_1p_1+\lambda_2p_2\\
        \textbf{s.t.}&\quad w_i\geq\frac{x_i}{1-\rho_i},\quad i=1,2,\\
        &\quad \rho_1w_1+\rho_2w_2\geq\frac{\rho_1x_1+\rho_2x_2}{1-\rho_1-\rho_2},\\
        &\quad \rho_1+\rho_2<1,\\
        &\quad V(x_1)-\theta_1w_1-p_1\geq V(x_2)-\theta_1w_2-p_2,\\
        &\quad V(x_1)-\theta_2w_1-p_1\leq V(x_2)-\theta_2w_2-p_2,\\
        &\quad V(x_i)-\theta_iw_i-p_i\geq 0,\quad i=1,2.
    \end{aligned}
\end{equation}

\begin{proposition}
    \label{proposition:revenue_maximization}
    The revenue maximization problem is equivalent to Problem~\eqref{eq:welfare_maximization_decentralized} with virtual delay sensitivities $\varphi(\theta_1)=\theta_1+\frac{\lambda_2}{\lambda_1}(\theta_1-\theta_2)$, $\varphi(\theta_2)=\theta_2$.  
\end{proposition}

In the continuous-type model, the revenue maximization problem is
\begin{equation}
    \label{equation:continuous_types_revenue_maximization}
    \begin{aligned}
        \max_{x(\cdot), w(\cdot),p(\cdot)}\quad&\lambda\int_{\underline{\theta}}^{\bar{\theta}}p(\theta)dF(\theta)\\
        \text{s.t.}\quad&\lambda\int_\mathcal{S} x(\theta)w(\theta)f(\theta)d\theta\geq \frac{\lambda\int_\mathcal{S} x^2(\theta)f(\theta)d\theta}{1-\lambda\int_\mathcal{S} x(\theta)f(\theta)d\theta}, \quad \forall\mathcal{S}\subseteq [\underline{\theta},\bar{\theta}],\\
        &V(x(\theta))-\theta w(\theta)-p(\theta)\geq V(x(\hat\theta))-\theta w(\hat\theta)-p(\hat\theta),\quad \forall \theta,\hat\theta\in[\underline{\theta},\bar{\theta}],\\
        &V(x(\theta))-\theta w(\theta)-p(\theta)\geq 0,\quad \forall \theta\in[\underline{\theta},\bar{\theta}].
    \end{aligned}
\end{equation}

We impose an additional assumption that the virtual delay sensitivities are monotone in $\theta$. This is a common assumption in the mechanism design literature \citep{afeche2016optimal,yang2017trading}. Under this regularity assumption, the ordering of virtual delay sensitivities agrees with the ordering of true delay sensitivities. Therefore, the structural analysis for the continuous-type welfare problem can be applied after replacing $\theta$ by virtual delay sensitivities.
\begin{assumption}
    \label{assumption:continuous_types_revenue_regular}
    The probability density function $f$ is continuously differentiable on $[\underline{\theta},\bar{\theta}]$. Moreover, the virtual delay sensitivity
    \begin{equation*}
        \varphi(\theta):=\theta+\frac{F(\theta)}{f(\theta)}
    \end{equation*}
    is a strictly increasing function in $\theta$.
\end{assumption}

\begin{proposition}
    \label{proposition:continuous_type_revenue_reduction}
    Suppose Assumption~\ref{assumption:continuous_types_revenue_regular} holds. For any operationally feasible and incentive compatible allocation $(x(\cdot),w(\cdot))$, the maximum long-term average revenue attainable by payments is
    \begin{equation*}
        \lambda\int_{\underline{\theta}}^{\bar{\theta}}\left(V(x(\theta))-\varphi(\theta) w(\theta)\right)dF(\theta),
    \end{equation*}
    where $\varphi(\theta)=\theta+F(\theta)/f(\theta)$. Consequently, Problem~\eqref{equation:continuous_types_revenue_maximization} is equivalent to Problem~\eqref{equation:continuous_types_final_problem} with virtual delay sensitivities.
\end{proposition}

\subsection{Randomized Service Depth in the Two-Type Model}
\label{subsection:randomized}

In Section~\ref{section:two-type setting} and Section~\ref{section:continuous-type setting}, we restrict our attention to policies that assign a deterministic service depth to each type of customers. This restriction creates a welfare gap with two types but not with a continuum of types. The difference stems from the decision maker's ability to tailor service depth assignments across customer types. In the two-type model, each type represents a customer group of positive mass. As a result, changing the service depth assigned to a type necessarily changes the operational role of an entire customer type. When the patient type is sufficiently large, the decision maker may prefer to assign it a smaller service depth and move it ahead in the queue. Although this reduces congestion, it reverses the sojourn time ordering required for incentive compatibility.

The continuous-type model provides substantially greater flexibility. Because types vary continuously, the decision maker can adjust service depth assignments smoothly across the type space. Rather than moving less delay-sensitive customers in front of more delay-sensitive customers in the queue, the decision maker assigns service depths monotonically, so that customers with slightly greater delay sensitivity receive slightly smaller service depth. The queue is still organized according to the same $c\mu$ principle, but the service depth assignment can be chosen so that the $c\mu$ ordering and the incentive compatible ordering coincide.

Motivated by this, we consider an approach that uses randomized service depths within one type of customer to close the welfare gap. In the two type case with $\theta_1>\theta_2>0$, instead of assigning a fixed service depth $x_i$ to each type of customer, we assign a service depth distribution $\mathcal{G}_i$ to each type of customer. When a type-$i$ customer arrives, a service depth $x$ is drawn from $\mathcal{G}_i$, and the realized service time for the customer is still an exponential distribution with mean $x$. The priority rule and the expected sojourn time will be based on the realized service depths. Proposition~\ref{proposition:randomized} states that under this setting, the optimal randomized service depth assignment has a monotone structure similar to the continuous-type model: the patient type receives weakly higher service depth than the impatient type.

\begin{proposition}
    \label{proposition:randomized}
    Fix any feasible randomized service depth assignment $(\mathcal{G}_1,\mathcal{G}_2)$ and any priority rule. There exists a feasible randomized service depth assignment $(\tilde{\mathcal{G}}_1,\tilde{\mathcal{G}}_2)$ such that
    \begin{equation*}
        \sup\operatorname{supp}(\tilde{\mathcal{G}}_1)\leq \inf\operatorname{supp}(\tilde{\mathcal{G}}_2),
    \end{equation*}
    and which, under the $c\mu$-rule, achieves weakly higher long-term average social welfare. Under $(\tilde{\mathcal{G}}_1,\tilde{\mathcal{G}}_2)$ the $c\mu$-rule gives priority to smaller realized service depth, so all the impatient customers have priority over patient customers.
\end{proposition}

In the decentralized setting, we implement a menu $\mathcal{M}=\{(p(x),x,w(x)):x\in\mathcal{X}\}$ which may contain more than two options. $\mathcal{X}$ is the set of offered depths, $p(x)$ is the price, and $w(x)$ is the expected sojourn time of a customer who selects depth $x$. The provider implements the randomized assignment through contracts: a customer who chooses contract $i$ is assigned an option $x\in\mathcal{X}$ drawn from $\mathcal{G}_i$. We require incentive compatibility to hold after the assignment. That is, for every $x\in\operatorname{supp}(\mathcal{G}_i)$, a type-$i$ customer weakly prefers the assigned option $(p(x),x,w(x))$ to every other option in $\mathcal{M}$ and obtains a nonnegative surplus. As before, the announced sojourn times must coincide with the steady-state sojourn times induced by these choices.

\begin{lemma}
    \label{lemma:randomized_ic}
    A feasible randomized assignment $(\mathcal{G}_1,\mathcal{G}_2)$ with sojourn times $w(\cdot)$ is implementable if and only if
    \begin{equation*}
        w(x_1)\leq w(x_2),\qquad \forall x_1\in\operatorname{supp}(\mathcal{G}_1),\ x_2\in\operatorname{supp}(\mathcal{G}_2).
    \end{equation*}
    In this case, letting $\bar w_1:=\sup\{w(x):x\in\operatorname{supp}(\mathcal{G}_1)\}$, one implementing price rule is
    \begin{equation*}
        p(x)=
        \begin{cases}
            V(x)-\theta_1 w(x), & x\in\operatorname{supp}(\mathcal{G}_1),\\
            V(x)-\theta_2 w(x)-(\theta_1-\theta_2)\bar w_1, & x\in\operatorname{supp}(\mathcal{G}_2).
        \end{cases}
    \end{equation*}
\end{lemma}

Lemma~\ref{lemma:randomized_ic} is the counterpart of Lemma~\ref{lemma:welfare_maximization_incentive_compatibility}: every option chosen by the impatient type must have a weakly shorter sojourn time than every option chosen by the patient type. By Proposition~\ref{proposition:randomized}, the decision maker can restrict attention to assignments with exactly this property, which gives the main result of this subsection.
\begin{proposition}
    \label{proposition:randomized_no_gap}
    With randomized service depths, any welfare achievable by the centralized decision maker is also achievable by a price menu.
\end{proposition}

\subsection{Dynamic Policies}
\label{subsection:dynamic}

In practice, a service system may adjust the service depth to congestion, for example by providing shallower service when the queue is long. A fully state-dependent model, in which the service depth of each arriving customer may depend on the types and assigned depths of all customers in the system, is substantially more complex. We therefore consider a simpler class of threshold policies in the two-type model that captures the main effect of congestion-dependent control. The decision maker chooses service depths $(x_1,x_2)$, a priority rule, and a threshold $N$. A customer who arrives when the expected workload in the system is $n_1x_1+n_2x_2<N$ receives service depth $x_i$; otherwise the customer leaves immediately. The static model corresponds to $N=\infty$. Under a threshold policy, the numbers of customers $(n_1,n_2)$ form a Markov chain on a finite state space, so the induced steady-state quantities are well defined for any $(x_1,x_2)$.

Let $\alpha$ denote the steady-state probability that an arrival is admitted, which is the same for both types since they are blocked in the same states, and let $w_i$ denote the expected sojourn time of admitted type-$i$ customers. The social welfare is $\alpha\sum_{i}\lambda_i\left(V(x_i)-\theta_iw_i\right)$. Since $\alpha$ and $w_i$ have no closed form for general $N$, we compare the two models numerically.

Table~\ref{tab:threshold_policy} compares the optimal static and threshold policies across $\lambda_2/\lambda_1$. Threshold control reduces the congestion of the system and improves welfare, but more importantly, threshold control reduces the welfare gap. The gap vanishes at $\lambda_2/\lambda_1=2$, where the static policy already prioritizes patient customers, and stays below $0.3\%$ for larger ratios, compared with up to $2.2\%$ under the static policy. With the threshold control, the decision maker is able to keep both sojourn times short and the incentive conflict of Section~\ref{section:two-type setting} becomes less costly.

\begin{table}[htbp]
    \centering
    \caption{Static and threshold policies across $\lambda_2/\lambda_1$}
    \label{tab:threshold_policy}
    \renewcommand{\arraystretch}{1.1}

    \begin{tabular}{lccccccc}
    \toprule
    & \multicolumn{7}{c}{$\lambda_2/\lambda_1$} \\
    \cmidrule(lr){2-8}
    & 0.5 & 1.0 & 2.0 & 4.0 & 8.0 & 16.0 & 32.0 \\
    \midrule
    Welfare gap, static policy & 0.0\% & 0.0\% & 0.4\% & 1.7\% & 2.2\% & 2.0\% & 1.5\%\\
    Welfare gap, threshold policy & 0.0\% & 0.0\% & 0.0\% & 0.1\% & 0.2\% & 0.2\% & 0.2\%\\
    Welfare improvement, centralized & 13.2\% & 13.9\% & 14.0\% & 13.1\% & 13.0\% & 13.4\% & 14.0\%\\
    Welfare improvement, decentralized & 13.2\% & 13.9\% & 14.5\% & 15.0\% & 15.3\% & 15.4\% & 15.5\%\\
    \bottomrule
    \\
    \end{tabular}
    \begin{minipage}{0.75\textwidth}
    \footnotesize
    Note: The parameters are $\lambda=1.0$, $\theta_1=0.105$, $\theta_2=0.1$, and $V(x)=\log(1+x)$. The welfare improvement is the relative increase in optimal welfare from the static policy to the threshold policy, for the centralized decision maker and for the decentralized mechanism respectively.
    \end{minipage}
\end{table}

\section{Conclusion}
\label{section:conclusion}

This paper studies the joint design of service depth and service priority in a single-server queue with heterogeneous customers. We characterize how the structure of the optimal policy depends on customers’ delay sensitivities and the population sizes of different customer types, and analyze the difference between the centralized decision maker’s solution and the optimal mechanism under private information. In the two-type setting, we show that the welfare-maximizing policy can have a surprising structure: the decision maker may optimally prioritize the more patient type, which violates the incentive compatibility constraints and cannot be implemented by a decentralized menu. In the continuum-of-types setting, in contrast, the price menu always matches the decision maker's policy. Our results highlight the importance of accounting for customer population sizes and allowing sufficient personalization of the policy when designing service systems.

Our work shows that increasing flexibility fundamentally changes the nature of queueing systems. In particular, when both service depths and priorities are treated as decision variables, the structure of the optimal policy differs from that in traditional queueing models. This implies that, in modern service environments such as agentic systems, it is important to develop new modeling frameworks and policies rather than relying on classical approaches. For practitioners, our results highlight a fundamental trade-off in service system design: simple mechanisms are easier to develop and implement, whereas more personalized mechanisms can substantially improve system efficiency but are typically more complex to design and operate.

Several promising directions remain for future research. On the theoretical side, it would be valuable to move beyond this framework and study more general settings, such as non-exponential service requirements, multi-server systems, and state-dependent policies. It would also be interesting to study different kinds of heterogeneity. For example, customers may differ not only in their delay sensitivities but also in the types of tasks they require, which may generate different utilities. Similarly, servers may possess varying levels of expertise across different classes of tasks, leading to heterogeneous service capabilities. Another natural extension is to allow unsuccessful or low-quality service to trigger follow up requests, as when an insufficient AI response prompts the user to ask again; service depth would then affect not only value and congestion but also the arrival rate itself, creating a feedback loop between depth choices and demand. On the application side, it remains challenging to apply such policies to real AI-agent systems. It will be meaningful to validate the theoretical insights in practical settings, develop reliable measures of AI service quality, and better understand the relationship between processing time, service quality, and system congestion in real-world agentic systems.

\bibliographystyle{ormsv080} 
\bibliography{references}

@article{lin2023wait,
  title={Wait time--based pricing for queues with customer-chosen service times},
  author={Lin, Chen-An and Shang, Kevin and Sun, Peng},
  journal={Management Science},
  volume={69},
  number={4},
  pages={2127--2146},
  year={2023},
  publisher={INFORMS}
}

@article{lin2024dynamic,
  title={Dynamic Pricing and Release Time Control for Service Systems},
  author={Lin, Chen-An Andy and Shang, Kevin and Sun, Peng},
  journal={Available at SSRN 5222063},
  year={2024}
}

@article{afeche2013incentive,
  title={Incentive-compatible revenue management in queueing systems: Optimal strategic delay},
  author={Af{\`e}che, Philipp},
  journal={Manufacturing \& Service Operations Management},
  volume={15},
  number={3},
  pages={423--443},
  year={2013},
  publisher={INFORMS}
}

@article{afeche2016optimal,
  title={Optimal price/lead-time menus for queues with customer choice: Segmentation, pooling, and strategic delay},
  author={Af{\`e}che, Philipp and Pavlin, J Michael},
  journal={Management Science},
  volume={62},
  number={8},
  pages={2412--2436},
  year={2016},
  publisher={INFORMS}
}

@article{yang2017trading,
  title={Trading time in a congested environment},
  author={Yang, Luyi and Debo, Laurens and Gupta, Varun},
  journal={Management Science},
  volume={63},
  number={7},
  pages={2377--2395},
  year={2017},
  publisher={INFORMS}
}

@article{mendelson1990optimal,
  title={Optimal incentive-compatible priority pricing for the M/M/1 queue},
  author={Mendelson, Haim and Whang, Seungjin},
  journal={Operations Research},
  volume={38},
  number={5},
  pages={870--883},
  year={1990},
  publisher={INFORMS}
}

@article{kleinrock1967optimum,
  title={Optimum bribing for queue position},
  author={Kleinrock, Leonard},
  journal={Operations Research},
  volume={15},
  number={2},
  pages={304--318},
  year={1967},
  publisher={INFORMS}
}

@article{lui1985equilibrium,
  title={An equilibrium queuing model of bribery},
  author={Lui, Francis T},
  journal={Journal of Political Economy},
  volume={93},
  number={4},
  pages={760--781},
  year={1985},
  publisher={The University of Chicago Press}
}

@article{coffman1980characterization,
  title={A characterization of waiting time performance realizable by single-server queues},
  author={Coffman Jr, Edward G and Mitrani, Isi},
  journal={Operations Research},
  volume={28},
  number={3-part-ii},
  pages={810--821},
  year={1980},
  publisher={INFORMS}
}

@inproceedings{scully2021gittins,
  title={The Gittins policy in the M/G/1 queue},
  author={Scully, Ziv and Harchol-Balter, Mor},
  booktitle={2021 19th International Symposium on Modeling and Optimization in Mobile, Ad hoc, and Wireless Networks (WiOpt)},
  pages={1--8},
  year={2021},
  organization={IEEE}
}

@article{kittsteiner2005priority,
  title={Priority auctions and queue disciplines that depend on processing time},
  author={Kittsteiner, Thomas and Moldovanu, Benny},
  journal={Management Science},
  volume={51},
  number={2},
  pages={236--248},
  year={2005},
  publisher={INFORMS}
}

@article{weber1987optimal,
  title={Optimal control of service rates in networks of queues},
  author={Weber, Richard R and Stidham Jr, Shaler},
  journal={Advances in applied probability},
  volume={19},
  number={1},
  pages={202--218},
  year={1987},
  publisher={Cambridge University Press}
}

@article{stidham1989monotonic,
  title={Monotonic and insensitive optimal policies for control of queues with undiscounted costs},
  author={Stidham Jr, Shaler and Weber, Richard R},
  journal={Operations Research},
  volume={37},
  number={4},
  pages={611--625},
  year={1989},
  publisher={INFORMS}
}

@article{veatch1992monotone,
  title={Monotone control of queueing networks},
  author={Veatch, Michael H and Wein, Lawrence M},
  journal={Queueing Systems},
  volume={12},
  number={3},
  pages={391--408},
  year={1992},
  publisher={Springer}
}

@article{george2001dynamic,
  title={Dynamic control of a queue with adjustable service rate},
  author={George, Jennifer M and Harrison, J Michael},
  journal={Operations Research},
  volume={49},
  number={5},
  pages={720--731},
  year={2001},
  publisher={INFORMS}
}

@article{ata2006dynamic,
  title={Dynamic control of an M/M/1 service system with adjustable arrival and service rates},
  author={Ata, Bari{\c{s}} and Shneorson, Shiri},
  journal={Management Science},
  volume={52},
  number={11},
  pages={1778--1791},
  year={2006},
  publisher={INFORMS}
}

@article{ha1998incentive,
  title={Incentive-compatible pricing for a service facility with joint production and congestion externalities},
  author={Ha, Albert Y},
  journal={Management Science},
  volume={44},
  number={12-part-1},
  pages={1623--1636},
  year={1998},
  publisher={INFORMS}
}

@article{ha2001optimal,
  title={Optimal pricing that coordinates queues with customer-chosen service requirements},
  author={Ha, Albert Y},
  journal={Management Science},
  volume={47},
  number={7},
  pages={915--930},
  year={2001},
  publisher={INFORMS}
}

@article{afeche2004pricing,
  title={Pricing and priority auctions in queueing systems with a generalized delay cost structure},
  author={Af{\`e}che, Philipp and Mendelson, Haim},
  journal={Management Science},
  volume={50},
  number={7},
  pages={869--882},
  year={2004},
  publisher={INFORMS}
}

@article{smith1956various,
  title={Various optimizers for single-stage production},
  author={Smith, Wayne E},
  journal={Naval Research Logistics Quarterly},
  volume={3},
  number={1-2},
  pages={59--66},
  year={1956},
  publisher={Wiley Online Library}
}

@article{klimov1975time,
  title={Time-sharing service systems. I},
  author={Klimov, Gennadi P},
  journal={Theory of Probability \& Its Applications},
  volume={19},
  number={3},
  pages={532--551},
  year={1975},
  publisher={SIAM}
}

@article{van1995dynamic,
  title={Dynamic scheduling with convex delay costs: The generalized c| mu rule},
  author={Van Mieghem, Jan A},
  journal={The Annals of Applied Probability},
  pages={809--833},
  year={1995},
  publisher={JSTOR}
}

@article{cobham1954priority,
  title={Priority assignment in waiting line problems},
  author={Cobham, Alan},
  journal={Journal of the Operations Research Society of America},
  volume={2},
  number={1},
  pages={70--76},
  year={1954},
  publisher={INFORMS}
}

@article{kleinrock1965conservation,
  title={A conservation law for a wide class of queueing disciplines},
  author={Kleinrock, Leonard},
  journal={Naval Research Logistics Quarterly},
  volume={12},
  number={2},
  pages={181--192},
  year={1965},
  publisher={Wiley Online Library}
}

@article{shanthikumar1992multiclass,
  title={Multiclass queueing systems: Polymatroidal structure and optimal scheduling control},
  author={Shanthikumar, J George and Yao, David D},
  journal={Operations Research},
  volume={40},
  number={3-supplement-2},
  pages={S293--S299},
  year={1992},
  publisher={INFORMS}
}

@article{li2025throughput,
  title={Throughput-optimal scheduling algorithms for llm inference and ai agents},
  author={Li, Yueying and Dai, Jim and Peng, Tianyi},
  journal={arXiv preprint arXiv:2504.07347},
  year={2025}
}

@article{mitzenmacher2025queueing,
  title={Queueing, predictions, and large language models: Challenges and open problems},
  author={Mitzenmacher, Michael and Shahout, Rana},
  journal={Stochastic Systems},
  volume={15},
  number={3},
  pages={195--219},
  year={2025},
  publisher={INFORMS}
}

@article{ao2025optimizing,
  title={Optimizing llm inference: Fluid-guided online scheduling with memory constraints},
  author={Ao, Ruicheng and Luo, Gan and Simchi-Levi, David and Wang, Xinshang},
  journal={arXiv preprint arXiv:2504.11320},
  year={2025}
}

@article{lee2024design,
  title={Design and Scheduling of an AI-based Queueing System},
  author={Lee, Jiung and Namkoong, Hongseok and Zeng, Yibo},
  journal={arXiv preprint arXiv:2406.06855},
  year={2024}
}

@article{legros2026human,
  title={Human-AI Interaction in Congested Service Systems: When It Improves Performance, and When It Doesn't},
  author={Legros, Benjamin and de V{\'e}ricourt, Francis},
  journal={Available at SSRN 5401875},
  year={2026}
}

@article{yang2026design,
  title={On the Design of an AI-Human Service System with Strategic Customers: Should AI be mandated?},
  author={Yang, Guanling and Roet-Green, Ricky},
  journal={Available at SSRN 6475258},
  year={2026}
}

@article{sapkota2025ai,
  title={Ai agents vs. agentic ai: A conceptual taxonomy, applications and challenges},
  author={Sapkota, Ranjan and Roumeliotis, Konstantinos I and Karkee, Manoj},
  journal={Information Fusion},
  pages={103599},
  year={2025},
  publisher={Elsevier}
}

@article{huang2024understanding,
  title={Understanding the planning of llm agents: A survey},
  author={Huang, Xu and Liu, Weiwen and Chen, Xiaolong and Wang, Xingmei and Wang, Hao and Lian, Defu and Wang, Yasheng and Tang, Ruiming and Chen, Enhong},
  journal={arXiv preprint arXiv:2402.02716},
  year={2024}
}

@article{wei2022chain,
  title={Chain-of-thought prompting elicits reasoning in large language models},
  author={Wei, Jason and Wang, Xuezhi and Schuurmans, Dale and Bosma, Maarten and Xia, Fei and Chi, Ed and Le, Quoc V and Zhou, Denny and others},
  journal={Advances in neural information processing systems},
  volume={35},
  pages={24824--24837},
  year={2022}
}

@inproceedings{snell2025scaling,
  title={Scaling LLM test-time compute optimally can be more effective than scaling parameters for reasoning},
  author={Snell, Charlie and Lee, Jaehoon and Xu, Kelvin and Kumar, Aviral},
  booktitle={International Conference on Learning Representations},
  volume={2025},
  pages={10131--10165},
  year={2025}
}

@article{zheng2023response,
  title={Response length perception and sequence scheduling: An llm-empowered llm inference pipeline},
  author={Zheng, Zangwei and Ren, Xiaozhe and Xue, Fuzhao and Luo, Yang and Jiang, Xin and You, Yang},
  journal={Advances in Neural Information Processing Systems},
  volume={36},
  pages={65517--65530},
  year={2023}
}

@article{fu2024efficient,
  title={Efficient llm scheduling by learning to rank},
  author={Fu, Yichao and Zhu, Siqi and Su, Runlong and Qiao, Aurick and Stoica, Ion and Zhang, Hao},
  journal={Advances in Neural Information Processing Systems},
  volume={37},
  pages={59006--59029},
  year={2024}
}

@article{cox1961queues,
  title={Queues},
  author={Cox, DR and Smith, Walter L},
  journal={Methuen, London},
  year={1961}
}

@article{anand2011quality,
  title={Quality--speed conundrum: Trade-offs in customer-intensive services},
  author={Anand, Krishnan S and Pa{\c{c}}, M Faz{\i}l and Veeraraghavan, Senthil},
  journal={Management Science},
  volume={57},
  number={1},
  pages={40--56},
  year={2011},
  publisher={INFORMS}
}

@article{deshmukh1977capacity,
  title={Capacity design and service quality control in a queuing system},
  author={Deshmukh, SD and Jain, Suresh},
  journal={Operations Research},
  volume={25},
  number={4},
  pages={651--661},
  year={1977},
  publisher={INFORMS}
}

@article{abdolmaleki2026pricing,
  title={Pricing Delayed Agentic AI Services: When High-Value Jobs Wait Longer},
  author={Abdolmaleki, Mojtaba and Duenyas, Izak and Kapuscinski, Roman},
  journal={Available at SSRN 6721378},
  year={2026}
}

@article{guo2026congestion,
  title={Congestion-Aware Static LLM Cascades: Analysis of a Steady-State Framework},
  author={Guo, Yuan and Jasin, Stefanus and Lin, Chen-An Andy},
  journal={Available at SSRN 6759238},
  year={2026}
}

@article{van2000price,
  title={Price and service discrimination in queuing systems: Incentive compatibility of Gc $\mu$ scheduling},
  author={Van Mieghem, Jan A},
  journal={Management Science},
  volume={46},
  number={9},
  pages={1249--1267},
  year={2000},
  publisher={INFORMS}
}

@article{afeche2019pricing,
  title={Pricing and prioritizing time-sensitive customers with heterogeneous demand rates},
  author={Af{\`e}che, Philipp and Baron, Opher and Milner, Joseph and Roet-Green, Ricky},
  journal={Operations Research},
  volume={67},
  number={4},
  pages={1184--1208},
  year={2019},
  publisher={INFORMS}
}

@article{freund2024dedicated,
  title={The dedicated docket in us immigration courts: An analysis of fairness and efficiency properties},
  author={Freund, Daniel and Weng, Wentao},
  journal={Available at SSRN 4785713},
  year={2024}
}

@article{bari2025optimal,
  title={Optimal Scheduling Algorithms for LLM Inference: Theory and Practice},
  author={Bari, Agrim and Hegde, Parikshit and de Veciana, Gustavo},
  journal={Proceedings of the ACM on Measurement and Analysis of Computing Systems},
  volume={9},
  number={3},
  pages={1--43},
  year={2025},
  publisher={ACM New York, NY, USA}
}

@article{lobel2026pricing,
  title={Pricing AI as a Service},
  author={Lobel, Ilan and Sun, Yanwei and Wu, Jiahua},
  journal={Available at SSRN 6254459},
  year={2026}
}

@article{feldman2026pricing,
  title={Pricing Human-AI Services: The Role of Observability and Congestion},
  author={Feldman, Pnina and Segev, Ella and Sekar, Shreyas},
  journal={Available at SSRN},
  year={2026}
}

@article{huang2026deepswe,
  title={DeepSWE: Measuring Frontier Coding Agents on Original, Long-Horizon Engineering Tasks},
  author={Huang, Wenqi and Lee, Charley and Tng, Leonard and Ge, Serena},
  journal={arXiv preprint arXiv:2607.07946},
  year={2026}
}

\newpage
\begin{APPENDIX}{}

\section{Proofs for Section~\ref{section:two-type setting}}
\label{appendix:section_3}

\subsection{Proof of Proposition~\ref{proposition:welfare_maximization_centralized_case}}

By Lemma \ref{lemma:cmu_rule}, we only have to consider two priority rules: prioritize type $1$ customers or prioritize type $2$ customers. We first reformulate the welfare objective in terms of residual capacities. We then establish the exclusion regime and characterize the candidate allocations when both types receive service. Comparing the optimized welfare under the two priority orders yields a unique switching threshold. Finally, we establish the threshold’s dependence on $a$.

\textbf{Reformulation.} 

Suppose that the decision maker gives priority to type $1$. The resulting problem is
    \begin{equation*}
        \max_{x_1,x_2}\quad
        \lambda_1\left(ax_1-\theta_1\frac{x_1}{1-\lambda_1x_1}\right)
        +\lambda_2\left(
            ax_2-\theta_2\left(
                \frac{x_2}{1-\lambda_1x_1-\lambda_2x_2}
                +\frac{\lambda_1x_1^2}{(1-\lambda_1x_1)(1-\lambda_1x_1-\lambda_2x_2)}
            \right)
        \right).
    \end{equation*}
    Similarly, if the decision maker gives priority to type $2$, the resulting problem is  
    \begin{equation*}
        \max_{x_1,x_2}\quad
        \lambda_1\left(
            ax_1-\theta_1\left(
                \frac{x_1}{1-\lambda_1x_1-\lambda_2x_2}
                +\frac{\lambda_2x_2^2}{(1-\lambda_2x_2)(1-\lambda_1x_1-\lambda_2x_2)}
            \right)
        \right)
        +\lambda_2\left(
            ax_2-\theta_2\frac{x_2}{1-\lambda_2x_2}
        \right).
    \end{equation*}
    
    Denote $y_1=1-\lambda_1x_1$, $y_2=1-\lambda_2x_2$, and $y_0=1-\lambda_1x_1-\lambda_2x_2$. Denote $r=\frac{\lambda_2}{\lambda_1}$. If type $1$ is prioritized, the decision maker's problem reduces to
    \begin{equation}
        \label{equation:welfare_maximization_1first}
        \begin{aligned}
            \max_{y_1, y_0}\quad& a(1-y_0)-\theta_1\frac{1-y_1}{y_1}-\theta_2\frac{y_1-y_0}{y_0}-\theta_2r\frac{(1-y_1)^2}{y_1y_0}\\
            \textbf{s.t.}\quad& 0<y_0\leq y_1\leq 1.
        \end{aligned}
    \end{equation}
    If type $2$ is prioritized, the decision maker's problem reduces to
    \begin{equation}
        \label{equation:welfare_maximization_2first}
        \begin{aligned}
            \max_{y_2, y_0}\quad& a(1-y_0)-\theta_1\frac{y_2-y_0}{y_0}-\theta_1\frac{1}{r}\frac{(1-y_2)^2}{y_0y_2}-\theta_2\frac{1-y_2}{y_2}\\
            \textbf{s.t.}\quad& 0<y_0\leq y_2\leq 1.
        \end{aligned}
    \end{equation}

    Denote $\text{Obj}_1(r)$ as the optimal value of problem \eqref{equation:welfare_maximization_1first}, and $\text{Obj}_2(r)$ as the optimal value of problem \eqref{equation:welfare_maximization_2first}. 
    
    \textbf{Optimization under a fixed priority rule.}

    Since the optimal solution of original problem is given by one of the two priority rules, we first analyze problem \eqref{equation:welfare_maximization_1first} and problem \eqref{equation:welfare_maximization_2first} separately. Lemma~\ref{lemma:1priority_firstcase} and Lemma~\ref{lemma:2priority_firstcase} give characterizations of the optimal solutions of problem \eqref{equation:welfare_maximization_1first} and \eqref{equation:welfare_maximization_2first}.

    \begin{lemma}
        \label{lemma:1priority_firstcase}
        If $\theta_1^2\geq a\theta_2$, the optimal solution of problem \eqref{equation:welfare_maximization_1first} is given by $(y_1,y_0)=(1,\sqrt{\theta_2/a})$, and $\text{Obj}_1(r)=a+\theta_2-2\sqrt{a\theta_2}$. If $\theta_1^2< a\theta_2$, the optimal solution of problem \eqref{equation:welfare_maximization_1first} is given by \begin{equation*}
            \begin{aligned}
                &a\theta_2y_1\left(\left(1+r\right)y_1^2-r\right)^2=\theta_1^2\left(r(1-y_1)^2+y_1^2\right),\\
                &y_0=\sqrt{\frac{\theta_2}{a}\frac{(y_1)^2+r(1-y_1)^2}{y_1}},
            \end{aligned}
        \end{equation*}
        and $0<y_0<y_1<1$.
    \end{lemma}
    \begin{proof}{Proof of Lemma~\ref{lemma:1priority_firstcase}.}
        When type $1$ customers are prioritized, problem \eqref{equation:welfare_maximization_1first} is equivalent to 
    \begin{equation*}
        \min_{0<y_0\le y_1\le 1}\quad
        ay_0+\frac{\theta_1}{y_1}
        +\frac{\theta_2y_1}{y_0}
        +\theta_2r\frac{(1-y_1)^2}{y_1y_0}.
    \end{equation*}
    
    By the AM-GM inequality,
    \begin{equation*}
        ay_0+\frac{\theta_1}{y_1}+\frac{\theta_2y_1}{y_0}+\theta_2r\frac{(1-y_1)^2}{y_1y_0}\geq \frac{\theta_1}{y_1}+2\sqrt{a\theta_2\frac{y_1^2+r(1-y_1)^2}{y_1}}.
    \end{equation*}
    Denote $g_1(y)=\frac{\theta_1}{y}+2\sqrt{a\theta_2\frac{y^2+r(1-y)^2}{y}}$, then a direct calculation gives
    \begin{equation*}
    g_1'(y)=\frac{1}{y^2}\left(\frac{\sqrt{a\theta_2}\left(\left(1+r\right)y^2-r\right)}{\sqrt{\frac{y^2+r(1-y)^2}{y}}}-\theta_1\right).
\end{equation*}
    
    If $\theta_1^2\geq a\theta_2$, for all $y\in(0,1]$,
    \begin{equation*}
        \left(1+r\right)y^2-r\le 1,
    \end{equation*}
    and 
    \begin{equation*}
        \left(1+r\right)y^2-r\le y^2+r(1-y)^2\le \frac{y^2+r(1-y)^2}{y}.
    \end{equation*}
    Hence
    \begin{equation*}
        g_1'(y)\le \frac{\sqrt{a\theta_2}-\theta_1}{y^2}\le 0,
    \end{equation*}
    where the last inequality uses $\theta_1^2\ge a\theta_2$. Therefore $g_1$ is non-increasing on $(0,1]$, so
    \begin{equation*}
        g_1(y)\geq g_1(1)=\theta_1+2\sqrt{a\theta_2}.
    \end{equation*}
    Equality is attained at $y_1=1$ and $y_0=\sqrt{\theta_2/a}$, which yields
    \begin{equation*}
        x_1=0,\quad x_2=\frac{1-\sqrt{\theta_2/a}}{\lambda_2}.
    \end{equation*}

    If $\theta_1^2<a\theta_2$, notice that $g_1'(y)$ has the same sign as
    \begin{equation*}
        \tilde{g}_1(y)=y^2g_1'(y)=\frac{\sqrt{a\theta_2}\left(\left(1+r\right)y^2-r\right)}{\sqrt{\frac{y^2+r(1-y)^2}{y}}}-\theta_1.
    \end{equation*}
    For all $y<\sqrt{\frac{r}{1+r}}$, we have $\tilde g(y)<0$. Taking the derivative of $\tilde{g}_1(y)$ gives
    \begin{equation*}
        \tilde{g}_1'(y)=\frac{2\sqrt{a\theta_2}\left(1+r\right)y\sqrt{\frac{y^2+r(1-y)^2}{y}}-\sqrt{a\theta_2}\left(\left(1+r\right)y^2-r\right)\frac{\left(\left(1+r\right)-r\frac{1}{y^2}\right)}{2\sqrt{\frac{y^2+r(1-y)^2}{y}}}}{\frac{y^2+r(1-y)^2}{y}}.
    \end{equation*}
    Since
    \begin{equation*}
        4(1+r)>1+r-r\frac{1}{y^2} 
    \end{equation*}
    and
    \begin{equation*}
        y^2+r(1-y)^2\geq \left(1+r\right)y^2-r
    \end{equation*}
    for all $y\in(\sqrt{\frac{r}{1+r}},1]$, $\tilde{g}_1'(y)> 0$. Thus, $\tilde{g}_1(y)$ is a strictly increasing function. Since $\tilde{g}_1(\sqrt{\frac{r}{1+r}})=-\theta_1<0$ and $\tilde{g}_1(1)=\sqrt{a\theta_2}-\theta_1>0$, There exists a unique $y_1^*\in(0,1)$ such that $\tilde{g}_1(y_1^*)=0$.

    Thus, $g_1(y)$ is minimized when $y=y_1^*$, and the optimal solution of problem \eqref{equation:welfare_maximization_1first} is $(y_0, y_1)=(\sqrt{\frac{\theta_2}{a}\frac{(y_1^*)^2+r(1-y_1^*)^2}{y_1^*}}, y_1^*)$. In this case,
    \begin{equation*}
        y_0=\sqrt{\frac{\theta_2}{a}\frac{(y_1^*)^2+r(1-y_1^*)^2}{y_1^*}}=\frac{\theta_2}{\theta_1}\left(\left(1+r\right)(y_1^*)^2-r\right)< y_1,
    \end{equation*}
    so $0<y_0<y_1<1$.
    \hfill{\Halmos}
    \end{proof}

    \begin{lemma}
        \label{lemma:2priority_firstcase}
        If $\theta_1^2\geq a\theta_2$, we have $\text{Obj}_2(r)<\text{Obj}_1(r)$. If $\theta_1^2< a\theta_2$ and $\text{Obj}_2(r)>\text{Obj}_1(r)$, then the optimal solution of problem \eqref{equation:welfare_maximization_2first} is given by
        \begin{equation*}
            \begin{aligned}
                &a\theta_1y_2\left(\left(1+\frac{1}{r}\right)y_2^2-\frac{1}{r}\right)^2=\theta_2^2\left(\frac{1}{r}(1-y_2)^2+y_2^2\right),\\
                &y_0=\sqrt{\frac{\theta_1}{a}\frac{(y_2)^2+\frac{1}{r}(1-y_2)^2}{y_2}}.
            \end{aligned}
        \end{equation*}
    \end{lemma}
    \begin{proof}{Proof of Lemma~\ref{lemma:2priority_firstcase}.}
        We first consider the case $\theta_1^2\geq a\theta_2$. If $y_2=1$, the objective of problem \eqref{equation:welfare_maximization_2first} is bounded by
        \begin{equation*}
            \begin{aligned}
                a(1-y_0)-\theta_1\frac{y_2-y_0}{y_0}-\theta_1\frac{1}{r}\frac{(1-y_2)^2}{y_0y_2}-\theta_2\frac{1-y_2}{y_2}=&a(1-y_0)-\theta_1\frac{1-y_0}{y_0}\\
                &\leq a+\theta_1-2\sqrt{a\theta_1}\\
                &< a+\theta_2-2\sqrt{a\theta_2},
            \end{aligned}
        \end{equation*}
        where the first inequality holds by AM-GM inequality and the second inequality holds because $a>\theta_1>\theta_2$.
        
        If $y_2<1$, notice that
    \begin{equation*}
        -\theta_1\frac{1}{r}\frac{(1-y_2)^2}{y_0y_2}<0,
    \end{equation*}
    so the objective of problem \eqref{equation:welfare_maximization_2first} is strictly bounded above by
    \begin{equation*}
        \begin{aligned}
            \max_{y_2, y_0}\quad& a(1-y_0)-\theta_1\frac{y_2-y_0}{y_0}-\theta_2\frac{1-y_2}{y_2}\\
            \textbf{s.t.}\quad& 0<y_0\leq y_2\leq 1.
        \end{aligned}
    \end{equation*}
    
    If $y_0\geq \frac{\theta_2}{\theta_1}$,
    \begin{equation*}
        \begin{aligned}
            a(1-y_0)-\theta_1\frac{y_2-y_0}{y_0}-\theta_2\frac{1-y_2}{y_2}\leq& a(1-y_0)-\theta_2\frac{1-y_0}{y_0}\\
            \leq & a+\theta_2-2\sqrt{a\theta_2},
        \end{aligned}
    \end{equation*}
    where the first inequality is because the left hand side is decreasing with respect to $y_2$ on $[\sqrt{\frac{\theta_2}{\theta_1}y_0},+\infty)$ and the constraint $y_2\geq y_0$.
    
    If $y_0< \frac{\theta_2}{\theta_1}$,
    \begin{equation*}
        \begin{aligned}
            a(1-y_0)-\theta_1\frac{y_2-y_0}{y_0}-\theta_2\frac{1-y_2}{y_2}\leq& a(1-y_0)+\theta_1+\theta_2-2\sqrt{\frac{\theta_1\theta_2}{y_0}}\\
            \leq & a(1-\frac{\theta_2}{\theta_1})-\theta_1+\theta_2\\
            \leq & a+\theta_2-2\sqrt{a\theta_2},
        \end{aligned}
    \end{equation*}
    where the second inequality is due to $\theta_1^2\geq a\theta_2$.

    Therefore, $\text{Obj}_2(r)<a+\theta_2-2\sqrt{a\theta_2}$.

    Next, we consider the case $\theta_1^2<a\theta_2$. When type $2$ customers are prioritized, problem \eqref{equation:welfare_maximization_2first} is equivalent to
    \begin{equation*}
        \min_{0<y_0\le y_2\le 1}\quad
        ay_0+\frac{\theta_2}{y_2}
        +\frac{\theta_1y_2}{y_0}
        +\theta_1\frac{1}{r}\frac{(1-y_2)^2}{y_2y_0}.
    \end{equation*}
    
    If $y_2=1$, the optimal solution for $y_0$ is $y_0=\sqrt{\frac{\theta_1}{a}}$ with objective value $2\sqrt{a\theta_1}+\theta_2$. Consider solution $(y_0,y_2)=(\sqrt{\frac{\theta_1}{a}},1-\epsilon)$, the objective value
    \begin{equation*}
        \begin{aligned}
            (2-\epsilon)\sqrt{a\theta_1}+\frac{\theta_2}{1-\epsilon}+\theta_1\frac{1}{r}\frac{\epsilon^2}{(1-\epsilon)\sqrt{\frac{\theta_1}{a}}}<2\sqrt{a\theta_1}+\theta_2
        \end{aligned}
    \end{equation*}
    when $\epsilon>0$ is sufficiently small since $\theta_2<\sqrt{a\theta_1}$. Thus, $y_2=1$ is suboptimal.

    If $y_2=y_0$, type $1$ customers receive zero service depth. In this case, prioritizing type $1$ gives strictly better social welfare. Therefore, $y_2=y_0$ is suboptimal.

    In conclusion, the optimal policy satisfies $0<y_0<y_2<1$, so both types of customers receive strictly positive and finite service depth. Moreover, since the welfare objective goes to $-\infty$ uniformly as $y_0\rightarrow 0$, the optimal solution of problem \eqref{equation:welfare_maximization_2first} is interior and satisfies the first order condition. Similar to the case where type $1$ customers are prioritized, the optimal solution of problem \eqref{equation:welfare_maximization_2first} satisfies
    \begin{equation*}
            \begin{aligned}
                &a\theta_1y_2\left(\left(1+\frac{1}{r}\right)y_2^2-\frac{1}{r}\right)^2=\theta_2^2\left(\frac{1}{r}(1-y_2)^2+y_2^2\right),\\
                &y_0=\sqrt{\frac{\theta_1}{a}\frac{(y_2)^2+\frac{1}{r}(1-y_2)^2}{y_2}}.
            \end{aligned}
        \end{equation*}
    \hfill{\Halmos}
    \end{proof}
    
    \textbf{Properties of the optimal solution.} 
    
    In order to compare the two optimal solutions, we first show the following properties of $\text{Obj}_1(r)$ and $\text{Obj}_2(r)$.

    \begin{lemma}
        \label{lemma:objective_monotonicity}
        Suppose $\theta_1^2<a\theta_2$. $\text{Obj}_1(r)$ is a strictly decreasing function and $\text{Obj}_2(r)$ is a strictly increasing function. That is, for $r_1< r_2$, 
        \begin{equation*}
            \text{Obj}_1(r_1)> \text{Obj}_1(r_2),\quad \text{Obj}_2(r_1)< \text{Obj}_2(r_2).
        \end{equation*}
        Moreover, $\text{Obj}_1(r)$, $\text{Obj}_2(r)$ are continuous functions with respect to $r$.
    \end{lemma}
    \begin{proof}{Proof of Lemma~\ref{lemma:objective_monotonicity}.}
        Denote $\left(y_0(r), y_1(r)\right)$ as the optimal solution of problem \eqref{equation:welfare_maximization_1first}. Then for $r_1<r_2$,
        \begin{equation*}
            \begin{aligned}
                \text{Obj}_1(r_2)=&a(1-y_0(r_2))-\theta_1\frac{1-y_1(r_2)}{y_1(r_2)}-\theta_2\frac{y_1(r_2)-y_0(r_2)}{y_0(r_2)}-\theta_2r_2\frac{(1-y_1(r_2))^2}{y_1(r_2)y_0(r_2)}\\
                <&a(1-y_0(r_2))-\theta_1\frac{1-y_1(r_2)}{y_1(r_2)}-\theta_2\frac{y_1(r_2)-y_0(r_2)}{y_0(r_2)}-\theta_2r_1\frac{(1-y_1(r_2))^2}{y_1(r_2)y_0(r_2)}\\
                \leq&\max_{0<y_0\leq y_1\leq 1}\quad \left[a(1-y_0)-\theta_1\frac{1-y_1}{y_1}-\theta_2\frac{y_1-y_0}{y_0}-\theta_2r_1\frac{(1-y_1)^2}{y_1y_0}\right]\\
                =& \text{Obj}_1(r_1).
            \end{aligned}
        \end{equation*}
        The first inequality is strict because $y_1(r_2)$ is strictly less than $1$. Similarly, we can show $\text{Obj}_2(r_1)< \text{Obj}_2(r_2)$.

        For $r_1<r_2$, $|r_1-r_2|<\epsilon$,
        \begin{equation*}
            \begin{aligned}
                \text{Obj}_1(r_1)=&a(1-y_0(r_1))-\theta_1\frac{1-y_1(r_1)}{y_1(r_1)}-\theta_2\frac{y_1(r_1)-y_0(r_1)}{y_0(r_1)}-\theta_2r_1\frac{(1-y_1(r_1))^2}{y_1(r_1)y_0(r_1)}\\
                =&a(1-y_0(r_1))-\theta_1\frac{1-y_1(r_1)}{y_1(r_1)}-\theta_2\frac{y_1(r_1)-y_0(r_1)}{y_0(r_1)}-\theta_2r_2\frac{(1-y_1(r_1))^2}{y_1(r_1)y_0(r_1)}+\theta_2(r_1-r_2)\frac{(1-y_1(r_1))^2}{y_1(r_1)y_0(r_1)}\\
                \leq&\max_{0<y_0\leq y_1\leq 1}\quad \left[a(1-y_0)-\theta_1\frac{1-y_1}{y_1}-\theta_2\frac{y_1-y_0}{y_0}-\theta_2r_2\frac{(1-y_1)^2}{y_1y_0}\right]+\epsilon\theta_2\frac{(1-y_1(r_1))^2}{y_1(r_1)y_0(r_1)}\\
                =&\text{Obj}_1(r_2)+\epsilon\theta_2\frac{(1-y_1(r_1))^2}{y_1(r_1)y_0(r_1)}.
            \end{aligned}
        \end{equation*}
        When $\epsilon\rightarrow0$, $\epsilon\theta_2\frac{(1-y_1(r_1))^2}{y_1(r_1)y_0(r_1)}\rightarrow0$. Therefore, $\text{Obj}_1(r)$ is continuous. Similarly, $\text{Obj}_2(r)$ is continuous.
        \hfill{\Halmos}
    \end{proof}
    \begin{lemma}
        \label{lemma:objectve_limit}
        If $\theta_1^2<a\theta_2$, we have
        \begin{equation*}
            \begin{aligned}
                &\lim_{r\rightarrow 0} \text{Obj}_1(r)=a+\theta_1+\theta_2-3\sqrt[3]{a\theta_1\theta_2},\\
                &\lim_{r\rightarrow +\infty} \text{Obj}_1(r)=a+\theta_2-2\sqrt{a\theta_2},\\
                &\lim_{r\rightarrow 0} \text{Obj}_2(r)=a+\theta_1-2\sqrt{a\theta_1},\\
                &\lim_{r\rightarrow +\infty} \text{Obj}_2(r)=a+\theta_1+\theta_2-3\sqrt[3]{a\theta_1\theta_2}.
            \end{aligned}
        \end{equation*}
    \end{lemma}
    \begin{proof}{Proof of Lemma~\ref{lemma:objectve_limit}.}
        For problem \eqref{equation:welfare_maximization_1first}, the objective is upper bounded by
        \begin{equation*}
            \begin{aligned}
                a(1-y_0)-\theta_1\frac{1-y_1}{y_1}-\theta_2\frac{y_1-y_0}{y_0}-\theta_2r\frac{(1-y_1)^2}{y_1y_0}&\leq a(1-y_0)-\theta_1\frac{1-y_1}{y_1}-\theta_2\frac{y_1-y_0}{y_0}\\
                &\leq a+\theta_1+\theta_2-3\sqrt[3]{a\theta_1\theta_2}.
            \end{aligned}
        \end{equation*}
        When $r\rightarrow 0$, solution $(y_0,y_1)=(\sqrt[3]{\frac{\theta_1\theta_2}{a^2}},\sqrt[3]{\frac{\theta_1^2}{a\theta_2}})$ gives social welfare
        \begin{equation*}
            a+\theta_1+\theta_2-3\sqrt[3]{a\theta_1\theta_2}-\theta_2r\frac{a\left(1-\sqrt[3]{\frac{\theta_1^2}{a\theta_2}}\right)^2}{\theta_1}\rightarrow  a+\theta_1+\theta_2-3\sqrt[3]{a\theta_1\theta_2}.
        \end{equation*}
        Therefore, $\lim_{r\rightarrow 0} \text{Obj}_1(r)=a+\theta_1+\theta_2-3\sqrt[3]{a\theta_1\theta_2}$.

        On the other hand, when $r\rightarrow\infty$, $y_1\rightarrow 1$, otherwise $-\theta_2r\frac{(1-y_1)^2}{y_1y_0}\rightarrow-\infty$. In this case, the objective
        \begin{equation*}
            a(1-y_0)-\theta_1\frac{1-y_1}{y_1}-\theta_2\frac{y_1-y_0}{y_0}-\theta_2r\frac{(1-y_1)^2}{y_1y_0}\rightarrow a(1-y_0)-\theta_2\frac{1-y_0}{y_0}\leq a+\theta_2-2\sqrt{a\theta_2}.
        \end{equation*}
        The inequality is tight if $(y_0,y_1)=(\sqrt\frac{\theta_2}{a},1)$. Therefore, $\lim_{r\rightarrow +\infty} \text{Obj}_1(r)=a+\theta_2-2\sqrt{a\theta_2}$.
        \hfill{\Halmos}
    \end{proof}

Given the useful lemmas above, we are able to prove the full Proposition~\ref{proposition:welfare_maximization_centralized_case}.
\begin{proof}{Proof of Proposition~\ref{proposition:welfare_maximization_centralized_case}.}
    \noindent\textbf{Part~1: the case $\theta_1^2\ge a\theta_2$.}
    
    When $\theta_1^2\ge a\theta_2$, $(x_1^*, x_2^*)$ gives long-term average social welfare
    \begin{equation}
        \lambda_2\left(V(x_2)-\theta_2\frac{x_2}{1-\lambda_2x_2}\right)=a+\theta_2-2\sqrt{a\theta_2}.
    \end{equation}
    
    By Lemma~\ref{lemma:1priority_firstcase} and Lemma~\ref{lemma:2priority_firstcase}, $\text{Obj}_2(r)<\text{Obj}_1(r)=a+\theta_2-2\sqrt{a\theta_2}$, thus no policy with type-$2$ priority can outperform the candidate found above, proving Part~1.

    \noindent\textbf{Part~2: the case $\theta_1^2<a\theta_2$.}
    
    First, we show that both types of customers receive a strictly positive and finite service depth.

    By Lemma~\ref{lemma:1priority_firstcase}, if type $1$ customers are prioritized, the optimal solution can be written as
    \begin{equation*}
            \begin{aligned}
                &x_1^* \quad\text{satisfies}\quad a\theta_2(1-\lambda_1x_1^*)\left(\left(1+\frac{\lambda_2}{\lambda_1}\right)(1-\lambda_1x_1^*)^2-\frac{\lambda_2}{\lambda_1}\right)^2=\theta_1^2\left(\lambda_1\lambda_2{x_1^*}^2+(1-\lambda_1x_1^*)^2\right),\\
                &x_2^*=\frac{1}{\lambda_2}\left[1-\frac{\theta_2}{\theta_1}+\left(2\frac{\theta_2}{\theta_1}(\lambda_1+\lambda_2)-\lambda_1\right)x_1^*-\frac{\theta_2}{\theta_1}\lambda_1(\lambda_1+\lambda_2){x_1^*}^2\right].
            \end{aligned}
        \end{equation*}
        Since $0<y_0<y_1<1$, we have $0<x_1^*,x_2^*<\infty$. Similarly, if prioritizing type $2$ customers is optimal, by Lemma~\ref{lemma:2priority_firstcase} we have
    \begin{equation*}
            \begin{aligned}
                &x_2^* \quad\text{satisfies}\quad a\theta_1(1-\lambda_2x_2^*)\left(\left(1+\frac{\lambda_1}{\lambda_2}\right)(1-\lambda_2x_2^*)^2-\frac{\lambda_1}{\lambda_2}\right)^2=\theta_2^2\left(\lambda_1\lambda_2{x_2^*}^2+(1-\lambda_2x_2^*)^2\right),\\
                &x_1^*=\frac{1}{\lambda_1}\left[1-\frac{\theta_1}{\theta_2}+\left(2\frac{\theta_1}{\theta_2}(\lambda_1+\lambda_2)-\lambda_2\right)x_2^*-\frac{\theta_1}{\theta_2}\lambda_2(\lambda_1+\lambda_2){x_2^*}^2\right],
            \end{aligned}
        \end{equation*}
        and $0<y_0<y_2<1$, so $0<x_1^*,x_2^*<\infty$.
    
    Next, we characterize the threshold for which priority rule is optimal. Denote $h(r)=\text{Obj}_2(r)-\text{Obj}_1(r)$. By Lemma~\ref{lemma:objective_monotonicity}, $h(r)$ is a continuous and strictly increasing function. By Lemma~\ref{lemma:objectve_limit}, $h(0)=3\sqrt[3]{a\theta_1\theta_2}-2\sqrt{a\theta_1}-\theta_2<0$, $h(+\infty)=2\sqrt{a\theta_2}+\theta_1-3\sqrt[3]{a\theta_1\theta_2}>0$. Therefore, there exists a unique value $c>0$ such that $h(c)=0$. It follows that type $1$ is always optimally prioritized when $r< c$, whereas type $2$ is always optimally prioritized when $r>c$. When $r=c$, the two different solutions (one prioritizes type $1$ customers and the other prioritizes type $2$ customers) gives the same maximum social welfare.

    Lastly, we characterize the structure of $c$. When $r=c$, problem \eqref{equation:welfare_maximization_1first} and problem \eqref{equation:welfare_maximization_2first} gives the same optimal value:
    \begin{equation*}
        \min_{0<y_0\le y_1\le 1}\quad
        ay_0+\frac{\theta_1}{y_1}
        +\frac{\theta_2y_1}{y_0}
        +\theta_2c\frac{(1-y_1)^2}{y_1y_0}=\min_{0<y_0\le y_2\le 1}\quad
        ay_0+\frac{\theta_2}{y_2}
        +\frac{\theta_1y_2}{y_0}
        +\theta_1\frac{1}{c}\frac{(1-y_2)^2}{y_2y_0}.
    \end{equation*}
    Consider the first order condition of $y_0$ on both sides, we have
    \begin{equation*}
        \min_{0<y_1\le 1}\quad
        \frac{\theta_1}{y_1}+2\sqrt{a\theta_2\frac{y_1^2+c(1-y_1)^2}{y_1}}=\min_{0<y_2\le 1}\quad
        \frac{\theta_2}{y_2}+2\sqrt{a\theta_1\frac{y_2^2+\frac{1}{c}(1-y_2)^2}{y_2}}.
    \end{equation*}
    Denote $u=\frac{\theta_1^2}{a\theta_2}$ and $v=\frac{\theta_1}{\theta_2}$, we have
    \begin{equation*}
        \min_{0<y_1\le 1}\quad
        \frac{\sqrt{u}}{y_1}+2\sqrt{\frac{y_1^2+c(1-y_1)^2}{y_1}}=\min_{0<y_2\le 1}\quad
        \frac{\sqrt{u}}{vy_2}+2\sqrt{v\frac{y_2^2+\frac{1}{c}(1-y_2)^2}{y_2}}.
    \end{equation*}
    Thus, $c$ is only determined by $u$ and $v$. 
    
    Denote $C(u,v)$ as the function that maps $u$ and $v$ to $c$, we can further show $C(u,v)$ is increasing with respect to $u$. Define
    \begin{equation*}
        \mathcal{H}(u,v,r)=\min_{0<y_1\le 1}\,
        \frac{\sqrt{u}}{y_1}+2\sqrt{\frac{y_1^2+r(1-y_1)^2}{y_1}}-\min_{0<y_2\le 1}\,
        \frac{\sqrt{u}}{vy_2}+2\sqrt{v\frac{y_2^2+\frac{1}{r}(1-y_2)^2}{y_2}},
    \end{equation*}
    then $\mathcal{H}(u,v,C(u,v))=0$. Moreover, given $u=\frac{\theta_1^2}{a\theta_2}$ and $v=\frac{\theta_1}{\theta_2}$, $\mathcal{H}(u,v,r)=h(r)/\sqrt{a\theta_2}$. 
    % By the implicit-function theorem
    % \begin{equation*}
    %     \frac{\partial C(u,v)}{\partial u}=-\frac{\mathcal H_u(u,v,C(u,v))}{\mathcal H_r(u,v,C(u,v))}.
    % \end{equation*}
    Let $\tilde y_1$ denote the optimal solution of $\min_{0<y_1\le 1}\,\phi_1(y_1;c):=\frac{\sqrt{u}}{y_1}+2\sqrt{\frac{y_1^2+c(1-y_1)^2}{y_1}}$, and $\tilde y_2$ denote the optimal solution of $\min_{0<y_2\le 1}\,\phi_2(y_2;c):=\frac{\sqrt{u}}{vy_2}+2\sqrt{v\frac{y_2^2+\frac{1}{c}(1-y_2)^2}{y_2}}$. By the envelope theorem,
    \begin{equation*}
        \begin{aligned}
            &\frac{\partial\mathcal{H}(u,v,r)}{\partial u}=\frac{1}{2\tilde y_1\sqrt{u}}-\frac{1}{2v\tilde y_2\sqrt{u}}.%\\
            %&\frac{\partial\mathcal{H}(u,v,r)}{\partial r}=\frac{1}{\sqrt{a\theta_2}}\frac{dh(r)}{dr}>0,
        \end{aligned}
    \end{equation*}
    If $\frac{\partial\mathcal{H}(u,v,r)}{\partial u}<0$, then for $u<u'$, we have
    \begin{equation*}
        \mathcal{H}(u',v,C(u',v))=0=\mathcal{H}(u,v,C(u,v))>\mathcal{H}(u',v,C(u,v)),
    \end{equation*}
    so $C(u',v)>C(u,v)$.
    It remains to show that
    \begin{equation*}
        \tilde y_1>v\tilde y_2.
    \end{equation*}
    Let $z=vy_2$. The second minimization problem can be written as
    \begin{equation*}
        \min_{0<z\le v}\quad\frac{\sqrt u}{z}+2\sqrt{\frac{z^2+r^{-1}(v-z)^2}{z}}.
    \end{equation*}
    Let $z=v\tilde y_2$. At $c=C(u,v)$, the two minimized values are equal. Since the first objective evaluated at $y_1=1$ equals $\sqrt u+2$, the common minimized value is upper bounded by $\sqrt u+2$. If $z \geq 1$, then
    \begin{equation*}
        \frac{\sqrt u}{\tilde z}+2\sqrt{\frac{\tilde z^2+c^{-1}(v-\tilde z)^2}{\tilde z}}>\sqrt u+2,
    \end{equation*}
    therefore, $z<1$.
    
    For $t\in(0,1]$,
   \begin{equation*}
       \phi_1(t;c)-\phi_2\left(\frac{t}{v};c\right)=2\left(\sqrt{\frac{t^2+c(1-t)^2}{t}}-\sqrt{\frac{t^2+c^{-1}(v-t)^2}{t}}\right),
   \end{equation*}
    and thus
    \begin{equation*}
        \operatorname{sign}\left\{\phi_1(t;c)-\phi_2\left(\frac{t}{v};c\right)\right\}=\operatorname{sign}\{c(1-t)-(v-t)\}.
    \end{equation*}
    Optimality gives
    \begin{equation*}
        \phi_1(\tilde y_1;c)-\phi_2\left(\frac{\tilde y_1}{v};c\right)\le0,\quad \phi_1(\tilde z;c)-\phi_2\left(\frac{\tilde z}{v};c\right)\ge0,
    \end{equation*}
    which imply $c>v$. Hence $t_0:=\frac{c-v}{c-1}\in(0,1)$ is the unique crossing point of $\phi_1(t;c)-\phi_2\left(\frac{t}{v};c\right)$ on $(0,1]$. The sign pattern is positive for $t<t_0$ and negative for $t>t_0$. Therefore,
    \begin{equation*}
        \tilde z\le t_0\le \tilde y_1.
    \end{equation*}
    If $\tilde z=\tilde y_1$, then by the first order condition we have
    \begin{equation*}
        \frac{d\phi_1(\tilde y_1;c)}{dy_1}=\frac{d\phi_2\left(\tilde y_2;c\right)}{dy_2}=0
    \end{equation*}
    However,
    \begin{equation*}
        \frac{d\left(\phi_1(t;c)-\phi_2\left(\frac{t}{v};c\right)\right)}{dt}<0,
    \end{equation*}
    which contradicts the first order condition at $\tilde y_1$ and $\tilde z=v\tilde y_2$. Thus,
    \begin{equation*}
        \tilde y_1>\tilde z=v\tilde y_2.
    \end{equation*}
    Consequently,
    \begin{equation*}
        \frac{\partial \mathcal H}{\partial u}=\frac{1}{2\sqrt u}\left(\frac{1}{\tilde y_1}-\frac{1}{v\tilde y_2}\right)<0.
    \end{equation*}
    % which yields
    % \begin{equation*}
    %     \frac{\partial C(u,v)}{\partial u}=-\frac{\mathcal H_u(u,v,C(u,v))}{\mathcal H_r(u,v,C(u,v))}>0.
    % \end{equation*}

    On the other hand, for any $r\in[0,\infty)$,
    \begin{equation*}
        \lim_{u\rightarrow1}\min_{0<y_1\le 1}\,
        \frac{\sqrt{u}}{y_1}+2\sqrt{\frac{y_1^2+r(1-y_1)^2}{y_1}}=\min_{0<y_1\le 1}\,
        \frac{1}{y_1}+2\sqrt{\frac{y_1^2+r(1-y_1)^2}{y_1}}\geq \min_{0<y_1\le 1}\,
        \frac{1}{y_1}+2\sqrt{y_1}=3,
    \end{equation*}
    and the inequality is tight when $y_1=1$. Similarly,
    \begin{equation*}
        \lim_{u\rightarrow1}\min_{0<y_2\le 1}\,
        \frac{\sqrt{u}}{vy_2}+2\sqrt{v\frac{y_2^2+\frac{1}{r}(1-y_2)^2}{y_2}}=\min_{0<y_2\le 1}\,
        \frac{1}{vy_2}+2\sqrt{v\frac{y_2^2+\frac{1}{r}(1-y_2)^2}{y_2}}\geq \min_{0<y_2\le 1}\,
        \frac{1}{vy_2}+2\sqrt{vy_2}=3.
    \end{equation*}
    However, in this case, the lower bound is not achievable, since it requires $y_2=1$ and $v=1$. Hence,
    \begin{equation*}
         \lim_{u\rightarrow1}\min_{0<y_2\le 1}\,
        \frac{\sqrt{u}}{vy_2}+2\sqrt{v\frac{y_2^2+\frac{1}{r}(1-y_2)^2}{y_2}}>3.
    \end{equation*}
    Therefore, there exists $u_0<1$ such that
    \begin{equation*}
        \min_{0<y_2\le 1}\,
        \frac{\sqrt{u_0}}{vy_2}+2\sqrt{v\frac{y_2^2+\frac{1}{r}(1-y_2)^2}{y_2}}>\min_{0<y_1\le 1}\,
        \frac{\sqrt{u_0}}{y_1}+2\sqrt{\frac{y_1^2+r(1-y_1)^2}{y_1}},
    \end{equation*}
    which is equivalent to $h(r)<0$. This implies that $C(u_0,v)>r$, so $C(u,v)$ is unbounded. In other words, $\lim_{u\rightarrow 1}C(u,v)=\infty$.
    \hfill{\Halmos}
\end{proof}

\subsection{Proof of Proposition~\ref{proposition:welfare_maximization_centralized_case_general_u}}
\begin{proof}{Proof of Proposition~\ref{proposition:welfare_maximization_centralized_case_general_u}.}
Denote $r=\frac{\lambda_2}{\lambda_1}$ and let $\alpha=\lim_{x\rightarrow\infty}\frac{V(x)}{x}$. Since $V$ is concave and $V(0)=0$, $V(x)/x$ is non-increasing in $x$, so the limit $\alpha$ exists. Let $\text{Obj}_1(r)$ and $\text{Obj}_2(r)$ denote the optimal social welfare when type $1$ and type $2$ customers are prioritized, respectively. We first characterize the limiting value of $\text{Obj}_1(r)$ as $r\rightarrow\infty$.

Suppose type $1$ customers are prioritized. Using $y_1=1-\lambda_1x_1$ and $y_0=1-\lambda_1x_1-\lambda_2x_2$, the welfare can be written as
\begin{equation*}
\begin{aligned}
W_1(r;y_0,y_1)
:=\lambda_1V\left(\frac{1-y_1}{\lambda_1}\right)+\lambda_2V\left(\frac{y_1-y_0}{\lambda_2}\right)
-\theta_1\frac{1-y_1}{y_1}-\theta_2\frac{y_1-y_0}{y_0}
-\theta_2r\frac{(1-y_1)^2}{y_0y_1},
\end{aligned}
\end{equation*}
Define
\begin{equation*}
    S_2(\Lambda)=\max_{0\leq \rho<1}\left\{\Lambda V\left(\frac{\rho}{\Lambda}\right)-\theta_2\frac{\rho}{1-\rho}\right\}.
\end{equation*}
For any feasible $(y_0,y_1)$,
\begin{equation*}
\begin{aligned}
W_1(r;y_0,y_1)-S_2(\lambda_2)
\leq&\lambda_1V\left(\frac{1-y_1}{\lambda_1}\right)
-\theta_2r\frac{(1-y_1)^2}{y_0y_1}\\
\leq&\lambda_1V\left(\frac{1-y_1}{\lambda_1}\right)-\theta_2r(1-y_1)^2.
\end{aligned}
\end{equation*}
By concavity and monotonicity of $V$, for all $x\geq 0$,
\begin{equation*}
    V(x)\leq V(1)(1+x).
\end{equation*}
\begin{equation*}
    \lambda_1V\left(\frac{1-y_1}{\lambda_1}\right)\leq V(1)\lambda_1+V(1)(1-y_1).
\end{equation*}
It follows that
\begin{equation*}
    W_1(r;y_0,y_1)-S_2(\lambda_2)\leq V(1)\lambda_1+V(1)(1-y_1)-\theta_2r(1-y_1)^2\leq V(1)\lambda_1+\frac{V(1)^2}{4\theta_2r}.
\end{equation*}
Taking the supremum over all feasible $(y_0,y_1)$ gives
\begin{equation*}
    \text{Obj}_1(r)\leq S_2(\lambda_2)+V(1)\lambda_1+\frac{V(1)^2}{4\theta_2r}.
\end{equation*}
On the other hand, setting $y_1=1$ is feasible, so $\text{Obj}_1(r)\geq S_2(\lambda_2)$. Since $\lambda_2\rightarrow\lambda$ as $r\rightarrow\infty$, and $S_2(\Lambda)$ is continuous at $\Lambda=\lambda$, we obtain
\begin{equation*}
\lim_{r\rightarrow\infty}\text{Obj}_1(r)=S_2(\lambda).
\end{equation*}

Next, suppose type $2$ customers are prioritized. Using $y_2=1-\lambda_2x_2$ and $y_0=1-\lambda_1x_1-\lambda_2x_2$, the welfare can be written as
\begin{equation*}
\begin{aligned}
W_2(r;y_0,y_2)
:=\lambda_1V\left(\frac{y_2-y_0}{\lambda_1}\right)+\lambda_2V\left(\frac{1-y_2}{\lambda_2}\right)
-\theta_1\frac{y_2-y_0}{y_0}-\theta_1\frac{1}{r}\frac{(1-y_2)^2}{y_0y_2}-\theta_2\frac{1-y_2}{y_2}.
\end{aligned}
\end{equation*}
For any feasible $(y_0,y_2)$,
\begin{equation*}
    \lim_{r\rightarrow\infty}\lambda_1V\left(\frac{y_2-y_0}{\lambda_1}\right)\geq \alpha(y_2-y_0),
\end{equation*}
so
\begin{equation*}
    \liminf_{r\rightarrow\infty}\text{Obj}_2(r)\geq \alpha(y_2-y_0)+\lambda V\left(\frac{1-y_2}{\lambda}\right)-\theta_1\frac{y_2-y_0}{y_0}-\theta_2\frac{1-y_2}{y_2}.
\end{equation*}
Optimizing over $y_0\in(0,y_2]$ yields
\begin{equation*}
    \liminf_{r\rightarrow\infty}\text{Obj}_2(r)\geq\sup_{0<y_2\leq 1}\left\{\lambda V\left(\frac{1-y_2}{\lambda}\right)-\theta_2\frac{1-y_2}{y_2}+\left(\sqrt{\alpha y_2}-\sqrt{\theta_1}\right)_+^2\right\}
\end{equation*}
Let $y_2=1-\hat{\rho}$, then we have
\begin{equation*}
    \liminf_{r\rightarrow\infty}\text{Obj}_2(r)\geq S_2(\lambda)+\left(\sqrt{\alpha (1-\hat{\rho})}-\sqrt{\theta_1}\right)^2>S_2(\lambda),
\end{equation*}
thus
\begin{equation*}
    \liminf_{r\rightarrow\infty}\text{Obj}_2(r)>\lim_{r\rightarrow\infty} \text{Obj}_1(r).
\end{equation*}
Therefore, there exists $c>0$ such that type $2$ customers are prioritized if $r>c$.
\hfill{\Halmos}
\end{proof}

\begin{proof}{Proof of Lemma~\ref{lemma:welfare_maximization_incentive_compatibility}.}
    Incentive compatibility constraints require
    \begin{equation*}
        \begin{aligned}
            V(x_1)-\theta_1w_1-p_1\geq V(x_2)-\theta_1w_2-p_2,\\
            V(x_1)-\theta_2w_1-p_1\leq V(x_2)-\theta_2w_2-p_2.
        \end{aligned}
    \end{equation*}
    Therefore,
    \begin{equation*}
        (\theta_1-\theta_2)(w_1-w_2)\leq 0,
    \end{equation*}
    which yields $w_1\leq w_2$.

    On the other hand, if $w_1\leq w_2$, set
    \begin{equation*}
        \begin{aligned}
            &p_1=V(x_1)-\theta_1w_1,\\
            &p_2=V(x_2)-\theta_2w_2-(\theta_1-\theta_2)w_1,
        \end{aligned}
    \end{equation*}
    the resulting menu satisfies incentive compatibility constraints.
    \hfill{\Halmos}
\end{proof}

\subsection{Proof of Proposition~\ref{proposition:welfare_maximization_decentralized_case}}
\begin{proof}{Proof of Proposition~\ref{proposition:welfare_maximization_decentralized_case}.}
    \textbf{Part 1: the case $\theta_1^2\geq a \theta_2$.}

    Consider menu $\mathcal{M}=\{(0,0,0), (ax_2^*-\theta_2\frac{x_2^*}{1-\lambda_2x_2^*},x_2^*,\frac{x_2^*}{1-\lambda_2x_2^*})\}$. Under this menu, type $1$ customers will choose not to enter the system, and type $2$ customers will choose to enter the system. As a result, the long-term average social welfare matches that of the centralized optimum.

    \textbf{Part 2: the case $\theta_1^2< a \theta_2$ and $\frac{\lambda_2}{\lambda_1}\leq c$.}

    By Lemma \ref{lemma:welfare_maximization_incentive_compatibility}, we only have to prove the optimal centralized policy satisfies $w_1\leq w_2$. When $r=\frac{\lambda_2}{\lambda_1}\leq c$, type $1$ customers are prioritized. Therefore, the optimal policy corresponds to the optimal solution of problem \eqref{equation:welfare_maximization_1first}, which is
    \begin{equation*}
        y_0=\sqrt{\frac{\theta_2}{a}\frac{(y_1^*)^2+r(1-y_1^*)^2}{y_1^*}},\quad y_1=y_1^*,
    \end{equation*}
    where $y_1^*$ is the unique solution in $(0,1]$ such that
    \begin{equation*}
        \tilde{g}_1(y)=\frac{\sqrt{a\theta_2}\left(\left(1+r\right)y^2-r\right)}{\sqrt{\frac{y^2+r(1-y)^2}{y}}}-\theta_1=0.
    \end{equation*}
    In this case,
    \begin{equation*}
        \begin{aligned}
            &w_1=\frac{x_1}{1-\lambda_1x_1}=\frac{1}{\lambda_1}\frac{1-y_1}{y_1},\\
            &w_2=\frac{x_2}{1-\lambda_1x_1-\lambda_2x_2}
                +\frac{\lambda_1x_1^2}{(1-\lambda_1x_1)(1-\lambda_1x_1-\lambda_2x_2)}=\frac{1}{\lambda_2}\frac{y_1-y_0}{y_0}+\frac{1}{\lambda_1}\frac{(1-y_1)^2}{y_1y_0}.
        \end{aligned}
    \end{equation*}
    Notice that $y_0=\frac{\theta_2}{\theta_1}\left((1+r)y_1^2-r\right)<(1+r)y_1^2-r$, then we have
    \begin{equation*}
        \begin{aligned}
            w_2-w_1=&\frac{1}{\lambda_2y_0y_1}\left(y_1(y_1-y_0)+r(1-y_1)^2-ry_0(1-y_1)\right)\\
            =&\frac{1}{\lambda_2y_0y_1}\left(y_1^2+r(1-y_1)^2-y_0\left(y_1+r(1-y_1)\right)\right)\\
            >&\frac{1}{\lambda_2y_0y_1}\left(y_1^2+r(1-y_1)^2-\left((1+r)y_1^2-r\right)\left(y_1+r(1-y_1)\right)\right)\\
            =&\frac{1}{\lambda_2y_0y_1}\left(2r(1-y_1)+\left((1+r)y_1^2-r\right)\left(1-y_1-r(1-y_1)\right)\right)\\
            =&\frac{1-y_1}{\lambda_2y_0y_1}\left(2r+\left((1+r)y_1^2-r\right)\left(1-r\right)\right)\\
            >&\frac{1-y_1}{\lambda_2y_0y_1}\left(r+\left((1+r)y_1^2-r\right)\right)\\
            >&0.
        \end{aligned}
    \end{equation*}

    \textbf{Part 3: the case $\theta_1^2< a \theta_2$ and $\frac{\lambda_2}{\lambda_1}> c$.}

    By Lemma \ref{lemma:welfare_maximization_incentive_compatibility}, we first prove the optimal centralized policy satisfies $w_1> w_2$, so that it is not incentive compatible. When $r=\frac{\lambda_2}{\lambda_1}> c$, type $2$ customers are prioritized for the optimal centralized policy, so we have
    \begin{equation*} 
        \begin{aligned}
            &w_1=\frac{x_1}{1-\lambda_1x_1-\lambda_2x_2}
                +\frac{\lambda_2x_2^2}{(1-\lambda_2x_2)(1-\lambda_1x_1-\lambda_2x_2)},\\
            &w_2=\frac{x_2}{1-\lambda_2x_2}.
        \end{aligned}     
    \end{equation*}
    By Lemma \ref{lemma:cmu_rule}, $\frac{x_1}{\theta_1}>\frac{x_2}{\theta_2}$, so $x_1>x_2$.
    Then,
    \begin{equation*}
        w_1-w_2\geq \frac{x_1}{1-\lambda_1x_1-\lambda_2x_2}-\frac{x_2}{1-\lambda_2x_2}\geq \frac{x_1-x_2}{1-\lambda_2x_2}>0.
    \end{equation*}
    Therefore, the optimal centralized policy does not satisfy the incentive compatibility constraints.
    
    It remains to show that in this case, type $1$ customers are prioritized. Equivalently, the optimal solution of \eqref{eq:welfare_maximization_decentralized} corresponds to the optimal solution of \eqref{equation:welfare_maximization_1first}. For a given service depth assignment $(x_1, x_2)$, there are two different cases.

    If $\frac{x_1}{\theta_1}\leq \frac{x_2}{\theta_2}$, from Lemma \ref{lemma:cmu_rule} the statement already holds.

    If $\frac{x_1}{\theta_1}> \frac{x_2}{\theta_2}$, then $x_1>x_2$. The objective
    \begin{align*}
        &\lambda_1\left(V(x_1)-\theta_1w_1\right)+\lambda_2\left(V(x_2)-\theta_2w_2\right) \\
        &\qquad =
        \lambda_1V(x_1)+\lambda_2V(x_2)
        -\frac{\theta_1}{x_1}\rho_1w_1
        -\frac{\theta_2}{x_2}\rho_2w_2 \\
        &\qquad \leq
        \lambda_1V(x_1)+\lambda_2V(x_2)
        -\frac{\theta_1}{x_1}\frac{\rho_1x_1+\rho_2x_2}{1-\rho_1-\rho_2}+\left(\frac{\theta_1}{x_1}-\frac{\theta_2}{x_2}\right)\rho_2w_2,
    \end{align*}
    so the long-term average social welfare is maximized when $w_2$ is minimized. By incentive compatibility, $w_2\geq w_1$. Therefore,
    \begin{equation*}
        w_2\geq \frac{\rho_1x_1+\rho_2x_2}{(1-\rho_1-\rho_2)(\rho_1+\rho_2)}\geq \frac{x_2}{1-\rho_2}\quad \text{and}\quad w_2\geq \frac{x_1}{1-\rho_1},
    \end{equation*}
    so the optimal priority rule is to set
    \begin{equation}
        \label{equation:w_2=w_1}
        w_2=w_1=\max\{\frac{x_1}{1-\rho_1},\frac{\rho_1x_1+\rho_2x_2}{(1-\rho_1-\rho_2)(\rho_1+\rho_2)}\}
    \end{equation}

    In this case, the objective
    \begin{equation*}
        \begin{aligned}
            &\lambda_1\left(V(x_1)-\theta_1w_1\right)+\lambda_2\left(V(x_2)-\theta_2w_2\right)\\
            &\qquad \leq a(1-y_0)-(\lambda_1\theta_1+\lambda_2\theta_2)\frac{\rho_1x_1+\rho_2x_2}{y_0(1-y_0)}\\
            &\qquad = a(1-y_0)-(\theta_1+ r\theta_2)\frac{(1-y_1)^2+\frac{1}{r}(y_1-y_0)^2}{y_0(1-y_0)}\\
            &\qquad = a(1-y_0)-(\theta_1+ r\theta_2)\frac{(1+\frac{1}{r})y_1^2-2(1+\frac{y_0}{r})y_1+1+\frac{y_0^2}{r}}{y_0(1-y_0)}\\
            &\qquad \leq a(1-y_0)-(\theta_1+ r\theta_2)\frac{(1+\frac{1}{r})(\frac{y_0+r}{1+r})^2-2(1+\frac{y_0}{r})\frac{y_0+r}{1+r}+1+\frac{y_0^2}{r}}{y_0(1-y_0)}\\  
            &\qquad = a(1-y_0)-\frac{\theta_1+ r\theta_2}{1+r}\frac{1-y_0}{y_0}\\
            &\qquad < a(1-y_0)-\theta_2\frac{1-y_0}{y_0}.
        \end{aligned}
    \end{equation*}
    The first inequality follows from \eqref{equation:w_2=w_1}, the second inequality holds because the quadratic function with respect to $y_1$ is maximized when $y_1=\frac{y_0+r}{1+r}$, and the third inequality holds because $\theta_1>\theta_2$.
    
    However, the optimal solution of \eqref{equation:welfare_maximization_1first} gives long-term average social welfare at least $a(1-y_ 0)-\theta_2\frac{1-y_0}{y_0}$ when we set $y_1=1$. Thus, the allocation policy is suboptimal.

    Therefore, the optimal decentralized policy will never choose $(x_1,x_2)$ such that $\frac{x_1}{\theta_1}>\frac{x_2}{\theta_2}$, and will give priority to type $1$ customers. In this case, the long-term average social welfare
    \begin{equation*}
        \mathcal{W}_D=\text{Obj}_1(r)<\text{Obj}_2(r)=\mathcal{W}_C.
    \end{equation*}
    Consequently, the optimal menu of the decentralized problem gives strictly lower social welfare than the centralized case.

    Lastly, we show that the payment rule
    \begin{equation*}
    \begin{aligned}
        &p_1=ax_1-\theta_1w_1,\\
        &p_2=ax_2-\theta_2w_2-(\theta_1-\theta_2)w_1.
    \end{aligned}
    \end{equation*}
    is non-negative.

    If $\theta_1^2\geq a\theta_2$, then
    \begin{equation*}
    \begin{aligned}
        &p_1=0,\\
        &p_2=ax_2^*-\theta_2\frac{x_2^*}{1-\lambda_2x_2^*}=\frac{1}{\lambda_2}\left(\sqrt{a}-\sqrt{\theta_2}\right)^2>0.
    \end{aligned}
    \end{equation*}

    If $\theta_1^2<a\theta_2$, we have
    \begin{equation*}
        \begin{aligned}
            \tilde g_1\left(\frac{\theta_1}{a}\right)&=\frac{\sqrt{a\theta_2}\left(\left(1+r\right)\left(\frac{\theta_1}{a}\right)^2-r\right)}{\sqrt{\frac{\left(\frac{\theta_1}{a}\right)^2+r\left(1-\frac{\theta_1}{a}\right)^2}{\frac{\theta_1}{a}}}}-\theta_1\\
            &=\sqrt{\theta_1\theta_2}\frac{\left(1+r\right)\left(\frac{\theta_1}{a}\right)^2-r}{\sqrt{\left(\frac{\theta_1}{a}\right)^2+r\left(1-\frac{\theta_1}{a}\right)^2}}-\theta_1\\
            &<\sqrt{\theta_1\theta_2}\sqrt\frac{\left(1+r\right)\left(\frac{\theta_1}{a}\right)^2-r}{\left(\frac{\theta_1}{a}\right)^2+r\left(1-\frac{\theta_1}{a}\right)^2}-\theta_1\\
            &<\sqrt{\theta_1\theta_2}-\theta_1<0.
        \end{aligned}
    \end{equation*}
    Therefore, the optimal solution of \eqref{equation:welfare_maximization_1first} satisfies $y_1>\frac{\theta_1}{a}$. Then
    \begin{equation*}
        p_1=\frac{1-y_1}{\lambda_1y_1}(ay_1-\theta_1)>0.
    \end{equation*}
    Regarding the second price,
    \begin{equation*}
        p_2=\frac{1}{\lambda_2}\left(a(y_1-y_0)-\theta_2\left(\frac{y_1-y_0}{y_0}+\frac{r(1-y_1)^2}{y_0y_1}\right)-(\theta_1-\theta_2)\frac{r(1-y_1)}{y_1}\right).
    \end{equation*}
    Notice that
    \begin{equation*}
        \frac{a}{\theta_2}y_0^2y_1=y_1^2+r(1-y_1)^2 \quad \text{and}\quad y_0=\frac{\theta_2}{\theta_1}\left((1+r)y_1^2-r\right),
    \end{equation*}
    then
    \begin{equation*}
        \begin{aligned}
            p_2=\frac{\theta_2}{\lambda_2y_1y_0^2}\left(\left(y_1^2+r(1-y_1)^2\right)\left(y_1-y_0\right)-y_1y_0(y_1-y_0)-ry_0(1-y_1)^2-ry_0(1-y_1)\left((1+r)y_1^2-r-y_0\right)\right)
        \end{aligned}
    \end{equation*}
    Denote
    \begin{equation*}
        A_1=(1+r)y_1^2-r,\quad A_2=y_1^2+r(1-y_1)^2,\quad A_3=y_1-A_1,\quad A_4=A_1-y_0,
    \end{equation*}
    then
    \begin{equation*}
        \begin{aligned}
            &\left(y_1^2+r(1-y_1)^2\right)\left(y_1-y_0\right)-y_1y_0(y_1-y_0)-ry_0(1-y_1)^2-ry_0(1-y_1)\left((1+r)y_1^2-r-y_0\right)\\
            =\,&A_2(A_3+A_4)-y_1(A_1-A_4)(A_3+A_4)-r(1-y_1)(A_1-A_4)(1-y_1+A_4)\\
            =\,&A_3(A_2-y_1A_1)-r(1-y_1)^2A_1+A_4(A_2-y_1A_1+y_1A_3-r(1-y_1)A_1+r(1-y_1)^2)+A_4^2(r(1-y_1)+y_1)\\
            =\,&(1-y_1)^2(r^2y_1^3+r^2y_1+2r^2+2ry_1^3+ry_1+y_1^3)\\&\qquad+A_4(1-y_1)(r^2(1-y_1^2)+ry_1^2+2r+2y_1^2)+A_4^2(r(1-y_1)+y_1)\\
            >\,&0
        \end{aligned}
    \end{equation*}
    Therefore, $p_2\geq 0$.
    \hfill{\Halmos}
\end{proof}

\subsection{Proof of Proposition~\ref{proposition:social_welfare_gap}}
\begin{proof}{Proof of Proposition~\ref{proposition:social_welfare_gap}.}
    If $\theta_1^2\geq a\theta_2$, $\mathcal{W}_{D}=\mathcal{W}_{C}$.

    If $\theta_1^2< a\theta_2$, we have
    \begin{align*}
        \mathcal{W}_{C}\leq a+\theta_1+\theta_2-3\sqrt[3]{a\theta_1\theta_2},\\
        \mathcal{W}_{D}\geq a+\theta_2-2\sqrt{a\theta_2}.
    \end{align*}
    Then we only have to prove
    \begin{align*}
        a+\theta_2-2\sqrt{a\theta_2}\geq \frac{3}{4}(a+\theta_1+\theta_2-3\sqrt[3]{a\theta_1\theta_2}),
    \end{align*}
    which is equivalent to
    \begin{equation*}
        2\sqrt{a\theta_2}+\theta_1-3\sqrt[3]{a\theta_1\theta_2}\leq \frac{1}{3}(\sqrt{a}-\sqrt{\theta_2})^2.
    \end{equation*}
    Denote $k=\frac{\sqrt{a\theta_2}}{\theta_1}$, then $k>1$ and $\sqrt{a}\geq k\sqrt{\theta_1}$. Then
    \begin{align*}
        (\sqrt{a}-\sqrt{\theta_2})^2\geq (\sqrt{a}-\sqrt{\theta_1})^2\geq (k\sqrt{\theta_1}-\sqrt{\theta_1})^2=(k-1)^2\theta_1,\\
        2\sqrt{a\theta_2}+\theta_1-3\sqrt[3]{a\theta_1\theta_2}=(2k+1-3k^{\frac{2}{3}})\theta_1.
    \end{align*}
    Notice that
    \begin{align*}
        \frac{2k+1-3k^{\frac{2}{3}}}{(k-1)^2}&=\frac{(k^{\frac{1}{3}}-1)^2(2k^{\frac{1}{3}}+1)}{(k-1)^2}\\
        &=\frac{2k^{\frac{1}{3}}+1}{(k^{\frac{2}{3}}+k^{\frac{1}{3}}+1)^2}\\
        &\leq \frac{1}{k^{\frac{2}{3}}+k^{\frac{1}{3}}+1}\\
        &\leq \frac{1}{3},
    \end{align*}
    therefore,
    \begin{equation*}
        2\sqrt{a\theta_2}+\theta_1-3\sqrt[3]{a\theta_1\theta_2}\leq \frac{1}{3}(\sqrt{a}-\sqrt{\theta_2})^2.
    \end{equation*}
    On the other hand, let $\frac{\lambda_2}{\lambda_1}\rightarrow\infty$, $a=1$, $\theta_1=(1-e)^2(1+e^2)$, $\theta_2=(1-e)^2$ and $e\rightarrow0$, then
    \begin{equation*}
        \mathcal{W}_{D}\rightarrow \frac{3}{4}\mathcal{W}_{C},
    \end{equation*}
    so the bound is tight.
    \hfill{\Halmos}
\end{proof}

\section{Proofs for Section~\ref{section:continuous-type setting}}
\label{appendix:section_4}

\subsection{Proof of Lemma~\ref{lemma:feasible_region_continuous}}
\begin{proof}{Proof of Lemma~\ref{lemma:feasible_region_continuous}.}
    We first prove necessity. Fix any measurable subset of customers $\mathcal S\subseteq[\underline{\theta},\bar{\theta}]$. First consider a work-conserving  priority rule without holding delays. Remove all customers outside $\mathcal S$, while keeping the arrival times, realized service requirements, and the relative priority rule for customers in $\mathcal S$ fixed. Under this coupling, every customer in $\mathcal S$ completes no later in the isolated subsystem than in the original system.

    The isolated subsystem is stable because $\lambda\int_{\mathcal S}x(\theta)f(\theta)d\theta\leq\lambda\int_{\underline{\theta}}^{\bar{\theta}}x(\theta)f(\theta)d\theta<1$. By Kleinrock's conservation law, every work-conserving  priority rule satisfies
    \begin{equation*}
       \lambda\int_\mathcal{S} x(\theta)w(\theta)f(\theta)d\theta= \frac{\lambda\int_\mathcal{S} x^2(\theta)f(\theta)d\theta}{1-\lambda\int_\mathcal{S} x(\theta)f(\theta)d\theta}.
    \end{equation*}

    Since removing customers outside $\mathcal S$ cannot increase the sojourn time of customers in $\mathcal S$, the original system satisfies
    \begin{equation*}
       \lambda\int_{\mathcal S}
       x(\theta)w(\theta)f(\theta)d\theta
       \geq
       \frac{
       \lambda\int_{\mathcal S}x^2(\theta)f(\theta)d\theta
       }{
       1-\lambda\int_{\mathcal S}x(\theta)f(\theta)d\theta
       }.
    \end{equation*}
    The same inequality holds for randomized priority rules by conditioning on the server's randomization and then taking expectations. Non-work-conserving policies and nonnegative type-dependent holding delays can only increase the sojourn times and therefore preserve the inequality.

    Finally, the inequality is preserved under the $L^1(\nu_x)$ closure, where $\nu_x(d\theta)=\lambda x(\theta)dF(\theta)$. Indeed, for any two sojourn-time functions $w(\cdot)$ and $\tilde w(\cdot)$,
    \begin{equation*}
        \left|
        \lambda\int_{\mathcal S}
        x(\theta)\bigl(w(\theta)-\tilde w(\theta)\bigr)
        f(\theta)d\theta
        \right|
        \leq
        \lambda\int_{\underline{\theta}}^{\bar{\theta}}
        x(\theta)|w(\theta)-\tilde w(\theta)|
        f(\theta)d\theta
        =
        \|w-\tilde w\|_x.
    \end{equation*}
    Thus, the left-hand side of the required inequality is continuous in $L^1(\nu_x)$, which proves necessity for every operationally achievable sojourn-time function.

    Next we prove sufficiency. Let $\mathcal A$ be the set of operationally achievable sojourn-time functions; equivalently, $\mathcal A$ is the $L^1(\nu_x)$-closure of the mean-sojourn-time functions generated by randomized  priority rules, together with nonnegative type-dependent holding delays. This set is nonempty and closed by definition. It is convex because the server can randomize, independently of the arrival and service processes, over any two priority rules and holding-delay policies. It is upward closed because the server can strategically delay any measurable subset of customers by any nonnegative holding time.

    Now suppose $w(\cdot)$ satisfies all the inequalities for the given $x(\cdot)$. Suppose, toward a contradiction, that $w(\cdot)\notin\mathcal A$. Since $\mathcal A$ is a nonempty closed convex subset of $L^1(\nu_x)$, the separating hyperplane theorem and the duality between $L^1(\nu_x)$ and $L^\infty(\nu_x)$ imply that there exists an essentially bounded measurable function $a(\cdot)$ such that
    \begin{equation}
        \label{equation:separation_continuous}
        \lambda\int_{\underline{\theta}}^{\bar{\theta}}
        a(\theta)x(\theta)w(\theta)f(\theta)d\theta
        <
        \inf_{\tilde w\in\mathcal A}
        \lambda\int_{\underline{\theta}}^{\bar{\theta}}
        a(\theta)x(\theta)\tilde w(\theta)f(\theta)d\theta.
    \end{equation}
    Since $\mathcal A$ is upward closed, the separating function must be nonnegative $\nu_x$-almost everywhere. Indeed, if $a(\theta)<0$ on a set of positive $\nu_x$-measure, then there exists a subset of positive $\nu_x$-measure on which $a(\theta)$ is bounded above by a strictly negative constant. Starting from any element of $\mathcal A$ and adding an arbitrarily large holding delay on this subset would drive the right-hand side of \eqref{equation:separation_continuous} to $-\infty$, which contradicts the strict separation. Thus, after modifying $a(\cdot)$ on a $\nu_x$-null set, assume that $a(\theta)\geq 0$ everywhere.

    Define
    \begin{equation*}
        \mathcal S_t:=\{\theta:a(\theta)>t\},\quad t\geq 0.
    \end{equation*}
    Since $a(\cdot)$, $x(\cdot)$, and $w(\cdot)$ are nonnegative,
    \begin{equation*}
        \begin{aligned}
            \lambda\int_{\underline{\theta}}^{\bar{\theta}}
            a(\theta)x(\theta)w(\theta)f(\theta)d\theta
            &=
            \lambda\int_{0}^{\infty}
            \int_{\mathcal S_t}
            x(\theta)w(\theta)f(\theta)d\theta dt\\
            &\geq
            \int_{0}^{\infty}
            \frac{
            \lambda\int_{\mathcal S_t}
            x^2(\theta)f(\theta)d\theta
            }{
            1-\lambda\int_{\mathcal S_t}
            x(\theta)f(\theta)d\theta
            }dt.
        \end{aligned}
    \end{equation*}
    The denominators are strictly positive because for every $t\geq 0$,
    \begin{equation*}
        \lambda\int_{\mathcal S_t}x(\theta)f(\theta)d\theta
        \leq
        \lambda\int_{\underline{\theta}}^{\bar{\theta}}
        x(\theta)f(\theta)d\theta
        <1.
    \end{equation*}

    On the other hand, consider the work-conserving  priority policy that gives priority to customers with larger $a(\theta)$, with ties broken according to any fixed rule. Let $\tilde w(\cdot)$ be the sojourn-time function induced by this policy. For every $t\geq 0$, all customers in $\mathcal S_t$ have strict priority over all customers outside $\mathcal S_t$. Therefore, whenever at least one customer in $\mathcal S_t$ is present, the server works on a customer in $\mathcal S_t$. Consequently, the aggregate workload process of customers in $\mathcal S_t$ is exactly the workload process of the isolated work-conserving subsystem containing only customers in $\mathcal S_t$.

    By Kleinrock's conservation law, for every $t\geq 0$,
    \begin{equation*}
        \lambda\int_{\mathcal S_t}
        x(\theta)\tilde w(\theta)f(\theta)d\theta
        =
        \frac{
        \lambda\int_{\mathcal S_t}
        x^2(\theta)f(\theta)d\theta
        }{
        1-\lambda\int_{\mathcal S_t}
        x(\theta)f(\theta)d\theta
        }.
    \end{equation*}
    Moreover, applying the same argument to the full type space gives
    \begin{equation*}
        \lambda\int_{\underline{\theta}}^{\bar{\theta}}
        x(\theta)\tilde w(\theta)f(\theta)d\theta
        =
        \frac{
        \lambda\int_{\underline{\theta}}^{\bar{\theta}}
        x^2(\theta)f(\theta)d\theta
        }{
        1-\lambda\int_{\underline{\theta}}^{\bar{\theta}}
        x(\theta)f(\theta)d\theta
        }
        <\infty,
    \end{equation*}
    where the inequality follows from Assumption~\ref{assumption:continuous_types}. Hence $\tilde w(\cdot)\in L^1(\nu_x)$, and because it is generated by an admissible  priority policy, $\tilde w(\cdot)\in\mathcal A$. Then we have
    \begin{equation*}
        \begin{aligned}
            \lambda\int_{\underline{\theta}}^{\bar{\theta}}
            a(\theta)x(\theta)\tilde w(\theta)f(\theta)d\theta
            &=
            \lambda\int_{0}^{\infty}
            \int_{\mathcal S_t}
            x(\theta)\tilde w(\theta)f(\theta)d\theta dt\\
            &=
            \int_{0}^{\infty}
            \frac{
            \lambda\int_{\mathcal S_t}
            x^2(\theta)f(\theta)d\theta
            }{
            1-\lambda\int_{\mathcal S_t}
            x(\theta)f(\theta)d\theta
            }dt.
        \end{aligned}
    \end{equation*}
    Therefore,
    \begin{equation*}
        \lambda\int_{\underline{\theta}}^{\bar{\theta}}
        a(\theta)x(\theta)w(\theta)f(\theta)d\theta
        \geq
        \lambda\int_{\underline{\theta}}^{\bar{\theta}}
        a(\theta)x(\theta)\tilde w(\theta)f(\theta)d\theta
        \geq
        \inf_{\tilde w\in\mathcal A}
        \lambda\int_{\underline{\theta}}^{\bar{\theta}}
        a(\theta)x(\theta)\tilde w(\theta)f(\theta)d\theta.
    \end{equation*}
    This contradicts \eqref{equation:separation_continuous}. Therefore,
    $w(\cdot)\in\mathcal A$, which proves sufficiency.
    \hfill{\Halmos}
\end{proof}
\subsection{Proof of Proposition~\ref{proposition:continuous_types_welfare_maximization}}
\begin{proof}{Proof of Proposition~\ref{proposition:continuous_types_welfare_maximization}.}
    Fix any admissible assignment $x(\cdot)$ and any operationally feasible sojourn time $w(\cdot)$. Let
    \begin{equation*}
        G(z):=\mathbb{P}_{\theta\sim F}\left[x(\theta)\leq z\right]
    \end{equation*}
    denote the aggregate distribution of assigned service depth across all arrivals, and let
    \begin{equation*}
        G^{-1}(u):=\inf\{z\geq 0:G(z)\geq u\},\quad u\in[0,1],
    \end{equation*}
    be its generalized inverse. Because $F$ is continuous, the quantile map
    \begin{equation*}
        \tilde x(\theta):=G^{-1}(1-F(\theta))
    \end{equation*}
    is measurable and non-increasing in $\theta$. Moreover, the pushforward of $F$ under $\tilde x(\cdot)$ is exactly $G$. Therefore,
    \begin{equation*}
        \lambda\int_{\underline{\theta}}^{\bar{\theta}} V(\tilde x(\theta))\,dF(\theta)
        =\lambda\int V(x)\,dG(x)
        =\lambda\int_{\underline{\theta}}^{\bar{\theta}} V(x(\theta))\,dF(\theta).
    \end{equation*}
    Hence the value term in \eqref{equation:continuous_types_welfare_maximization} is unchanged.

    On the other hand, if $\lambda\int_{\underline{\theta}}^{\bar{\theta}}\theta w(\theta)dF(\theta)<\infty$ (otherwise the result is immediate), the long-term average delay cost can be written as
    \begin{equation*}
        \begin{aligned}
            \lambda\int_{\underline{\theta}}^{\bar{\theta}}\theta w(\theta)dF(\theta)= \lambda\underline{\theta}\int_{\underline{\theta}}^{\bar{\theta}}w(\theta)dF(\theta)+ \int_{\underline{\theta}}^{\bar{\theta}}\left(\lambda\int_{c}^{\bar{\theta}}w(\theta)dF(\theta)\right)dc.
        \end{aligned}
    \end{equation*}
    
    Let $\tilde w(\theta)$ denote the expected sojourn time under the deterministic assignment $\tilde x(\theta)$ under the priority rule that gives higher priority to larger $\theta$. Because $\tilde x(\theta)$ is non-increasing, the ratio $\tilde x(\theta)/\theta$ is also non-increasing in $\theta$. Thus this priority rule is consistent with the $c\mu$-rule.

    For $c\in[\underline{\theta},\bar{\theta}]$, define
    \begin{equation*}
        S(c):=\lambda\int_{c}^{\bar{\theta}}w(\theta)dF(\theta)
        \qquad\text{and}\qquad
        \tilde S(c):=\lambda\int_{c}^{\bar{\theta}} \tilde w(\theta)\,dF(\theta).
    \end{equation*}
    Under the assignment $\tilde x(\cdot)$, the set of types $[c,\bar{\theta}]$ receives exactly the $\alpha(c)=1-F(c)$ fraction of smallest service depths in the aggregate distribution $G$. Therefore,
    \begin{equation*}
        \mathbb{P}_{\theta\sim F}[x(\theta)\leq z,\theta\in[c, \bar{\theta}]]\leq \mathbb{P}_{\theta\sim F}[\tilde x(\theta)\leq z]=G(z)
    \end{equation*}
    Denote $G_c(z)=\mathbb{P}_{\theta\sim F}[x(\theta)\leq z,\theta\in[c, \bar{\theta}]]$ and $G_c^{-1}(u)=\inf\{z\geq 0: G_c(z)\geq u\}$, then
    \begin{equation*}
        G^{-1}(u)\leq G_c^{-1}(u), \quad\forall u\in[0,\alpha(c)].
    \end{equation*}
    Then for all $\alpha\in[0,\alpha(c)]$,
    \begin{equation*}
        \begin{aligned}
            \lambda\int_{0}^{\alpha}(G^{-1}(u))^2du\leq \lambda\int_{0}^{\alpha}(G_c^{-1}(u))^2du\quad \text{and}\quad
            \lambda\int_{0}^{\alpha}G^{-1}(u)du\leq \lambda\int_{0}^{\alpha}G_c^{-1}(u)du.
        \end{aligned}
    \end{equation*}
    Thus, using the standard  formula for a continuum of priorities,
    \begin{equation}
        \label{equation:tail1}
        \begin{aligned}
            \tilde S(c) &= \lambda\int_c^{\bar{\theta}}\left[\frac{\lambda\int_{\theta}^{\bar{\theta}}\tilde x^2(c)dF(c)}{\left(1-\lambda\int_{\theta}^{\bar{\theta}}\tilde x(c)dF(c)\right)^2}+\frac{\tilde x(\theta)}{1-\lambda\int_{\theta}^{\bar{\theta}}\tilde x(c)dF(c)}\right]dF(\theta)\\
            &=\lambda\int_0^{\alpha(c)}\left[\frac{\lambda\int_{0}^{\alpha}(G^{-1}(u))^2du}{\left(1-\lambda\int_{0}^{\alpha}G^{-1}(u)du\right)^2}+\frac{G^{-1}(\alpha)}{1-\lambda\int_{0}^{\alpha}G^{-1}(u)du}\right]d\alpha\\
            &\leq \lambda\int_0^{\alpha(c)}\left[\frac{\lambda\int_{0}^{\alpha}(G_c^{-1}(u))^2du}{\left(1-\lambda\int_{0}^{\alpha}G_c^{-1}(u)du\right)^2}+\frac{G_c^{-1}(\alpha)}{1-\lambda\int_{0}^{\alpha}G_c^{-1}(u)du}\right]d\alpha,
        \end{aligned}
    \end{equation}
    where the inequality holds because all the denominators are positive by admissibility.
    
    On the other hand, consider a subsystem consisting only of jobs whose types lie in $[c,\bar{\theta}]$ with service depth means given by $x(\theta)$. Removing all customers outside $[c,\bar{\theta}]$ cannot increase the sojourn times of customers in $[c,\bar{\theta}]$, because those customers can only consume the server's capacity and increase congestion for customers in $[c,\bar{\theta}]$. To prove this, take any sample path and remove all the customers  outside $[c,\bar{\theta}]$. The service epochs assigned to the remaining jobs in the original sample path are feasible in the isolated subsystem, possibly with additional idleness. Hence the aggregate sojourn time of customers in $[c,\bar{\theta}]$ in the original system is at least the minimum aggregate sojourn time achievable in the isolated subsystem.

    In the isolated subsystem, the arrival rate is $\lambda_c=\alpha(c)\lambda$, and customers' aggregate service depth distribution has cdf $\frac{1}{\alpha(c)}G_c(z)$. Moreover, among customers in this isolated subsystem, the aggregate sojourn time $\lambda\int_c^{\bar{\theta}} w(\theta)dF(\theta)$ is minimized by the $c\mu$-rule with common delay cost, equivalently by giving priority to smaller service depths. The resulting minimum aggregate sojourn time is
    \begin{equation*}
        \lambda\int_0^{\alpha(c)}\left[\frac{\lambda\int_{0}^{\alpha}(G_c^{-1}(u))^2du}{\left(1-\lambda\int_{0}^{\alpha}G_c^{-1}(u)du\right)^2}+\frac{G_c^{-1}(\alpha)}{1-\lambda\int_{0}^{\alpha}G_c^{-1}(u)du}\right]d\alpha.
    \end{equation*}
    Therefore,
    \begin{equation}
        \label{equation:tail2}
        S(c)\geq \lambda\int_0^{\alpha(c)}\left[\frac{\lambda\int_{0}^{\alpha}(G_c^{-1}(u))^2du}{\left(1-\lambda\int_{0}^{\alpha}G_c^{-1}(u)du\right)^2}+\frac{G_c^{-1}(\alpha)}{1-\lambda\int_{0}^{\alpha}G_c^{-1}(u)du}\right]d\alpha.
    \end{equation}
    Combining \eqref{equation:tail1} and \eqref{equation:tail2}, we obtain the tail comparison
    \begin{equation}
        \label{equation:tail}
        \tilde S(c)\leq S(c),\qquad \forall c\in[\underline{\theta},\bar{\theta}].
    \end{equation}
    Using the result of \eqref{equation:tail} for every $c\in[\underline{\theta},\bar{\theta}]$, we have
    \begin{equation*}
        \begin{aligned}
            \lambda\int_{\underline{\theta}}^{\bar{\theta}} \theta \tilde w(\theta)\,dF(\theta)= \underline{\theta}\tilde S(\underline{\theta})+\int_{\underline{\theta}}^{\bar{\theta}} \tilde S(c)\,dc\leq \underline{\theta}S(\underline{\theta})+\int_{\underline{\theta}}^{\bar{\theta}} S(c)\,dc= \lambda\int_{\underline{\theta}}^{\bar{\theta}}\theta w(\theta)dF(\theta).
        \end{aligned}
    \end{equation*}
    Therefore the rearranged assignment $\tilde x(\cdot)$ weakly lowers the weighted delay-cost term while leaving the value term unchanged. Its welfare is thus weakly higher than that of the original assignment.

    Finally, because $\tilde x(\theta)$ is non-increasing, the $c\mu$-rule implies that the induced optimal priority rule is equivalent to giving strict priority to customers with larger delay sensitivities.
    \hfill{\Halmos}
\end{proof}

\subsection{Proof of Proposition~\ref{proposition:existence_of_optimality}}

In order to prove the proposition, we first state Lemma~\ref{lemma:bounded_service_time} that shows if the marginal service value $V'(M)$ is lower than the most patient customer's delay sensitivity, serving any customer longer than $M$ will be suboptimal.

\begin{lemma}
    \label{lemma:bounded_service_time}
    Suppose $V(x)$ is continuously differentiable and $V'(M)\leq\underline{\theta}$. For any admissible and non-increasing service depth assignment $x(\cdot)$, there exists a non-increasing function $\tilde x:[\underline{\theta}, \bar{\theta}]\mapsto [0,M]$ such that the assignment $\theta\mapsto \tilde x(\theta)$ achieves weakly higher welfare than $x(\cdot)$. 
\end{lemma}
\begin{proof}{Proof of Lemma~\ref{lemma:bounded_service_time}.}
    Fix an admissible service depth assignment $x(\cdot)$. Denote $\tilde x(\theta)= \min\{x(\theta), M\}$, then $\tilde x(\theta)$ is also admissible and non-increasing. For all $\theta\in[\underline{\theta}, \bar{\theta}]$, we have
    \begin{equation*}
        \int_{{\theta}}^{\bar{\theta}}x(c)dF(c)\geq\int_{{\theta}}^{\bar{\theta}}\tilde x(c)dF(c) \quad\text{and}\quad \int_{{\theta}}^{\bar{\theta}}x^2(c)dF(c)\geq\int_{{\theta}}^{\bar{\theta}}\tilde x^2(c)dF(c).
    \end{equation*}
    Therefore,
    \begin{equation*}
        \begin{aligned}
            &V(x(\theta))-\theta\left(\frac{\lambda\int_{\theta}^{\bar{\theta}}x^2(c)dF(c)}{\left(1-\lambda\int_{\theta}^{\bar{\theta}}x(c)dF(c)\right)^2}+\frac{x(\theta)}{1-\lambda\int_{\theta}^{\bar{\theta}}x(c)dF(c)}\right)\\
            &\qquad\leq V(\tilde x(\theta))+\underline{\theta}(x(\theta)-\tilde x(\theta))-\theta\left(\frac{\lambda\int_{\theta}^{\bar{\theta}}x^2(c)dF(c)}{\left(1-\lambda\int_{\theta}^{\bar{\theta}}x(c)dF(c)\right)^2}+\frac{x(\theta)}{1-\lambda\int_{\theta}^{\bar{\theta}}x(c)dF(c)}\right)\\
            &\qquad\leq V(\tilde x(\theta))-\theta\left(\frac{\lambda\int_{\theta}^{\bar{\theta}}x^2(c)dF(c)}{\left(1-\lambda\int_{\theta}^{\bar{\theta}}x(c)dF(c)\right)^2}+\frac{\tilde x(\theta)}{1-\lambda\int_{\theta}^{\bar{\theta}}x(c)dF(c)}\right)\\
            &\qquad\leq V(\tilde x(\theta))-\theta\left(\frac{\lambda\int_{\theta}^{\bar{\theta}}\tilde x^2(c)dF(c)}{\left(1-\lambda\int_{\theta}^{\bar{\theta}}\tilde x(c)dF(c)\right)^2}+\frac{\tilde x(\theta)}{1-\lambda\int_{\theta}^{\bar{\theta}}\tilde x(c)dF(c)}\right),
        \end{aligned}
    \end{equation*}
    which finishes the proof.
    \hfill{\Halmos}
\end{proof}

\begin{proof}{Proof of Proposition~\ref{proposition:existence_of_optimality}.}

\textbf{Part 1: the case $V(x)=ax$ and $a>\underline{\theta}$.}

In this case, the long-term average value generated by the server is
\begin{equation*}
    \lambda\int_{\underline{\theta}}^{\bar{\theta}}V(x(\theta))dF(\theta)=\lambda a\int_{\underline{\theta}}^{\bar{\theta}}x(\theta)dF(\theta),
\end{equation*}
and the long-term average delay cost can be bounded as
\begin{equation*}
    \begin{aligned}
        \lambda\int_{\underline{\theta}}^{\bar{\theta}}\theta\left(\frac{\lambda\int_{\theta}^{\bar{\theta}}x^2(c)dF(c)}{\left(1-\lambda\int_{\theta}^{\bar{\theta}}x(c)dF(c)\right)^2}+\frac{x(\theta)}{1-\lambda\int_{\theta}^{\bar{\theta}}x(c)dF(c)}\right)dF(\theta)&\geq  \lambda\underline{\theta}\int_{\underline{\theta}}^{\bar{\theta}}\frac{x(\theta)}{1-\lambda\int_{\theta}^{\bar{\theta}}x(c)dF(c)}dF(\theta)\\
        &=-\underline{\theta}\ln{\left(1-\lambda\int_{\underline{\theta}}^{\bar{\theta}}x(\theta)dF(\theta)\right)}.
    \end{aligned}
\end{equation*}
Therefore, the objective of problem \eqref{equation:continuous_types_final_problem} is upper bounded as
\begin{equation}
    \label{equation:prop5_1}
    a\lambda\int_{\underline{\theta}}^{\bar{\theta}}x(\theta)dF(\theta)+\underline{\theta}\ln{\left(1-\lambda\int_{\underline{\theta}}^{\bar{\theta}}x(\theta)dF(\theta)\right)}.
\end{equation}

Consider function $h(z)=az+\underline{\theta}\ln(1-z)$, $z\in(0,1)$ it is maximized when $z=1-\underline{\theta}/a$, thus \eqref{equation:prop5_1} is further bounded by
\begin{equation*}
    a-\underline{\theta}+\underline{\theta}\ln{\frac{\underline{\theta}}{a}}.
\end{equation*}
Notice that the equality holds only if $\frac{\lambda\int_{\theta}^{\bar{\theta}}x^2(c)dF(c)}{\left(1-\lambda\int_{\theta}^{\bar{\theta}}x(c)dF(c)\right)^2}=0$ with probability $1$, which is not achievable by any finite support function $x(\cdot)$. On the other hand, consider a sequence of functions $x_n(\cdot)$ such that 
\begin{equation*}
    x_n(\theta)=\left\{
    \begin{aligned}
        &\frac{1-\underline{\theta}/a}{\lambda nF(\theta)+\lambda e^{-n}}, &&\text{if} \, F(\theta)\in[0,\frac{1}{n}-\frac{e^{-n}}{n}],\\
        &0, &&\text{otherwise}.
    \end{aligned}
    \right.
\end{equation*}
Then
\begin{equation*}
    \lambda\int_{\underline{\theta}}^{\bar{\theta}}x_n(\theta)dF(\theta)=\lambda\int_{e^{-n}/n}^{1/n}\frac{1-\underline{\theta}/a}{\lambda nc}dc=1-\underline{\theta}/a,
\end{equation*}
and
\begin{equation*}
    \begin{aligned}
        \int_{\underline{\theta}}^{\bar{\theta}}\int_{\theta}^{\bar{\theta}}x_n^2(c)dF(c)dF(\theta)&=\int_{\underline{\theta}}^{\bar{\theta}}F(c)x_n^2(c)dF(c)\\
        &=\int_{e^{-n}/n}^{1/n}\left(c-\frac{e^{-n}}{n}\right)\frac{(1-\underline{\theta}/a)^2}{\lambda^2n^2c^2}dc\\
        &<\int_{e^{-n}/n}^{1/n}\frac{(1-\underline{\theta}/a)^2}{\lambda^2n^2c}dc\\
        &= \frac{(1-\underline{\theta}/a)^2}{\lambda^2n}.
    \end{aligned}
\end{equation*}

Thus, the social welfare under $x_n(\cdot)$ is bounded by
\begin{equation*}
    \begin{aligned}
        &\lambda\int_{\underline{\theta}}^{\bar{\theta}}\left[ax_n(\theta)-\theta\left(\frac{\lambda\int_{\theta}^{\bar{\theta}}x_n^2(c)dF(c)}{\left(1-\lambda\int_{\theta}^{\bar{\theta}}x_n(c)dF(c)\right)^2}+\frac{x_n(\theta)}{1-\lambda\int_{\theta}^{\bar{\theta}}x_n(c)dF(c)}\right)\right]dF(\theta)\\
        &\qquad = \left(a-\underline{\theta}+\underline{\theta}\ln{\frac{\underline{\theta}}{a}}\right)-\lambda\int_{\underline{\theta}}^{\bar{\theta}}\theta \frac{\lambda\int_{\theta}^{\bar{\theta}}x_n^2(c)dF(c)}{\left(1-\lambda\int_{\theta}^{\bar{\theta}}x_n(c)dF(c)\right)^2}dF(\theta)-\lambda\int_{\underline{\theta}}^{\bar{\theta}}(\theta-\underline{\theta})\frac{x_n(\theta)}{1-\lambda\int_{\theta}^{\bar{\theta}}x_n(c)dF(c)}dF(\theta)\\
        &\qquad \geq \left(a-\underline{\theta}+\underline{\theta}\ln{\frac{\underline{\theta}}{a}}\right)-\lambda^2\bar{\theta}\frac{a^2}{\underline{\theta}^2}\int_{\underline{\theta}}^{\bar{\theta}}\int_{\theta}^{\bar{\theta}}x_n^2(c)dF(c)dF(\theta)-(F^{-1}(\frac{1}{n})-\underline{\theta})\ln{\frac{a}{\underline{\theta}}}\\
        &\qquad \geq \left(a-\underline{\theta}+\underline{\theta}\ln{\frac{\underline{\theta}}{a}}\right)-\bar{\theta}\frac{a^2}{\underline{\theta}^2}\frac{(1-\underline{\theta}/a)^2}{n}-(F^{-1}(\frac{1}{n})-\underline{\theta})\ln{\frac{a}{\underline{\theta}}}\\
        &\qquad \rightarrow a-\underline{\theta}+\underline{\theta}\ln{\frac{\underline{\theta}}{a}} \qquad \text{as} \, n\rightarrow \infty.
    \end{aligned}
\end{equation*}

Therefore, the sequence of $x_n(\cdot)$ asymptotically achieves the upper bound of social welfare $a-\underline{\theta}+\underline{\theta}\ln{\frac{\underline{\theta}}{a}}$. As $n\rightarrow\infty$, the service depth assignment is degenerate, with $x(\underline{\theta})=\infty$ and $x(\theta)=0$ for all $\theta>\underline{\theta}$.

\textbf{Part 2: the case $V(x)$ is continuously differentiable and $\lim_{x\rightarrow\infty}V'(x)<\underline{\theta}$ .}

% By Lemma~\ref{lemma:bounded_service_time}, we can restrict our attention to the service depth assignments that is upper bounded by $M$. Let $\{x_n(\cdot)\}$ be a sequence that approaches the maximum welfare. Since $x_n(\cdot)$ are non-increasing functions, by Helly's selection theorem, there exists a subsequence of $\{x_n(\cdot)\}$ that converges pointwise to a non-increasing function $x^*(\cdot)$. Since the sequence is uniformly bounded, dominated convergence theorem implies convergence of the first and second workload integrals. Moreover, since $\{x_n(\cdot)\}$ is a welfare-maximizing sequence, 
% \begin{equation*}
%     \limsup_{n\rightarrow \infty}\lambda\int_{\underline{\theta}}^{\bar{\theta}}x_n(\theta)dF(\theta)<1.
% \end{equation*}
% Otherwise, the delay-cost term will be unbounded and $\{x_n(\cdot)\}$ can not be welfare-maximizing. Hence $\lambda\int_{\underline{\theta}}^{\bar{\theta}}x^*(\theta)dF(\theta)<1$. Then $x^*(\cdot)$ is an optimal service depth assignment of problem \eqref{equation:continuous_types_final_problem}.

Since $V(x)$ is weakly increasing and weakly concave, there exists $M\in(0,\infty)$ such that $V'(M)\leq\underline{\theta} $. By Proposition~\ref{proposition:continuous_types_welfare_maximization} and Lemma~\ref{lemma:bounded_service_time}, there exists a maximizing sequence $\{x_n(\cdot)\}$ such that every $x_n(\cdot)$ is non-increasing and satisfies $0\leq x_n(\theta)\leq M$. We first show that the stability constraints of this sequence are uniformly bounded away from equality. Since the zero assignment is feasible, the welfare supremum is nonnegative, so we may assume that the welfare under $x_n(\cdot)$ is lower bounded by $-1$. Moreover, it is upper bounded by
\begin{equation*}
\begin{aligned}
\lambda V(M)-\lambda\underline{\theta}\int_{\underline{\theta}}^{\bar{\theta}}\frac{x_n(\theta)}{1-\lambda\int_\theta^{\bar{\theta}}x_n(c)dF(c)}dF(\theta)
=\lambda V(M)+\underline{\theta}\ln\left(1-\lambda\int_{\underline{\theta}}^{\bar{\theta}}x_n(\theta)dF(\theta)\right).
\end{aligned}
\end{equation*}
Therefore,
\begin{equation*}
1-\lambda\int_{\underline{\theta}}^{\bar{\theta}}x_n(\theta)dF(\theta)\geq\exp\left(-\frac{1+\lambda V(M)}{\underline{\theta}}\right)=:\eta>0.
\end{equation*}

Since $\{x_n(\cdot)\}$ is a uniformly bounded sequence of non-increasing functions, Helly's selection theorem implies that there exists a subsequence, still denoted by $\{x_n(\cdot)\}$, and a non-increasing function $x^*(\cdot)$ such that $x_n(\theta)\rightarrow x^*(\theta)$ at every continuity point of $x^*(\cdot)$. A monotone function has at most countably many discontinuities, so the convergence holds for $F$-almost every $\theta$. By the dominated convergence theorem,
\begin{equation*}
\lambda\int_{\underline{\theta}}^{\bar{\theta}}x^*(\theta)dF(\theta)=\lim_{n\rightarrow\infty}\lambda\int_{\underline{\theta}}^{\bar{\theta}}x_n(\theta)dF(\theta)\leq1-\eta<1,
\end{equation*}
and $\int_{\underline{\theta}}^{\bar{\theta}}{x^*}^2(\theta)dF(\theta)\leq M^2<\infty$, so $x^*(\cdot)$ is admissible.

For every $\theta\in[\underline{\theta},\bar{\theta}]$ and $k\in\{1,2\}$, the dominated convergence theorem also gives
\begin{equation*}
\int_\theta^{\bar{\theta}}x_n^k(c)dF(c)\rightarrow\int_\theta^{\bar{\theta}}{x^*}^k(c)dF(c).
\end{equation*}
Furthermore,
\begin{equation*}
1-\lambda\int_\theta^{\bar{\theta}}x_n(c)dF(c)\geq\eta,
\end{equation*}
and the integrands in the welfare objective are uniformly bounded in absolute value by
\begin{equation*}
V(M)+\bar{\theta}\left(\frac{\lambda M^2}{\eta^2}+\frac{M}{\eta}\right).
\end{equation*}
Another application of the dominated convergence theorem therefore gives that the long-term average welfare of $x_n(\cdot)$ converges to the long-term average welfare of $x^*(\cdot)$. Hence $x^*(\cdot)$ is an optimal solution.

It remains to prove that $x^*(\cdot)$ has a continuous representative. Denote by $\mathcal{W}(x)$ the objective value of \eqref{equation:continuous_types_final_problem} for solution $x(\cdot)$. Define
\begin{equation*}
R(\theta)=1-\lambda\int_\theta^{\bar{\theta}}x^*(c)dF(c),\qquad Q(\theta)=\lambda\int_\theta^{\bar{\theta}}{x^*}^2(c)dF(c).
\end{equation*}
The functions $R(\cdot)$ and $Q(\cdot)$ are continuous and $R(\theta)\geq\eta$ for every $\theta$. For any bounded measurable function $h(\cdot)$ such that $x^*(\cdot)+\epsilon h(\cdot)$ is admissible and non-increasing for all sufficiently small $\epsilon>0$, define
\begin{equation*}
D\mathcal W(x^*)[h]=\lim_{\epsilon\downarrow0}\frac{\mathcal W(x^*+\epsilon h)-\mathcal W(x^*)}{\epsilon}.
\end{equation*}
Since $x^*(\cdot)$ and $h(\cdot)$ are bounded and all denominators remain uniformly bounded away from zero, differentiation under the integral sign is justified by the dominated convergence theorem. Applying Fubini's theorem gives
\begin{equation}
\label{equation:continuity_first_variation}
D\mathcal W(x^*)[h]=\lambda\int_{\underline{\theta}}^{\bar{\theta}}\Phi(\theta)h(\theta)dF(\theta),
\end{equation}
where
\begin{equation}
\label{equation:continuity_marginal_welfare}
\begin{aligned}
\Phi(\theta)=&V'(x^*(\theta))-\frac{\theta}{R(\theta)}-\lambda\int_{\underline{\theta}}^\theta c\left(\frac{x^*(c)}{R^2(c)}+\frac{2Q(c)}{R^3(c)}\right)dF(c)-2\lambda x^*(\theta)\int_{\underline{\theta}}^\theta\frac{c}{R^2(c)}dF(c).
\end{aligned}
\end{equation}

Suppose, toward a contradiction, that $x^*(\cdot)$ has a downward jump at some $\theta_0\in(\underline{\theta},\bar{\theta})$. Denote
\begin{equation*}
x_-=\lim_{\theta\uparrow\theta_0}x^*(\theta),\qquad x_+=\lim_{\theta\downarrow\theta_0}x^*(\theta),
\end{equation*}
where $x_->x_+$. Since $R(\cdot)$, $Q(\cdot)$ and all the cumulative integral terms in \eqref{equation:continuity_marginal_welfare} are continuous, $\Phi(\cdot)$ has one-sided limits at $\theta_0$ and
\begin{equation*}
\Phi(\theta_0+)-\Phi(\theta_0-)=V'(x_+)-V'(x_-)+2\lambda(x_--x_+)\int_{\underline{\theta}}^{\theta_0}\frac{c}{R^2(c)}dF(c)>0.
\end{equation*}
The inequality follows because concavity of $V$ implies $V'(x_+)\geq V'(x_-)$, while $f(c)>0$ and $c>0$ imply
\begin{equation*}
\int_{\underline{\theta}}^{\theta_0}\frac{c}{R^2(c)}dF(c)>0.
\end{equation*}

For $\delta\in(0,\min\{F(\theta_0),1-F(\theta_0)\})$, define
\begin{equation*}
h_\delta(c)=\mathds{1}_{(\theta_0,F^{-1}(F(\theta_0)+\delta))}(c)-\mathds{1}_{(F^{-1}(F(\theta_0)-\delta),\theta_0)}(c).
\end{equation*}
The two intervals have the same probability mass, so
\begin{equation*}
\int_{\underline{\theta}}^{\bar{\theta}}h_\delta(c)dF(c)=0.
\end{equation*}
Since $x^*(c)\geq x_-$ for $c<\theta_0$ and $x^*(c)\leq x_+$ for $c>\theta_0$, for every $\epsilon\in(0,(x_--x_+)/2)$ the function $x^*(\cdot)+\epsilon h_\delta(\cdot)$ is nonnegative and non-increasing after redefining its value at $\theta_0$, if necessary. This redefinition does not affect feasibility or welfare because $F$ has no atom at $\theta_0$. Moreover, its total workload is unchanged, so it is admissible.

Using \eqref{equation:continuity_first_variation}, we obtain
\begin{equation*}
\frac{D\mathcal W(x^*)[h_\delta]}{\lambda\delta}=\frac{1}{\delta}\int_{\theta_0}^{F^{-1}(F(\theta_0)+\delta)}\Phi(c)dF(c)-\frac{1}{\delta}\int_{F^{-1}(F(\theta_0)-\delta)}^{\theta_0}\Phi(c)dF(c).
\end{equation*}
Since $\Phi(\cdot)$ has one-sided limits at $\theta_0$,
\begin{equation*}
\lim_{\delta\downarrow0}\frac{D\mathcal W(x^*)[h_\delta]}{\lambda\delta}=\Phi(\theta_0+)-\Phi(\theta_0-)>0.
\end{equation*}
Thus, $D\mathcal W(x^*)[h_\delta]>0$ for all sufficiently small $\delta>0$. By the definition of the directional derivative, $\mathcal W(x^*+\epsilon h_\delta)>\mathcal W(x^*)$ for all sufficiently small $\epsilon>0$, which contradicts the optimality of $x^*(\cdot)$.

Therefore, $x^*(\cdot)$ has no discontinuity in $(\underline{\theta},\bar{\theta})$, because every discontinuity of a monotone function is a jump discontinuity. Since $x^*(\cdot)$ is bounded and monotone, its one-sided limits at $\underline{\theta}$ and $\bar{\theta}$ exist and are finite. Redefining $x^*(\underline{\theta})$ and $x^*(\bar{\theta})$ to equal these limits does not affect feasibility or social welfare. The resulting optimal solution is continuous and non-increasing on $[\underline{\theta},\bar{\theta}]$, which finishes the proof.
\hfill{\Halmos}

\end{proof}

\subsection{Proofs of Lemma~\ref{lemma:continuous_types_incentive_compatibility} and Proposition~\ref{proposition:continuous_types_welfare_maximization_decentralized_case}}
\begin{proof}{Proof of Lemma~\ref{lemma:continuous_types_incentive_compatibility}.}
    Let
    \begin{equation*}
        U(\theta):=V(x(\theta))-\theta w(\theta)-p(\theta)
    \end{equation*}
    denote the truthful utility of type $\theta$. Fix $\theta<\hat{\theta}$. Incentive compatibility implies
    \begin{equation*}
        \begin{aligned}
            U(\theta)&\geq V(x(\hat{\theta}))-\theta w(\hat{\theta})-p(\hat{\theta})
            =U(\hat{\theta})+(\hat{\theta}-\theta)w(\hat{\theta}),\\
            U(\hat{\theta})&\geq V(x(\theta))-\hat{\theta}w(\theta)-p(\theta)
            =U(\theta)-(\hat{\theta}-\theta)w(\theta).
        \end{aligned}
    \end{equation*}
    Hence
    \begin{equation*}
        (\hat{\theta}-\theta)w(\hat{\theta})\leq U(\theta)-U(\hat{\theta})\leq (\hat{\theta}-\theta)w(\theta).
    \end{equation*}
    In particular, $w(\hat{\theta})\leq w(\theta)$, so $w(\cdot)$ must be non-increasing.

    Now assume that $w(\cdot)$ is non-increasing. Define
    \begin{equation*}
        U(\theta):=\int_{\theta}^{\bar{\theta}}w(c)\,dc
        \quad\text{and}\quad
        p(\theta):=V(x(\theta))-\theta w(\theta)-U(\theta).
    \end{equation*}
    Then $U(\bar{\theta})=0$, so the highest type obtains zero utility and individual rationality will follow once incentive compatibility is established. Fix any $\theta$ and $\hat{\theta}$. If $\hat{\theta}\geq \theta$, then
    \begin{equation*}
        \begin{aligned}
            &U(\theta)-\left(V(x(\hat{\theta}))-\theta w(\hat{\theta})-p(\hat{\theta})\right)\\
            &\qquad =U(\theta)-U(\hat{\theta})-(\hat{\theta}-\theta)w(\hat{\theta})\\
            &\qquad =\int_{\theta}^{\hat{\theta}}\left(w(c)-w(\hat{\theta})\right)dc\geq 0.
        \end{aligned}
    \end{equation*}
    If instead $\hat{\theta}<\theta$, then
    \begin{equation*}
        \begin{aligned}
            &U(\theta)-\left(V(x(\hat{\theta}))-\theta w(\hat{\theta})-p(\hat{\theta})\right)\\
            &\qquad =U(\theta)-U(\hat{\theta})-(\hat{\theta}-\theta)w(\hat{\theta})\\
            &\qquad =\int_{\hat{\theta}}^{\theta}\left(w(\hat{\theta})-w(c)\right)dc\geq 0.
        \end{aligned}
    \end{equation*}
    Thus truthful reporting is optimal for every type, and the truthful utility is exactly $U(\theta)$. Since $U(\theta)\geq U(\bar{\theta})=0$, the mechanism is individually rational.

    Finally, consider any other incentive compatible and individually rational policy. Since
    \begin{equation*}
        (\hat{\theta}-\theta)w(\hat{\theta})\leq U(\theta)-U(\hat{\theta})\leq (\hat{\theta}-\theta)w(\theta),
    \end{equation*}
    letting $\hat \theta \rightarrow\theta$ implies $U'(\theta)=-w (\theta)$ for $\theta\in[\underline{\theta},\bar\theta]$ almost everywhere. Therefore,
    \begin{equation*}
        U(\theta)-U(\bar{\theta})=\int_{\theta}^{\bar{\theta}}w(c)\,dc,
    \end{equation*}
    Hence every incentive compatible payment schedule satisfies
    \begin{equation*}
        p(\theta)=V(x(\theta))-\theta w(\theta)-U(\bar{\theta})-\int_{\theta}^{\bar{\theta}} w(c)\,dc,
    \end{equation*}
    where individual rationality requires $U(\bar{\theta})\geq 0$.
    \hfill{\Halmos}
\end{proof}

\begin{proof}{Proof of Proposition~\ref{proposition:continuous_types_welfare_maximization_decentralized_case}.}

We only consider the case in which the centralized optimal policy $\left(x^*(\cdot), w^*(\cdot)\right)$ exists. Since $x^*(\cdot)$ is a non-increasing function, for $\theta_1>\theta_2$,
\begin{equation*}
    \begin{aligned}
        w^*(\theta_1)&=\frac{\lambda\int_{\theta_1}^{\bar{\theta}}{x^*}^2(c)dF(c)}{\left(1-\lambda\int_{\theta_1}^{\bar{\theta}}{x^*}(c)dF(c)\right)^2}+\frac{{x^*}(\theta_1)}{1-\lambda\int_{\theta_1}^{\bar{\theta}}{x^*}(c)dF(c)}\\
        &\leq \frac{\lambda\int_{\theta_2}^{\bar{\theta}}{x^*}^2(c)dF(c)}{\left(1-\lambda\int_{\theta_2}^{\bar{\theta}}{x^*}(c)dF(c)\right)^2}+\frac{{x^*}(\theta_2)}{1-\lambda\int_{\theta_2}^{\bar{\theta}}{x^*}(c)dF(c)}=w^*(\theta_2).
    \end{aligned}
\end{equation*}

Therefore, $w^*(\cdot)$ is a non-increasing function. By Lemma~\ref{lemma:continuous_types_incentive_compatibility}, there exists a price menu that is incentive compatible.

Next, we prove that $p^*(\theta)\geq0$. Take $x^*(\cdot)$ to be right-continuous on $[\underline{\theta},\bar{\theta}]$. Since $x^*(\cdot)$ is non-increasing and finite-valued, it is bounded, and $w^*(\cdot)$ given by \eqref{equation:continuous_types_sojourn_time} is also bounded and continuous.

Fix $\theta\in[\underline{\theta},\bar{\theta})$. For $\epsilon\in(0,1-F(\theta))$, consider an alternative service depth assignment
\begin{equation*}
    \tilde x(c)=\left\{
    \begin{aligned}
        &x^*(c),&c\leq\theta,\\
        &x^*\left(F^{-1}(F(c)+\epsilon)\right),&\theta<c\leq F^{-1}(1-\epsilon),\\
        &0,&c>F^{-1}(1-\epsilon).
    \end{aligned}
    \right.
\end{equation*}
Since $F^{-1}(F(c)+\epsilon)\geq c$ and $x^*(\cdot)$ is non-increasing, $\tilde x(\cdot)$ is non-increasing and $0\leq\tilde x(c)\leq x^*(c)$. Therefore, $\tilde x(\cdot)$ is admissible.

Denote by $\tilde w(\cdot)$ the corresponding sojourn time given by \eqref{equation:continuous_types_sojourn_time}. For any $c\in(\theta,F^{-1}(1-\epsilon)]$ and $k\in\{1,2\}$,
\begin{equation*}
\begin{aligned}
\int_c^{\bar{\theta}}\tilde x^k(s)dF(s)&=\int_c^{F^{-1}(1-\epsilon)}{x^*}^k\left(F^{-1}(F(s)+\epsilon)\right)dF(s)\\
&=\int_{F(c)+\epsilon}^{1}{x^*}^k(F^{-1}(u))du\\&
=\int_{F^{-1}(F(c)+\epsilon)}^{\bar{\theta}}{x^*}^k(s)dF(s).
\end{aligned}
\end{equation*}
It follows from \eqref{equation:continuous_types_sojourn_time} that $\tilde w(c)=w^*(F^{-1}(F(c)+\epsilon))$ for $\theta<c\leq F^{-1}(1-\epsilon)$, and $\tilde w(c)=0$ for $c>F^{-1}(1-\epsilon)$. Moreover, for $c\leq\theta$, the service depth is unchanged while both higher-priority workload integrals are weakly smaller, so $\tilde w(c)\leq w^*(c)$.

By the optimality of $x^*(\cdot)$,
\begin{equation*}
\begin{aligned}
0&\leq\int_{\underline{\theta}}^{\bar{\theta}}\left[V(x^*(c))-cw^*(c)-V(\tilde x(c))+c\tilde w(c)\right]dF(c)\\
&\leq\int_{\theta}^{\bar{\theta}}\left[V(x^*(c))-cw^*(c)-V(\tilde x(c))+c\tilde w(c)\right]dF(c),
\end{aligned}
\end{equation*}
where the second inequality follows because $\tilde x(c)=x^*(c)$ and $\tilde w(c)\leq w^*(c)$ for $c\leq\theta$.

Using the definitions of $\tilde x(\cdot)$ and $\tilde w(\cdot)$ and changing variables from $c$ to $u=F(c)$, we obtain
\begin{equation*}
\begin{aligned}
0\leq&\int_{F(\theta)}^{F(\theta)+\epsilon}\left[V(x^*(F^{-1}(u)))-F^{-1}(u)w^*(F^{-1}(u))\right]du\\&-\int_{F(\theta)+\epsilon}^{1}\left[F^{-1}(u)-F^{-1}(u-\epsilon)\right]w^*(F^{-1}(u))du.
\end{aligned}
\end{equation*}
Dividing by $\epsilon$ gives
\begin{equation}
\label{equation:payment_nonnegative}
\begin{aligned}
0\leq&\frac{1}{\epsilon}\int_{F(\theta)}^{F(\theta)+\epsilon}\left[V(x^*(F^{-1}(u)))-F^{-1}(u)w^*(F^{-1}(u))\right]du\\&-\frac{1}{\epsilon}\int_{F(\theta)+\epsilon}^{1}\left[F^{-1}(u)-F^{-1}(u-\epsilon)\right]w^*(F^{-1}(u))du.
\end{aligned}
\end{equation}

By the right-continuity of $x^*(\cdot)$ and $w^*(\cdot)$,
\begin{equation*}
\lim_{\epsilon\downarrow0}\frac{1}{\epsilon}\int_{F(\theta)}^{F(\theta)+\epsilon}\left[V(x^*(F^{-1}(u)))-F^{-1}(u)w^*(F^{-1}(u))\right]du=V(x^*(\theta))-\theta w^*(\theta).
\end{equation*}
Indeed, if $x^*(\theta)>0$, the result follows from the continuity of the finite concave function $V$ on $(0,\infty)$; if $x^*(\theta)=0$, monotonicity implies $x^*(c)=w^*(c)=0$ for every $c\geq\theta$, so the result also holds.

Since $f(c)>0$, $F$ is continuous and strictly increasing, and hence $F^{-1}$ is continuous. Using $F^{-1}(u)-F^{-1}(u-\epsilon)=\int_{(u-\epsilon,u]}dF^{-1}(v)$ and Fubini's theorem, we have
\begin{equation*}
\begin{aligned}
&
\frac{1}{\epsilon}\int_{F(\theta)+\epsilon}^{1}\left[F^{-1}(u)-F^{-1}(u-\epsilon)\right]w^*(F^{-1}(u))du
=\int_{(F(\theta),1]}\left[\frac{1}{\epsilon}\int_{\max\{F(\theta)+\epsilon,v\}}^{\min\{1,v+\epsilon\}}w^*(F^{-1}(u))du\right]dF^{-1}(v).
\end{aligned}
\end{equation*}
For every $v\in(F(\theta),1)$, the expression in brackets converges to $w^*(F^{-1}(v))$ by continuity, and it is bounded by $\sup_{c\in[\theta,\bar{\theta}]}w^*(c)=w^*(\theta)$. Therefore, the dominated convergence theorem implies
\begin{equation*}
\lim_{\epsilon\downarrow0}\frac{1}{\epsilon}\int_{F(\theta)+\epsilon}^{1}\left[F^{-1}(u)-F^{-1}(u-\epsilon)\right]w^*(F^{-1}(u))du=\int_{F(\theta)}^{1}w^*(F^{-1}(u))dF^{-1}(u)=\int_{\theta}^{\bar{\theta}}w^*(c)dc.
\end{equation*}

Letting $\epsilon\downarrow0$ in \eqref{equation:payment_nonnegative} gives
\begin{equation*}
p^*(\theta)=V(x^*(\theta))-\theta w^*(\theta)-\int_{\theta}^{\bar{\theta}}w^*(c)dc\geq0,\quad\theta\in[\underline{\theta},\bar{\theta}].
\end{equation*}
which finishes the proof.
\hfill{\Halmos}
\end{proof}

\section{Proofs for Section~\ref{section:extensions}}
\label{appendix:section_5}

%In the centralized setting, for any allocation that gives customers with delay sensitivity $\theta$ non-negative surplus, the decision maker can charge $p(\theta)=V(x(\theta))-\theta w(\theta)$ to extract the customer's entire surplus. Therefore, revenue maximization in the centralized setting coincides with social welfare maximization. It remains to discuss the revenue maximizing mechanism in the decentralized setting.
\subsection{Proofs of Proposition~\ref{proposition:revenue_maximization} and Proposition~\ref{proposition:continuous_type_revenue_reduction}}
\begin{proof}{Proof of Proposition~\ref{proposition:revenue_maximization}.}
Fix an operationally feasible allocation $(x_1,x_2,w_1,w_2)$. By Lemma~\ref{lemma:welfare_maximization_incentive_compatibility}, the allocation is implementable if and only if $w_1\leq w_2$. Let $U_i=V(x_i)-\theta_iw_i-p_i$ denote the truthful surplus of type $i$. The IC constraints are equivalent to
\begin{equation*}
    U_2-(\theta_1-\theta_2)w_2\leq U_1 \leq U_2-(\theta_1-\theta_2)w_1,
\end{equation*}
and the IR constraints require $U_i\geq 0$. Since revenue equals
\begin{equation*}
    \lambda_1(V(x_1)-\theta_1w_1-U_1)+\lambda_2(V(x_2)-\theta_2w_2-U_2),
\end{equation*}
the revenue-maximizing utilities for the fixed allocation are $U_1=0$ and $U_2=(\theta_1-\theta_2)w_1$. Therefore, the revenue-maximizing payments are
\begin{equation*}
    p_1=V(x_1)-\theta_1w_1,\quad p_2=V(x_2)-\theta_2w_2-(\theta_1-\theta_2)w_1.
\end{equation*}
Substituting these payments into the revenue objective gives
\begin{equation*}
    \begin{aligned}
    \lambda_1p_1+\lambda_2p_2=\lambda_1\left(V(x_1)-\left(\theta_1+\frac{\lambda_2}{\lambda_1}(\theta_1-\theta_2)\right)w_1\right)+\lambda_2\left(V(x_2)-\theta_2w_2\right).
    \end{aligned}
\end{equation*}
Thus the revenue-maximization problem has the same feasible set as the decentralized welfare problem, but with virtual delay sensitivities
\begin{equation*}
    \varphi(\theta_1)=\theta_1+\frac{\lambda_2}{\lambda_1}(\theta_1-\theta_2), \quad\varphi(\theta_2)=\theta_2.
\end{equation*}
\hfill{\Halmos}
\end{proof}

\begin{proof}{Proof of Proposition \ref{proposition:continuous_type_revenue_reduction}.}
    The revenue-maximizing payment schedule for the fixed allocation rule is
    \begin{equation*}
        p(\theta)=V(x(\theta))-\theta w(\theta)-\int_{\theta}^{\bar{\theta}} w(c)\,dc.
    \end{equation*}
    Therefore,
    \begin{equation*}
        \begin{aligned}
            \int_{\underline{\theta}}^{\bar{\theta}} p(\theta)\,dF(\theta)
            =\int_{\underline{\theta}}^{\bar{\theta}}\left(V(x(\theta))-\theta w(\theta)\right)dF(\theta)-\int_{\underline{\theta}}^{\bar{\theta}}\left(\int_{\theta}^{\bar{\theta}}w(c)\,dc\right)dF(\theta).
        \end{aligned}
    \end{equation*}
    Applying Fubini's theorem to the last term gives
    \begin{equation*}
        \begin{aligned}
            \int_{\underline{\theta}}^{\bar{\theta}}\left(\int_{\theta}^{\bar{\theta}}w(c)\,dc\right)dF(\theta)
            =\int_{\underline{\theta}}^{\bar{\theta}}F(c)w(c)\,dc
            =\int_{\underline{\theta}}^{\bar{\theta}}\frac{F(c)}{f(c)}w(c)\,dF(c).
        \end{aligned}
    \end{equation*}
    Thus, 
    \begin{equation*}
        \begin{aligned}
            \int_{\underline{\theta}}^{\bar{\theta}} p(\theta)\,dF(\theta)
            &=\int_{\underline{\theta}}^{\bar{\theta}}\left(V(x(\theta))-\theta w(\theta)\right)dF(\theta)-\int_{\underline{\theta}}^{\bar{\theta}}\frac{F(\theta)}{f(\theta)}w(\theta)\,dF(\theta)\\
            &=\int_{\underline{\theta}}^{\bar{\theta}}\left(V(x(\theta))-\varphi(\theta) w(\theta)\right)dF(\theta).
        \end{aligned}
    \end{equation*}
    Multiplying by $\lambda$ gives the long-term average revenue. The optimization problem is therefore equivalent to maximizing long-term social welfare under virtual delay sensitivities.
    \hfill{\Halmos}
\end{proof}

\subsection{Proof of Proposition~\ref{proposition:randomized}}
\begin{proof}{Proof of Proposition~\ref{proposition:randomized}.}
    Fix a feasible randomized service depth assignment $(\mathcal{G}_1,\mathcal{G}_2)$ and a priority rule, and let $w_i(x)$ denote the expected sojourn time of a type-$i$ customer with realized service depth $x$. Recall that $q=\lambda_1/\lambda$. Let
    \begin{equation*}
        G(z):=q\,\mathcal{G}_1(z)+(1-q)\,\mathcal{G}_2(z)
    \end{equation*}
    denote the aggregate distribution of service depth across all arrivals, and let $G^{-1}(u):=\inf\{z\geq 0:G(z)\geq u\}$ be its generalized inverse. Let $U$ be uniformly distributed on $[0,1]$. Define $\tilde{\mathcal{G}}_1$ as the distribution of $G^{-1}(U)$ conditional on $U\leq q$, and $\tilde{\mathcal{G}}_2$ as the distribution of $G^{-1}(U)$ conditional on $U>q$. That is, type~$1$ customers receive the fraction $q$ of smallest service depths in the aggregate distribution, and type~$2$ customers receive the rest. By construction,
    \begin{equation*}
        \sup\operatorname{supp}(\tilde{\mathcal{G}}_1)\leq G^{-1}(q)\leq\inf\operatorname{supp}(\tilde{\mathcal{G}}_2),
    \end{equation*}
    and the aggregate distribution is unchanged, i.e., $q\,\tilde{\mathcal{G}}_1+(1-q)\,\tilde{\mathcal{G}}_2=G$. Therefore, $(\tilde{\mathcal{G}}_1,\tilde{\mathcal{G}}_2)$ is feasible, and the value term
    \begin{equation*}
        \lambda_1\int V(x)\,d\tilde{\mathcal{G}}_1(x)+\lambda_2\int V(x)\,d\tilde{\mathcal{G}}_2(x)=\lambda\int V(x)\,dG(x)
    \end{equation*}
    is unchanged.

    Under $(\tilde{\mathcal{G}}_1,\tilde{\mathcal{G}}_2)$, type~$1$ customers have the larger delay sensitivity and weakly smaller service depths, so the $c\mu$-rule gives priority to smaller realized service depth, with type~$1$ customers served first when the depths are equal. Let $\tilde w_i(x)$ denote the expected sojourn times under this rule. Since a customer with a larger service depth has a longer service time and is preempted by more customers, $\tilde w_i(x)$ is non-decreasing in $x$. Hence every type~$1$ customer has a weakly shorter expected sojourn time than every type~$2$ customer.

    It remains to compare the delay costs. Let $S_i:=\lambda_i\int w_i(x)\,d\mathcal{G}_i(x)$ and $\tilde S_i:=\lambda_i\int \tilde w_i(x)\,d\tilde{\mathcal{G}}_i(x)$ denote the aggregate sojourn times of type~$i$ customers. The long-term average delay cost can be written as
    \begin{equation*}
        \theta_1S_1+\theta_2S_2=\theta_2(S_1+S_2)+(\theta_1-\theta_2)S_1.
    \end{equation*}
    First, both systems have the same arrival rate $\lambda$ and the same aggregate service depth distribution $G$. By the $c\mu$-rule with a common delay cost, the aggregate sojourn time of all customers is minimized by giving priority to smaller service depths. Therefore, $\tilde S_1+\tilde S_2\leq S_1+S_2$.

    Second, let $G_1(z):=q\,\mathcal{G}_1(z)$ and $G_1^{-1}(u):=\inf\{z\geq 0:G_1(z)\geq u\}$ for $u\in[0,q]$. Since $G_1(z)\leq G(z)$, we have $G^{-1}(u)\leq G_1^{-1}(u)$ for all $u\in[0,q]$. Under $(\tilde{\mathcal{G}}_1,\tilde{\mathcal{G}}_2)$, type~$1$ customers have priority over all type~$2$ customers, so by the same argument as in \eqref{equation:tail1},
    \begin{equation*}
        \begin{aligned}
            \tilde S_1&=\lambda\int_0^{q}\left[\frac{\lambda\int_{0}^{\alpha}(G^{-1}(u))^2du}{\left(1-\lambda\int_{0}^{\alpha}G^{-1}(u)du\right)^2}+\frac{G^{-1}(\alpha)}{1-\lambda\int_{0}^{\alpha}G^{-1}(u)du}\right]d\alpha\\
            &\leq \lambda\int_0^{q}\left[\frac{\lambda\int_{0}^{\alpha}(G_1^{-1}(u))^2du}{\left(1-\lambda\int_{0}^{\alpha}G_1^{-1}(u)du\right)^2}+\frac{G_1^{-1}(\alpha)}{1-\lambda\int_{0}^{\alpha}G_1^{-1}(u)du}\right]d\alpha.
        \end{aligned}
    \end{equation*}
    On the other hand, consider the subsystem consisting only of type~$1$ customers with service depth distribution $\mathcal{G}_1$. By the same argument as in \eqref{equation:tail2}, removing type~$2$ customers cannot increase the sojourn times of type~$1$ customers, and the minimum aggregate sojourn time in the isolated subsystem is exactly the right-hand side above. Therefore, $\tilde S_1\leq S_1$.

    Combining the two inequalities with $\theta_1>\theta_2>0$, the delay cost under $(\tilde{\mathcal{G}}_1,\tilde{\mathcal{G}}_2)$ is weakly lower, while the value term is unchanged. Thus the long-term average social welfare is weakly higher.
    \hfill{\Halmos}
\end{proof}

\subsection{Proofs of Lemma~\ref{lemma:randomized_ic} and Proposition~\ref{proposition:randomized_no_gap}}
\begin{proof}{Proof of Lemma~\ref{lemma:randomized_ic}.}
    Let $\mathcal{X}=\operatorname{supp}(\mathcal{G}_1)\cup\operatorname{supp}(\mathcal{G}_2)$. The assignment is implementable if there exist prices $p(\cdot)$ such that, for $i=1,2$, every $x\in\operatorname{supp}(\mathcal{G}_i)$ maximizes the surplus of type~$i$ customers over $\mathcal{X}$ and yields a non-negative surplus.

    We first prove necessity. Fix $x_1\in\operatorname{supp}(\mathcal{G}_1)$ and $x_2\in\operatorname{supp}(\mathcal{G}_2)$. Since type~$1$ customers weakly prefer $x_1$ and type~$2$ customers weakly prefer $x_2$,
    \begin{equation*}
        \begin{aligned}
            &V(x_1)-\theta_1w(x_1)-p(x_1)\geq V(x_2)-\theta_1w(x_2)-p(x_2),\\
            &V(x_2)-\theta_2w(x_2)-p(x_2)\geq V(x_1)-\theta_2w(x_1)-p(x_1).
        \end{aligned}
    \end{equation*}
    Adding the two inequalities gives $(\theta_1-\theta_2)(w(x_2)-w(x_1))\geq 0$, hence $w(x_1)\leq w(x_2)$.

    Next we prove sufficiency. Suppose $w(x_1)\leq w(x_2)$ for all $x_1\in\operatorname{supp}(\mathcal{G}_1)$ and $x_2\in\operatorname{supp}(\mathcal{G}_2)$, so that $w(x_1)\leq\bar w_1\leq w(x_2)$. If some $x$ belongs to both supports, then $w(x)=\bar w_1$ and the two expressions for $p(x)$ coincide, so the price rule is well defined. Under this price rule, a type~$1$ customer obtains surplus $0$ from any $x\in\operatorname{supp}(\mathcal{G}_1)$ and surplus
    \begin{equation*}
        V(x)-\theta_1w(x)-p(x)=(\theta_1-\theta_2)(\bar w_1-w(x))\leq 0
    \end{equation*}
    from any $x\in\operatorname{supp}(\mathcal{G}_2)$. A type~$2$ customer obtains surplus $(\theta_1-\theta_2)\bar w_1\geq 0$ from any $x\in\operatorname{supp}(\mathcal{G}_2)$ and surplus
    \begin{equation*}
        V(x)-\theta_2w(x)-p(x)=(\theta_1-\theta_2)w(x)\leq(\theta_1-\theta_2)\bar w_1
    \end{equation*}
    from any $x\in\operatorname{supp}(\mathcal{G}_1)$. Therefore, every option in $\operatorname{supp}(\mathcal{G}_i)$ is optimal for type~$i$ customers and gives a non-negative surplus, so the assignment is implementable.
    \hfill{\Halmos}
\end{proof}

\begin{proof}{Proof of Proposition~\ref{proposition:randomized_no_gap}.}
    Combining Proposition~\ref{proposition:randomized} and  Lemma~\ref{lemma:randomized_ic} directly gives the result.
    \hfill{\Halmos}
\end{proof}

\end{APPENDIX}

\end{document}